# ESA Science Programme Missions: Contributions and Exploitation – ESA Mission Publications

**Guido De Marchi · Arvind N. Parmar**

**Abstract** We examine over 68,000 refereed publications based on data from 25 missions in the ESA Science Programme and 11 additional missions in which ESA is involved as a junior partner. The publications cover the fields of astronomy, planetary science, and heliophysics and are spread over almost 50 years, spanning the period between the year a mission was launched and the end of 2021. We study the number of papers as a function of time and the evolution of several metrics, including citations and other indices. We also investigate the geographical distribution of the authors, and for ESA Member States we correlate the various indices with the level of financial contribution of the individual countries to the ESA Science Programme. We find that in general the involvement of the scientific communities in the various Member States follows the distribution expected from the countries' gross domestic products, with communities in some field and countries, both large and small, being particularly effective at turning data into scientific discoveries. We also analyse the differences between papers written by investigators directly involved in the provision of the payloads or in the definition of the scientific projects and those written by other scientists not directly involved in the process. We find that the latter, the so-called "archival papers", represent more than 50 % of the literature based on data from ESA Space Science missions, and have a similar impact on the literature in the respective fields, as judged by the number of citations. This highlights the importance of sharing and preserving the scientific data produced by the missions.

G. De Marchi
Directorate of Science, ESA, ESTEC
Noordwijk, The Netherlands.
E-mail: gdemarchi@esa.int

A. Parmar
Former Head of the Science Support Office
Directorate of Science, ESA, ESTEC
The Netherlands
*Present address:*
Department of Space and Climate Physics
MSSL/UCL
Dorking
UK

## 1 Introduction

A well-established method of assessing the scientific productivity of an astronomical facility is to perform a quantitative analysis of the publications that have made use of data from that facility. For a reliable comparison across facilities, it is necessary to use publication libraries created with uniform inclusion criteria. In this chapter, we present the analysis of the number and impact of the publications, with the aid of a number of tables and figures. The chapter is structured as follows: Sect. 2 addresses the definition and contents of the ESA publication libraries; Sect. 2.1 provides overall publication numbers and discusses the geographical distribution of authors' affiliations; Sect. 3 presents standard impact metrics, for the overall ESA Science Programme and for the individual missions, together with an analysis of their temporal evolution; Sect. 4 explores papers that use data from more than one mission; Sect. 5 discusses the concept of archival publications and provides statistics underlying their contribution to the Space Science literature. Conclusions are presented in Sect. 6.

## 2 Publication Libraries

The European Space Agency (ESA) mission publication libraries are found under: `https://www.cosmos.esa.int/web/guest/mission-publications`. There are 25 libraries covering all ESA-led missions from COS-B, launched in 1975, to Solar Orbiter, launched in 2020 (there are also libraries for the JUICE and Euclid missions, which are not covered in this work). The International Ultraviolet Explorer (IUE) was a collaboration between National Aeronautics and Space Administration (NASA), the Science and Engineering Research Council in the United Kingdom and ESA is included in these libraries as an ESA-led mission since ESA took over operational responsibility towards the end of the operational lifetime. In addition, there are 11 libraries for missions where ESA is a junior partner. The libraries are complete beyond the end of December 2023, but only publications appearing before the end of 2021 are included in this analysis. With the exception of Hubble Space Telescope (HST), each publication library is maintained by its project scientist, often with the support of Science Operations Centre (SOC) staff. For missions in their archival phase, a "contact scientist" is assigned who is responsible for maintaining the publication library. Space Telescope Science Institute (STScI) staff maintain a list of HST publications which can be accessed via a link from the above Uniform Resource Locator (URL). The libraries make use of the Smithsonian Astrophysical Observatory (SAO)/NASA Astrophysics Data Service (ADS) which is a digital library for researchers in physics and astronomy operated by the SAO under a NASA grant [1, 2, 3, 4]. An important advantage of using ADS is the extensive range of statistical analysis and database selection tools that are provided.

Refereed publications, published after launch, are included in the ESA libraries if they fulfil one or more of the following conditions:

1. Make direct use of data from a mission including from its primary catalogues.
2. Make quantitative predictions of results from a mission.
3. Describe a mission, its instruments, operations, software or calibrations.

Thus, publications that only refer to published results from a mission, and do not fulfil one of the criteria listed above, are not included in the mission publication libraries, but are counted as citations. Although the publications may vary in quality, depth and importance, their number provides an easily quantifiable (although not homogeneous qualitatively) assessment of the productivity of each of the Science Programme missions. Note that a publication will appear in multiple libraries if it makes direct use of data from more that one ESA mission.

Libraries of Doctor of Philosophy Degree (PhD)s from some of the missions are also available under the above web address. The PhD libraries include degrees awarded before launch so that PhDs describing developments prior to launch can be included.

Another very useful resource is the NASA's High-Energy Astrophysics Archive Centre (HEASARC) bibliographical database which includes links to ADS publication libraries for the high-energy astronomy and cosmic microwave background missions supported by the HEASARC (`https://heasarc.gsfc.nasa.gov/docs/heasarc/biblio/pubs/`).

**Table 1** The number of refereed publications per year for the ESA-led Science Programme missions using the inclusion criteria given in Sect. 2. The missions are arranged in order of launch date. For missions still being operated, only the launch date is given under "Operations". The total number of refereed publications until the end of 2021 for each mission is given under "Total Pubs".

| Mission | Operations | Year | | | | | | | | | | | |
|---|---|---|---|---|---|---|---|---|---|---|---|---|---|
| | | <2000 | 2000 | 2001 | 2002 | 2003 | 2004 | 2005 | 2006 | 2007 | 2008 | 2009 | 2010 |
| COS-B | 1975-1982 | 142 | 1 | | | | | | | | | | |
| IUE | 1978-1996 | 3713 | 115 | 115 | 98 | 85 | 80 | 95 | 60 | 63 | 60 | 51 | 65 |
| Exosat | 1983-1986 | 706 | 8 | 3 | 4 | | | 2 | 1 | 2 | 1 | 2 | |
| Giotto | 1985-1992 | 246 | 1 | | 3 | 4 | | | 2 | 2 | | 4 | |
| Hipparcos | 1989-1993 | 714 | 186 | 151 | 144 | 130 | 104 | 96 | 113 | 88 | 87 | 87 | 85 |
| Ulysses | 1990-2009 | 898 | 83 | 138 | 39 | 89 | 57 | 56 | 36 | 42 | 52 | 38 | 59 |
| ISO | 1995-1998 | 490 | 180 | 138 | 146 | 129 | 123 | 106 | 67 | 35 | 33 | 76 | 83 |
| SOHO | 1995- | 596 | 286 | 200 | 284 | 299 | 319 | 320 | 273 | 364 | 336 | 322 | 274 |
| Huygens | 1997-2005 | | 7 | 8 | 15 | 7 | 12 | 24 | 21 | 26 | 26 | 16 | 15 |
| XMM-Newton | 1999- | | 23 | 90 | 108 | 238 | 347 | 329 | 349 | 361 | 337 | 366 | 365 |
| Cluster | 2000- | 54 | 1 | 38 | 22 | 65 | 126 | 171 | 145 | 123 | 183 | 175 | 193 |
| INTEGRAL | 2002- | | | | 5 | 91 | 42 | 89 | 139 | 112 | 120 | 139 | 129 |
| SMART-1 | 2003-2006 | | | | | | 2 | 7 | 10 | 4 | 8 | 10 | 2 |
| Mars Express | 2003- | | | | | 4 | 21 | 49 | 87 | 75 | 83 | 90 | 110 |
| Rosetta | 2004-2016 | | | | | | 23 | 19 | 35 | 59 | 30 | 21 | 41 |
| Venus Express | 2005-2014 | | | | | | | | 20 | 39 | 72 | 44 | 38 |
| Herschel | 2009-2013 | | | | | | | | | | | | 228 |
| Planck | 2009-2013 | | | | | | | | | | | | |
| PROBA-2 | 2009- | | | | | | | | | | | | |
| Gaia | 2013- | | | | | | | | | | | | |
| LISA Pathfinder | 2015-2017 | | | | | | | | | | | | |
| ExoMars 2016 | 2016- | | | | | | | | | | | | |
| BepiColombo | 2018- | | | | | | | | | | | | |
| CHEOPS | 2019- | | | | | | | | | | | | |
| Solar Orbiter | 2020- | | | | | | | | | | | | |

| Mission | Operations | Year | | | | | | | | | | | Total |
|---|---|---|---|---|---|---|---|---|---|---|---|---|---|
| | | 2011 | 2012 | 2013 | 2014 | 2015 | 2016 | 2017 | 2018 | 2019 | 2020 | 2021 | Pubs |
| COS-B | 1975-1982 | | | | | 2 | 1 | | | | | | 146 |
| IUE | 1978-1996 | 44 | 49 | 54 | 49 | 55 | 46 | 41 | 39 | 47 | 45 | 36 | 5105 |
| Exosat | 1983-1986 | | 1 | 2 | 3 | 2 | 1 | 1 | 2 | 1 | 1 | | 743 |
| Giotto | 1985-1992 | 2 | | 1 | | 1 | | | | | | | 266 |
| Hipparcos | 1989-1993 | 67 | 88 | 71 | 58 | 60 | 54 | 57 | 71 | 62 | 42 | 77 | 2692 |
| Ulysses | 1990-2005 | 34 | 43 | 32 | 34 | 30 | 14 | 36 | 24 | 35 | 33 | 48 | 1950 |
| ISO | 1995-1998 | 77 | 48 | 41 | 58 | 35 | 26 | 38 | 42 | 28 | 31 | 32 | 2062 |
| SOHO | 1995- | 298 | 245 | 184 | 213 | 218 | 171 | 210 | 181 | 169 | 230 | 260 | 6252 |
| Huygens | 1997-2005 | 8 | 11 | 7 | 5 | | 6 | 3 | 1 | 1 | 1 | 2 | 222 |
| XMM-Newton | 1999- | 372 | 326 | 364 | 361 | 337 | 388 | 395 | 397 | 385 | 359 | 366 | 6963 |
| Cluster | 2000- | 234 | 187 | 174 | 151 | 175 | 159 | 137 | 127 | 108 | 88 | 126 | 2962 |
| INTEGRAL | 2002- | 93 | 114 | 78 | 101 | 95 | 93 | 105 | 76 | 84 | 84 | 115 | 1904 |
| SMART-1 | 2003-2006 | 5 | 6 | 2 | 4 | 2 | 1 | 2 | 4 | 1 | | | 70 |
| Mars Express | 2003- | 97 | 82 | 68 | 99 | 101 | 64 | 88 | 98 | 70 | 62 | 116 | 1464 |
| Rosetta | 2004-2016 | 26 | 53 | 32 | 50 | 157 | 188 | 177 | 102 | 131 | 100 | 70 | 1314 |
| Venus Express | 2005-2014 | 53 | 72 | 49 | 40 | 100 | 61 | 46 | 31 | 22 | 25 | 25 | 737 |
| Herschel | 2009-2013 | 109 | 256 | 323 | 347 | 301 | 305 | 276 | 248 | 234 | 250 | 407 | 3284 |
| Planck | 2009-2013 | 48 | 68 | 118 | 326 | 327 | 314 | 299 | 297 | 336 | 307 | 302 | 2742 |
| PROBA-2 | 2009- | 7 | 6 | 26 | 15 | 11 | 18 | 9 | 12 | 9 | 11 | 13 | 137 |
| Gaia | 2013- | | | | 31 | 37 | 63 | 403 | 881 | 1633 | 1692 | 1663 | 6403 |
| LISA Pathfinder | 2015-2017 | | | | | | 3 | 5 | 10 | 8 | 3 | 4 | 33 |
| ExoMars 2016 | 2016- | | | | | | 9 | 28 | 25 | 20 | 14 | 42 | 138 |
| BepiColombo | 2018- | | | | | | | | 4 | 11 | 28 | 28 | 71 |
| CHEOPS | 2019- | | | | | | | | | 11 | 13 | 14 | 38 |
| Solar Orbiter | 2020- | | | | | | | | | | 30 | 84 | 114 |

**Table 2** The number of refereed publications per year for the partner-led Science Programme missions using the inclusion criteria given in Sect. 2. The missions are arranged in order of launch date. For missions still being operated, only the launch date is given under "Operations". The total number of refereed publications until the end of 2021 for each mission is given under "Total Pubs".

| Mission | Operations | Year | | | | | | | | | | | |
|---|---|---|---|---|---|---|---|---|---|---|---|---|---|
| | | <2000 | 2000 | 2001 | 2002 | 2003 | 2004 | 2005 | 2006 | 2007 | 2008 | 2009 | 2010 |
| HST | 1990- | 2330 | 537 | 560 | 602 | 607 | 611 | 686 | 715 | 733 | 705 | 680 | 734 |
| Cassini | 1997-2017 | 22 | 1 | 16 | 18 | 11 | 50 | 109 | 149 | 157 | 172 | 201 | 234 |
| Double Star | 2003-2007 | | | | | | 1 | 26 | 6 | 12 | 40 | 11 | 14 |
| Suzaku | 2005-2015 | | | | | | | 19 | 49 | 78 | 139 | 92 | |
| AKARI | 2006-2011 | | | | | | | | 1 | 27 | 20 | 19 | 48 |
| Hinode | 2006- | | | | | | | | | 63 | 104 | 134 | 134 |
| CoRoT | 2006-2013 | | | | | | | | | 5 | 14 | 39 | 44 |
| Chandrayaan-1 | 2008-2009 | | | | | | | | | | 3 | 13 | 15 |
| IRIS | 2013- | | | | | | | | | | | | |
| Hitomi | 2016-2016 | | | | | | | | | | | | |
| MICROSCOPE | 2016-2018 | | | | | | | | | | | | |

| Mission | Operations | Year | | | | | | | | | | | Total |
|---|---|---|---|---|---|---|---|---|---|---|---|---|---|
| | | 2011 | 2012 | 2013 | 2014 | 2015 | 2016 | 2017 | 2018 | 2019 | 2020 | 2021 | Pubs |
| HST | 1990- | 789 | 848 | 785 | 820 | 853 | 883 | 914 | 968 | 1015 | 929 | 1029 | 19333 |
| Cassini | 1997-2017 | 184 | 177 | 134 | 127 | 83 | 97 | 56 | 72 | 71 | 40 | 46 | 2227 |
| Double Star | 2003-2007 | 12 | 6 | 4 | 5 | 4 | 4 | 2 | 1 | 3 | 3 | 7 | 161 |
| Suzaku | 2005-2015 | 109 | 103 | 95 | 82 | 85 | 97 | 62 | 46 | 38 | 26 | 19 | 1139 |
| AKARI | 2006-2011 | 89 | 115 | 94 | 72 | 64 | 69 | 129 | 89 | 99 | 76 | 75 | 1086 |
| Hinode | 2006- | 133 | 129 | 100 | 96 | 98 | 86 | 82 | 89 | 59 | 63 | 72 | 1442 |
| CoRoT | 2006-2013 | 18 | 21 | 20 | 15 | 15 | 7 | 10 | 12 | 8 | 7 | 3 | 238 |
| Chandrayaan-1 | 2008-2009 | 18 | 21 | 20 | 15 | 15 | 7 | 10 | 12 | 8 | 7 | 3 | 164 |
| IRIS | 2013- | | | 3 | 22 | 63 | 62 | 63 | 68 | 59 | 59 | 50 | 449 |
| Hitomi | 2016-2016 | | | | | | 14 | 5 | 45 | 4 | 4 | 2 | 74 |
| MICROSCOPE | 2016-2018 | | | | | | 2 | 1 | 6 | 5 | 1 | 2 | 17 |

## 2.1 Overall Publication Numbers and Authors Affiliations

Table 1 and Fig. 1 show the number of refereed publications for the ESA-led Science Programme missions for each year in the period considered here, and Table 2 and Fig. 2 the same for the partner-led missions. For missions launched before 2000, the number of refereed publications before this date are summed. The year of launch and end of operations (if applicable) are given for each mission. These numbers of publications can be obtained from the ESA libraries by selecting publications before the end of 2021 and also selecting on "refereed" as some of the libraries contain a small number of non-refereed publications. It can be seen that:

1. The ESA-led missions with the most publications are XMM-Newton [5], Solar and Heliospheric Observatory (SOHO) [6] and Gaia [7], all with >6000 refereed publications at the end of 2021. Two of these (SOHO and XMM-Newton) are long-lived (>20 years of operations), whereas the Gaia [7] publications in the table span only 8 years.
2. The partner-led mission publications are dominated by HST which shows an increase from ~500 publications per year to ~1000 per year between 2000 and 2021. Cassini [8] also contributes strongly to the partner-led publications.
3. Publications from a mission typically continue for many years after the end of the operational and any (typically 2 to 3 year) post-operational phases.

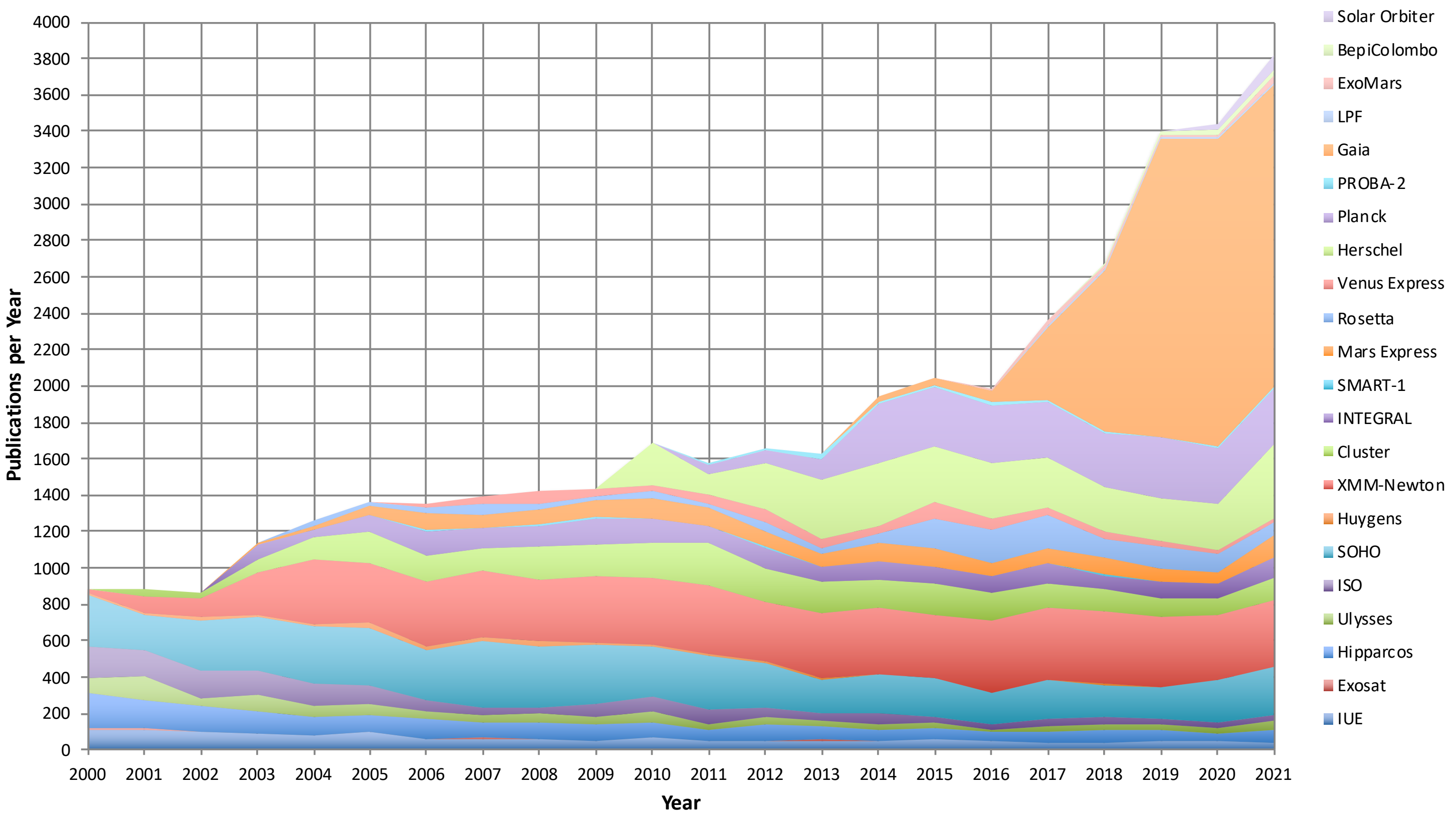

**Fig. 1** A "waterfall" diagram of the ESA Science Programme Mission refereed publications between 2000 and 2021. The missions are arranged in order of launch date with the earliest missions at the bottom. The large contribution from Gaia after 2016 (shown in light brown) is striking.

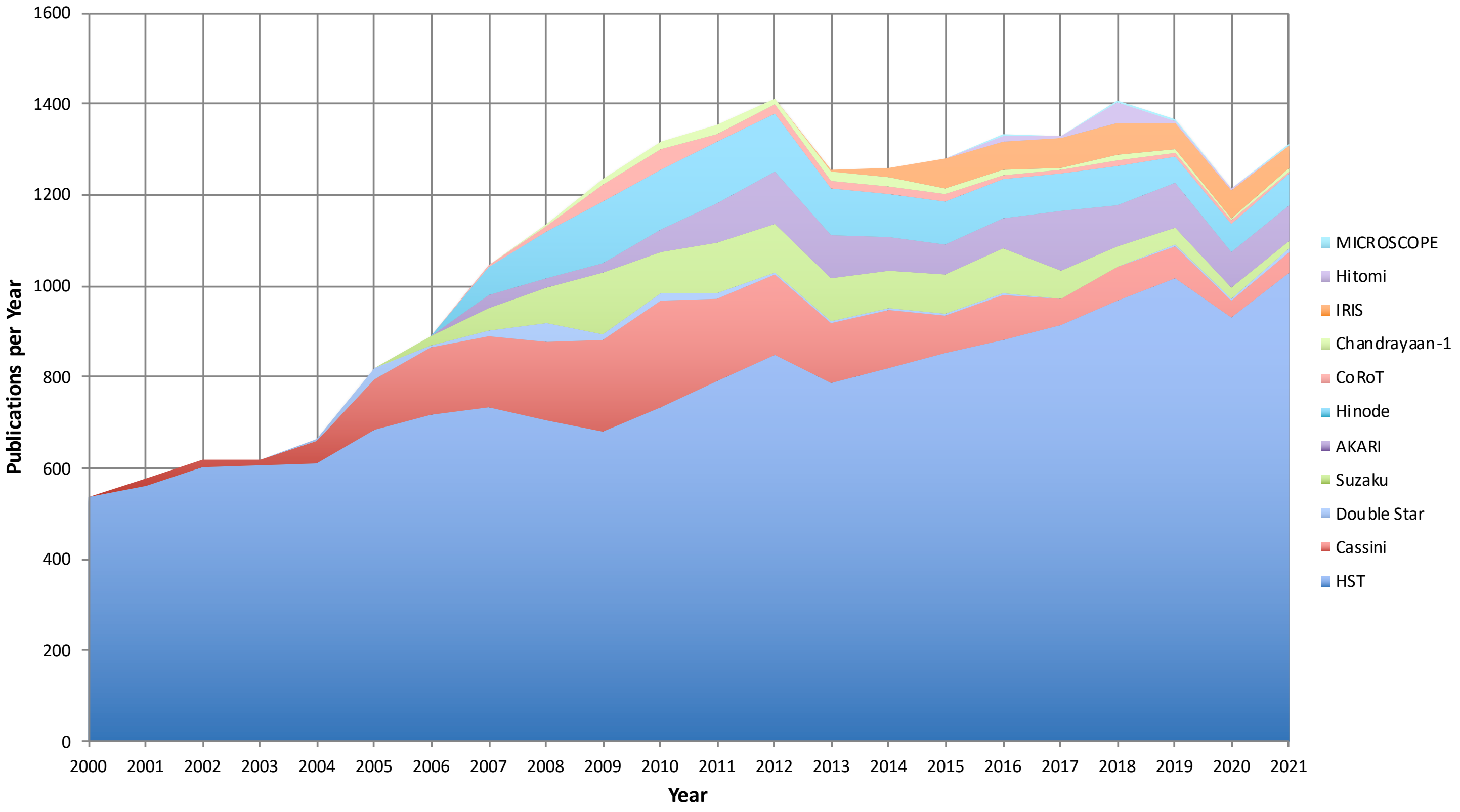

**Fig. 2** A "waterfall" diagram of the refereed publications between 2000 and 2021 for the partner-led missions in the ESA Science Programme. The missions are arranged in order of launch date with the earliest missions at the bottom.

4. There is a trend of an increase in the overall number of publications from the ESA-led missions with time, particularly between 2016 and 2019 where the large increase in Gaia publications contributed strongly to the total tally.
5. There is a core of long-lived Science Programme missions that have contributed significant numbers of publications for >20 years. These include Hipparcos [9], the SOHO [6], Cluster [10] and XMM-Newton [5].

It is important to stress that the missions in the ESA Science Programme cover a wide range of areas, they serve communities with rather different sizes, and have produced data over very different timelines, ranging from the few hours of Huygens, to the more than 30 years of HST. Therefore, it is not possible or meaningful to simply compare the number of scientific papers across missions or try and establish whether one is more or less successful than another. Instead, it is useful to explore the evolution of each mission, in terms of number of publications, but also how the scientific communities, in ESA Member States and elsewhere, have taken advantage of the research opportunities offered by the mission data.

We examined the institutes, or universities, in the 22 ESA Member States where the authors of the publications listed in Tables 1 and 2 are located. If available, these were extracted directly from the publications, or were derived using the names of the author's institute. For publications that were co-authored by a consortium, e.g., the Planck Collaboration, the first authors were assumed to have locations in proportion to those of all the authors listed in the consortium – e.g., a consortium with 50 members with 20 in country A and 30 in country B would have a first author count of 0.4 for country A and 0.6 for country B. Note that the countries used here are the affiliations listed in the publications and are not necessarily the same as the nationalities of the authors. For authors who indicated multiple affiliations, these were distributed pro-rata amongst the countries involved – e.g., if an author indicated affiliations in Greece and Germany, each country was counted as having 0.5 authors.
The authors' affiliations are presented in three ways:

1. Using the countries where the first authors of a publication are affiliated. This can provide an indication of the number of leadership positions for a particular mission in each included country.
2. Using the countries where *all* the authors of a publication are affiliated. This gives an indication of the total contribution of a country to a mission. We caution that this statistic may be affected by a small number of publications with large consortia often with many hundreds of members from other missions or facilities that may not have contributed directly to the specific ESA mission's results.
3. Using the countries of all authors as above, but normalising each publication to have a total of 1.0 authors. So if a publication has ten authors, the countries where each are affiliated would be counted as 0.1. This method prevents the typically few publications from large consortia from dominating the statistics and is our favoured method of investigating the *overall* contributions of different countries to the data exploitation of the missions.

We first present publication statistics in tabular form, using all three methods for ESA-led missions by authors located in the ESA Member States (Tables 3, 4 and 5), for partner-led missions by authors located in the ESA Member States (Tables 6, 7 and 8), and for all missions by authors located in some of the (non-ESA Member State) selected major space powers (Tables 9, 10 and 11). We then examine the implications for the utilisation of the science results from ESA's missions by various countries, which are summarised in Figs. 3, 4, 5, and 6.

**Table 3** The number of first authors for the ESA-led mission publications from institutes located in the ESA Member States. Refereed publications that appeared until the end of 2021 and meet the inclusion criteria listed in Sect. 2 are included. We indicate with "% Contrib." the percentage average financial contribution of a Member State to the Science Programme evaluated between 2000 and 2021 or between the year that a Member State joined ESA and 2021.

| Mission | AT | BE | CH | CZ | DE | DK | EE | ES | FI | FR | GB | GR | HU | IE | IT | LU | NL | NO | PL | PT | RO | SE |
|---|---|---|---|---|---|---|---|---|---|---|---|---|---|---|---|---|---|---|---|---|---|---|
| COS-B | 0 | 0 | 1 | 0 | 23 | 0 | 0 | 0 | 0 | 24 | 13 | 0 | 0 | 0 | 35 | 0 | 42 | 0 | 1 | 0 | 0 | 0 |
| IUE | 11 | 51 | 66 | 26 | 12 | 4 | 383 | 169 | 7 | 218 | 542 | 16 | 4 | 6 | 300 | 0 | 110 | 9 | 67 | 7 | 0 | 34 |
| Exosat | 1 | 1 | 2 | 3 | 117 | 8 | 0 | 0 | 4 | 21 | 164 | 3 | 0 | 2 | 94 | 0 | 139 | 0 | 6 | 0 | 0 | 2 |
| Giotto | 0 | 0 | 18 | 0 | 80 | 0 | 0 | 0 | 0 | 28 | 35 | 0 | 2 | 5 | 11 | 0 | 12 | 0 | 0 | 0 | 0 | 1 |
| Hipparcos | 37 | 73 | 35 | 37 | 216 | 46 | 1 | 121 | 11 | 262 | 120 | 4 | 10 | 2 | 119 | 0 | 59 | 1 | 39 | 5 | 2 | 43 |
| Ulysses | 7 | 13 | 41 | 3 | 201 | 2 | 0 | 7 | 9 | 116 | 150 | 31 | 17 | 1 | 63 | 0 | 53 | 4 | 17 | 4 | 5 | 2 |
| ISO | 26 | 53 | 8 | 1 | 230 | 10 | 0 | 118 | 24 | 251 | 152 | 3 | 16 | 11 | 118 | 0 | 169 | 0 | 9 | 2 | 1 | 42 |
| SOHO | 51 | 82 | 90 | 43 | 383 | 17 | 1 | 79 | 136 | 339 | 503 | 68 | 16 | 25 | 207 | 0 | 25 | 50 | 77 | 5 | 7 | 13 |
| Huygens | 8 | 1 | 1 | 0 | 18 | 0 | 0 | 12 | 4 | 47 | 10 | 2 | 0 | 0 | 6 | 0 | 9 | 0 | 1 | 1 | 0 | 0 |
| XMM-Newton | 19 | 90 | 103 | 22 | 751 | 14 | 5 | 306 | 30 | 293 | 840 | 70 | 10 | 18 | 936 | 0 | 308 | 0 | 37 | 4 | 0 | 16 |
| Cluster | 94 | 48 | 2 | 54 | 144 | 0 | 0 | 9 | 52 | 178 | 267 | 7 | 10 | 1 | 55 | 0 | 35 | 32 | 12 | 0 | 17 | 175 |
| INTEGRAL | 3 | 12 | 99 | 22 | 186 | 12 | 1 | 84 | 18 | 188 | 93 | 17 | 8 | 14 | 319 | 0 | 60 | 1 | 21 | 2 | 1 | 6 |
| SMART-1 | 0 | 0 | 2 | 0 | 6 | 0 | 0 | 2 | 8 | 2 | 18 | 0 | 0 | 0 | 2 | 0 | 13 | 0 | 0 | 0 | 0 | 2 |
| MEX | 5 | 30 | 15 | 12 | 158 | 3 | 0 | 29 | 22 | 213 | 84 | 1 | 7 | 5 | 105 | 0 | 22 | 3 | 5 | 2 | 1 | 72 |
| Rosetta | 21 | 13 | 71 | 2 | 219 | 1 | 0 | 39 | 9 | 196 | 51 | 2 | 8 | 2 | 128 | 0 | 27 | 18 | 26 | 1 | 2 | 59 |
| VEX | 29 | 39 | 7 | 1 | 68 | 0 | 0 | 30 | 14 | 67 | 58 | 0 | 2 | 2 | 28 | 0 | 13 | 0 | 1 | 8 | 2 | 25 |
| Herschel | 21 | 79 | 23 | 2 | 347 | 18 | 1 | 172 | 31 | 318 | 428 | 17 | 31 | 3 | 190 | 0 | 137 | 1 | 14 | 8 | 1 | 58 |
| Planck | 6 | 13 | 39 | 3 | 153 | 32 | 3 | 118 | 31 | 213 | 250 | 12 | 8 | 2 | 205 | 0 | 37 | 12 | 10 | 9 | 4 | 13 |
| PROBA-2 | 4 | 27 | 4 | 0 | 3 | 0 | 0 | 2 | 4 | 6 | 13 | 1 | 0 | 3 | 6 | 0 | 5 | 0 | 0 | 0 | 1 | 1 |
| Gaia | 32 | 69 | 85 | 47 | 475 | 54 | 6 | 307 | 12 | 279 | 535 | 19 | 56 | 15 | 294 | 0 | 150 | 0 | 86 | 21 | 1 | 92 |
| LISA PF | 0 | 0 | 0 | 0 | 1 | 0 | 0 | 9 | 0 | 0 | 1 | 0 | 0 | 0 | 9 | 0 | 8 | 0 | 0 | 0 | 0 | 0 |
| ExoMars | 0 | 14 | 4 | 0 | 4 | 0 | 0 | 11 | 0 | 10 | 13 | 0 | 0 | 0 | 10 | 0 | 2 | 0 | 0 | 0 | 0 | 1 |
| BepiColombo | 4 | 0 | 3 | 0 | 20 | 0 | 0 | 0 | 1 | 5 | 4 | 0 | 0 | 0 | 16 | 0 | 1 | 0 | 0 | 0 | 0 | 1 |
| CHEOPS | 4 | 2 | 7 | 0 | 0 | 0 | 0 | 0 | 0 | 2 | 3 | 0 | 1 | 0 | 1 | 0 | 0 | 0 | 0 | 2 | 0 | 0 |
| Solar Orbiter | 2 | 4 | 7 | 0 | 9 | 0 | 0 | 9 | 1 | 17 | 15 | 0 | 0 | 0 | 18 | 0 | 3 | 0 | 1 | 0 | 0 | 3 |
| Total | 385 | 714 | 733 | 278 | 4195 | 229 | 22 | 1633 | 428 | 3293 | 4362 | 273 | 206 | 117 | 3275 | 0 | 1439 | 131 | 430 | 81 | 45 | 661 |
| % MS Pubs | 1.68 | 3.11 | 3.20 | 1.21 | 18.29 | 1.00 | 0.10 | 7.12 | 1.87 | 14.36 | 19.02 | 1.19 | 0.90 | 0.51 | 14.28 | 0.00 | 6.28 | 0.57 | 1.88 | 0.35 | 0.20 | 2.88 |
| % Contrib. | 2.17 | 2.72 | 3.41 | 0.92 | 21.16 | 1.74 | 0.12 | 7.22 | 1.33 | 14.94 | 15.41 | 1.45 | 0.62 | 1.03 | 11.63 | 0.20 | 4.42 | 2.16 | 2.65 | 1.17 | 0.95 | 2.58 |
| Ratio | 0.77 | 1.14 | 0.94 | 1.31 | 0.86 | 0.57 | 0.81 | 0.99 | 1.40 | 0.96 | 1.23 | 0.82 | 1.45 | 0.50 | 1.23 | 0.00 | 1.42 | 0.26 | 0.71 | 0.30 | 0.21 | 1.12 |

**Table 4** Total number of authors for the ESA-led mission publications from institutes located in the ESA Member States. Refereed publications that appeared until the end of 2021 and meet the inclusion criteria listed in Sect. 2 are included. We indicate with "% Contrib." the percentage average financial contribution of a Member State to the Science Programme evaluated between 2000 and 2021 or between the year that a Member State joined ESA and 2021.

| Mission | AT | BE | CH | CZ | DE | DK | EE | ES | FI | FR | GB | GR | HU | IE | IT | LU | NL | NO | PL | PT | RO | SE |
|---|---|---|---|---|---|---|---|---|---|---|---|---|---|---|---|---|---|---|---|---|---|---|
| COS-B | 4 | 1 | 2 | 0 | 154 | 0 | 0 | 6 | 0 | 154 | 45 | 0 | 0 | 0 | 283 | 0 | 211 | 0 | 3 | 0 | 0 | 5 |
| IUE | 4 | 1 | 2 | 0 | 154 | 0 | 0 | 6 | 0 | 154 | 45 | 0 | 0 | 0 | 283 | 0 | 211 | 0 | 3 | 0 | 0 | 5 |
| Exosat | 4 | 3 | 7 | 10 | 448 | 26 | 0 | 16 | 12 | 88 | 540 | 4 | 4 | 4 | 414 | 0 | 493 | 2 | 21 | 2 | 3 | 8 |
| Giotto | 0 | 1 | 104 | 0 | 384 | 2 | 0 | 0 | 0 | 123 | 135 | 0 | 4 | 25 | 62 | 0 | 26 | 0 | 0 | 0 | 0 | 7 |
| Hipparcos | 140 | 374 | 388 | 170 | 1210 | 241 | 3 | 587 | 48 | 1655 | 665 | 20 | 62 | 15 | 759 | 0 | 283 | 8 | 180 | 117 | 9 | 166 |
| Ulysses | 43 | 72 | 227 | 19 | 1178 | 24 | 0 | 86 | 37 | 664 | 799 | 133 | 34 | 5 | 483 | 0 | 247 | 14 | 100 | 11 | 20 | 21 |
| ISO | 131 | 421 | 56 | 7 | 1700 | 64 | 0 | 957 | 106 | 2066 | 1339 | 45 | 109 | 62 | 844 | 0 | 1151 | 1 | 62 | 5 | 6 | 277 |
| SOHO | 414 | 537 | 570 | 211 | 2146 | 118 | 1 | 463 | 678 | 2095 | 2243 | 362 | 95 | 146 | 1506 | 0 | 214 | 222 | 262 | 65 | 61 | 85 |
| Huygens | 65 | 2 | 3 | 0 | 108 | 0 | 0 | 99 | 24 | 292 | 85 | 4 | 1 | 0 | 85 | 0 | 100 | 1 | 2 | 6 | 0 | 1 |
| XMM-Newton | 229 | 433 | 986 | 167 | 7799 | 299 | 27 | 3431 | 392 | 4818 | 6942 | 423 | 118 | 262 | 9419 | 0 | 2360 | 37 | 582 | 61 | 5 | 300 |
| Cluster | 808 | 236 | 35 | 263 | 1114 | 22 | 0 | 58 | 360 | 1950 | 2114 | 32 | 48 | 7 | 301 | 0 | 268 | 287 | 42 | 11 | 88 | 1121 |
| INTEGRAL | 76 | 133 | 840 | 177 | 3728 | 247 | 7 | 1503 | 315 | 2824 | 2206 | 79 | 130 | 200 | 5665 | 1 | 826 | 11 | 394 | 23 | 15 | 198 |
| SMART-1 | 1 | 0 | 24 | 0 | 35 | 0 | 0 | 32 | 54 | 44 | 131 | 0 | 0 | 0 | 16 | 0 | 100 | 0 | 0 | 2 | 0 | 13 |
| MEX | 59 | 201 | 138 | 25 | 1345 | 8 | 0 | 220 | 223 | 1831 | 630 | 5 | 27 | 40 | 890 | 0 | 192 | 19 | 51 | 15 | 3 | 660 |
| Rosetta | 163 | 195 | 1059 | 16 | 3697 | 13 | 0 | 1010 | 155 | 2708 | 530 | 7 | 127 | 13 | 2664 | 0 | 380 | 66 | 183 | 4 | 4 | 688 |
| VEX | 349 | 293 | 44 | 10 | 540 | 3 | 0 | 197 | 102 | 710 | 344 | 6 | 27 | 13 | 405 | 0 | 101 | 0 | 20 | 46 | 8 | 293 |
| Herschel | 256 | 1322 | 397 | 10 | 5989 | 310 | 12 | 3322 | 209 | 7738 | 8826 | 281 | 274 | 80 | 4060 | 0 | 2216 | 28 | 303 | 127 | 8 | 714 |
| Planck | 73 | 133 | 593 | 11 | 2638 | 762 | 28 | 3228 | 1070 | 9102 | 5994 | 69 | 74 | 150 | 7827 | 0 | 710 | 653 | 278 | 90 | 71 | 180 |
| PROBA-2 | 23 | 241 | 41 | 9 | 75 | 0 | 0 | 14 | 21 | 80 | 82 | 10 | 1 | 26 | 34 | 0 | 72 | 1 | 0 | 0 | 13 | 13 |
| Gaia | 418 | 1377 | 2473 | 379 | 7299 | 1008 | 43 | 6716 | 234 | 7009 | 7550 | 306 | 666 | 173 | 5994 | 0 | 1978 | 30 | 1192 | 838 | 13 | 994 |
| LISA PF | 0 | 0 | 105 | 1 | 362 | 0 | 0 | 314 | 0 | 125 | 264 | 0 | 0 | 0 | 409 | 0 | 102 | 0 | 0 | 0 | 0 | 0 |
| ExoMars | 1 | 242 | 79 | 2 | 25 | 1 | 0 | 155 | 10 | 186 | 125 | 0 | 7 | 0 | 145 | 0 | 32 | 0 | 10 | 0 | 2 | 5 |
| BepiColombo | 97 | 11 | 49 | 3 | 250 | 1 | 0 | 29 | 59 | 146 | 96 | 2 | 16 | 4 | 310 | 0 | 32 | 1 | 3 | 3 | 0 | 46 |
| CHEOPS | 91 | 48 | 326 | 0 | 89 | 1 | 0 | 75 | 0 | 103 | 118 | 0 | 37 | 1 | 127 | 0 | 51 | 0 | 0 | 64 | 0 | 52 |
| Solar Orbiter | 76 | 126 | 128 | 127 | 519 | 1 | 0 | 326 | 24 | 708 | 496 | 25 | 3 | 10 | 500 | 0 | 99 | 10 | 36 | 0 | 4 | 118 |
| Total | 3604 | 6679 | 8924 | 1728 | 44497 | 3213 | 135 | 23714 | 4170 | 48300 | 44395 | 1870 | 1881 | 1269 | 44491 | 1 | 12846 | 1425 | 3916 | 1536 | 333 | 6134 |
| % MS Pubs | 1.36 | 2.52 | 3.37 | 0.65 | 16.79 | 1.21 | 0.05 | 8.95 | 1.57 | 18.22 | 16.75 | 0.71 | 0.71 | 0.48 | 16.79 | 0.00 | 4.85 | 0.54 | 1.48 | 0.58 | 0.13 | 2.31 |
| % Contrib. | 2.17 | 2.72 | 3.41 | 0.92 | 21.16 | 1.74 | 0.12 | 7.22 | 1.33 | 14.94 | 15.41 | 1.45 | 0.62 | 1.03 | 11.63 | 0.20 | 4.42 | 2.16 | 2.65 | 1.17 | 0.95 | 2.58 |
| Ratio | 0.63 | 0.93 | 0.99 | 0.71 | 0.79 | 0.70 | 0.43 | 1.24 | 1.18 | 1.22 | 1.09 | 0.49 | 1.15 | 0.47 | 1.44 | 0.00 | 1.10 | 0.25 | 0.56 | 0.50 | 0.13 | 0.90 |

**Table 5** Number of authors for the ESA-led mission publications from institutes located in the ESA Member States normalised so that each publication contributes 1.0 authors. Refereed publications that appeared until the end of 2021 and meet the inclusion criteria listed in Sect. 2 are included. % Contrib. is the percentage average financial contribution of a Member State to the Science Programme evaluated between 2000 and 2021 or between the year that a Member State joined ESA and 2021.

| Mission | AT | BE | CH | CZ | DE | DK | EE | ES | FI | FR | GB | GR | HU | IE | IT | LU | NL | NO | PL | PT | RO | SE |
|---|---|---|---|---|---|---|---|---|---|---|---|---|---|---|---|---|---|---|---|---|---|---|
| COS-B | 0.0 | 0.1 | 0.3 | 0.0 | 24.7 | 0.0 | 0.0 | 0.4 | 0.0 | 22.3 | 12.2 | 0.0 | 0.0 | 0.0 | 37.7 | 0.0 | 36.1 | 0.0 | 0.9 | 0.0 | 0.0 | 0.0 |
| IUE | 14.3 | 45.8 | 68.3 | 19.0 | 375.4 | 13.6 | 4.7 | 178.2 | 7.1 | 199.7 | 521.9 | 14.0 | 3.6 | 7.5 | 294.8 | 0.0 | 116.5 | 10.9 | 58.5 | 7.4 | 0.0 | 35.7 |
| Exosat | 1.6 | 0.4 | 3.0 | 0.9 | 115.8 | 7.3 | 0.0 | 2.6 | 3.0 | 22.5 | 158.3 | 1.7 | 0.3 | 1.6 | 91.0 | 0.0 | 132.2 | 0.3 | 6.4 | 0.4 | 0.1 | 2.0 |
| Giotto | 0.0 | 0.2 | 16.8 | 0.0 | 82.3 | 0.1 | 0.0 | 0.0 | 0.0 | 25.6 | 32.5 | 0.0 | 1.2 | 6.9 | 12.6 | 0.0 | 10.6 | 0.0 | 0.0 | 0.0 | 0.0 | 1.8 |
| Hipparcos | 31.7 | 70.1 | 49.1 | 35.9 | 225.2 | 41.8 | 1.0 | 119.2 | 13.2 | 252.7 | 122.0 | 4.9 | 10.0 | 2.0 | 126.2 | 0.0 | 54.4 | 0.8 | 38.2 | 6.7 | 0.9 | 40.3 |
| Ulysses | 6.0 | 12.1 | 44.8 | 2.4 | 195.6 | 3.1 | 0.0 | 10.5 | 8.7 | 109.3 | 176.9 | 33.3 | 11.1 | 1.5 | 64.8 | 0.0 | 45.8 | 2.9 | 17.5 | 2.6 | 5.8 | 2.2 |
| ISO | 20.4 | 46.8 | 8.4 | 0.8 | 238.3 | 8.9 | 0.0 | 122.4 | 19.6 | 261.4 | 141.3 | 3.4 | 11.4 | 7.0 | 107.1 | 0.0 | 174.9 | 0.1 | 10.4 | 0.8 | 0.8 | 35.8 |
| SOHO | 60.8 | 85.6 | 79.8 | 41.9 | 416.8 | 19.6 | 0.3 | 70.4 | 123.0 | 363.4 | 490.0 | 61.1 | 19.4 | 24.4 | 199.5 | 0.0 | 21.7 | 47.6 | 65.9 | 5.1 | 5.9 | 13.0 |
| Huygens | 8.1 | 0.5 | 1.0 | 0.0 | 17.4 | 0.0 | 0.0 | 13.7 | 3.4 | 43.1 | 11.2 | 1.3 | 0.2 | 0.0 | 8.9 | 0.0 | 9.9 | 0.0 | 0.3 | 1.3 | 0.0 | 0.3 |
| XMM-Newton | 21.3 | 73.7 | 96.0 | 18.6 | 701.0 | 16.9 | 2.4 | 312.8 | 26.6 | 317.7 | 833.1 | 62.2 | 7.3 | 15.4 | 908.6 | 0.0 | 299.3 | 0.9 | 41.3 | 4.2 | 0.5 | 18.3 |
| Cluster | 100.1 | 44.3 | 3.9 | 46.0 | 158.3 | 2.2 | 0.0 | 7.0 | 46.8 | 253.6 | 305.3 | 5.4 | 7.0 | 1.5 | 50.5 | 0.0 | 34.9 | 33.0 | 8.2 | 1.3 | 13.4 | 166.0 |
| INTEGRAL | 3.4 | 9.8 | 88.4 | 23.6 | 179.5 | 15.1 | 1.1 | 85.9 | 19.2 | 185.9 | 107.4 | 14.3 | 5.4 | 11.8 | 298.1 | 0.1 | 57.1 | 0.4 | 22.7 | 1.4 | 0.5 | 6.9 |
| SMART-1 | 0.1 | 0.0 | 2.3 | 0.0 | 4.3 | 0.0 | 0.0 | 2.5 | 6.6 | 4.3 | 17.2 | 0.0 | 0.0 | 0.0 | 2.5 | 0.0 | 13.1 | 0.0 | 0.0 | 0.1 | 0.0 | 2.1 |
| MEX | 4.8 | 25.7 | 13.6 | 4.9 | 167.8 | 1.1 | 0.0 | 24.8 | 19.9 | 238.2 | 80.4 | 0.8 | 7.1 | 3.8 | 105.5 | 0.0 | 23.2 | 3.2 | 5.0 | 2.3 | 1.1 | 68.6 |
| Rosetta | 15.0 | 13.4 | 64.4 | 2.4 | 233.4 | 1.0 | 0.0 | 48.9 | 12.8 | 192.1 | 52.3 | 0.7 | 9.1 | 1.1 | 120.4 | 0.0 | 28.8 | 13.2 | 26.4 | 0.8 | 0.2 | 54.8 |
| VEX | 38.7 | 33.2 | 4.4 | 1.1 | 65.2 | 0.5 | 0.0 | 23.9 | 11.3 | 85.4 | 48.5 | 0.3 | 2.3 | 0.7 | 35.4 | 0.0 | 12.9 | 0.0 | 1.5 | 5.2 | 0.9 | 29.7 |
| Herschel | 20.5 | 71.4 | 16.8 | 1.7 | 361.5 | 15.4 | 0.6 | 174.7 | 20.9 | 356.4 | 396.8 | 18.8 | 21.3 | 3.6 | 197.8 | 0.0 | 125.0 | 0.4 | 17.0 | 6.9 | 1.2 | 49.5 |
| Planck | 8.0 | 13.5 | 39.2 | 2.5 | 145.2 | 31.6 | 4.9 | 124.1 | 25.5 | 245.3 | 257.7 | 13.3 | 7.0 | 2.1 | 222.2 | 0.0 | 37.7 | 15.4 | 14.5 | 11.0 | 3.4 | 12.6 |
| PROBA-2 | 3.6 | 31.1 | 3.7 | 1.4 | 3.9 | 0.0 | 0.0 | 1.9 | 2.6 | 9.7 | 10.2 | 0.7 | 0.1 | 3.1 | 5.7 | 0.0 | 2.9 | 0.0 | 0.0 | 0.0 | 0.9 | 0.9 |
| Gaia | 29.8 | 73.2 | 85.1 | 43.8 | 485.6 | 52.7 | 5.5 | 317.2 | 10.9 | 315.5 | 514.2 | 15.3 | 43.5 | 11.9 | 289.9 | 0.0 | 144.3 | 1.0 | 86.6 | 27.0 | 0.5 | 76.6 |
| LISA PF | 0.0 | 0.0 | 1.2 | 0.0 | 5.9 | 0.0 | 0.0 | 3.6 | 0.0 | 1.4 | 3.2 | 0.0 | 0.0 | 0.0 | 11.1 | 0.0 | 1.2 | 0.0 | 0.0 | 0.0 | 0.0 | 0.0 |
| ExoMars | 0.5 | 13.2 | 4.3 | 0.1 | 3.9 | 0.1 | 0.0 | 10.3 | 0.2 | 13.1 | 11.6 | 0.0 | 0.2 | 0.0 | 8.7 | 0.0 | 2.5 | 0.0 | 0.2 | 0.0 | 0.0 | 0.6 |
| BepiColombo | 3.4 | 0.5 | 1.8 | 0.1 | 20.1 | 0.0 | 0.0 | 1.1 | 1.4 | 5.7 | 3.2 | 0.0 | 0.3 | 0.0 | 14.1 | 0.0 | 1.1 | 0.1 | 0.3 | 0.0 | 0.0 | 1.3 |
| CHEOPS | 3.3 | 0.6 | 6.1 | 0.0 | 1.0 | 0.0 | 0.0 | 0.8 | 0.0 | 2.0 | 1.8 | 0.0 | 0.4 | 0.1 | 1.5 | 0.0 | 0.5 | 0.0 | 0.0 | 1.9 | 0.0 | 0.7 |
| Solar Orbiter | 2.7 | 3.8 | 5.1 | 2.6 | 12.7 | 0.0 | 0.0 | 8.5 | 0.8 | 16.1 | 15.9 | 0.3 | 0.2 | 0.1 | 13.7 | 0.0 | 2.1 | 0.7 | 1.2 | 0.0 | 0.1 | 2.9 |
| Total | 398.2 | 668.9 | 707.7 | 249.6 | 4240.5 | 231.0 | 20.5 | 1665.5 | 383.4 | 3542.6 | 4325.0 | 251.9 | 168.1 | 106.3 | 3228.3 | 0.1 | 1389.0 | 131.0 | 423.0 | 86.4 | 36.3 | 622.5 |
| % MS Pubs | 1.74 | 2.92 | 3.09 | 1.09 | 18.54 | 1.01 | 0.09 | 7.28 | 1.68 | 15.49 | 18.91 | 1.10 | 0.73 | 0.46 | 14.11 | 0.00 | 6.07 | 0.57 | 1.85 | 0.38 | 0.16 | 2.72 |
| % Contrib. | 2.17 | 2.72 | 3.41 | 0.92 | 21.16 | 1.74 | 0.12 | 7.22 | 1.33 | 14.94 | 15.41 | 1.45 | 0.62 | 1.03 | 11.63 | 0.20 | 4.42 | 2.16 | 2.65 | 1.17 | 0.95 | 2.58 |
| Ratio | 0.80 | 1.07 | 0.91 | 1.18 | 0.88 | 0.58 | 0.75 | 1.01 | 1.26 | 1.04 | 1.23 | 0.76 | 1.19 | 0.45 | 1.21 | 0.00 | 1.37 | 0.27 | 0.70 | 0.32 | 0.17 | 1.05 |

**Table 6** Number of first authors for the partner-led mission publications from institutes located in the ESA Member States. Refereed publications that appeared until the end of 2021 and meet the inclusion criteria listed in Sect. 2 are included.

| Mission | AT | BE | CH | CZ | DE | DK | EE | ES | FI | FR | GB | GR | HU | IE | IT | LU | NL | NO | PL | PT | RO | SE |
|---|---|---|---|---|---|---|---|---|---|---|---|---|---|---|---|---|---|---|---|---|---|---|
| HST | 53 | 108 | 196 | 13 | 1314 | 109 | 8 | 653 | 32 | 673 | 1567 | 35 | 15 | 39 | 1042 | 0 | 481 | 5 | 75 | 21 | 3 | 200 |
| Cassini | 12 | 29 | 3 | 10 | 142 | 0 | 0 | 29 | 2 | 174 | 254 | 14 | 8 | 0 | 58 | 0 | 6 | 3 | 2 | 1 | 0 | 36 |
| Double Star | 8 | 0 | 1 | 3 | 5 | 0 | 0 | 0 | 2 | 10 | 26 | 0 | 0 | 4 | 3 | 0 | 3 | 0 | 0 | 0 | 0 | 0 |
| Suzaku | 0 | 3 | 6 | 2 | 34 | 0 | 1 | 16 | 4 | 13 | 57 | 3 | 2 | 3 | 54 | 0 | 19 | 0 | 7 | 0 | 0 | 2 |
| AKARI | 6 | 26 | 1 | 6 | 54 | 3 | 0 | 40 | 6 | 38 | 69 | 6 | 15 | 1 | 21 | 0 | 12 | 0 | 28 | 0 | 0 | 4 |
| Hinode | 16 | 10 | 12 | 28 | 98 | 2 | 0 | 45 | 1 | 50 | 167 | 17 | 0 | 3 | 45 | 0 | 4 | 15 | 4 | 1 | 1 | 9 |
| CoRoT | 9 | 23 | 4 | 2 | 22 | 1 | 0 | 16 | 0 | 57 | 10 | 0 | 6 | 0 | 22 | 0 | 9 | 0 | 1 | 2 | 0 | 0 |
| Chandrayaan | 0 | 0 | 6 | 0 | 4 | 0 | 0 | 0 | 1 | 2 | 15 | 0 | 0 | 0 | 1 | 0 | 4 | 0 | 0 | 0 | 0 | 10 |
| IRIS | 0 | 1 | 12 | 15 | 14 | 0 | 0 | 4 | 0 | 11 | 41 | 4 | 0 | 0 | 11 | 0 | 1 | 38 | 5 | 0 | 0 | 9 |
| Hitomi | 0 | 0 | 0 | 0 | 1 | 0 | 0 | 0 | 0 | 0 | 2 | 0 | 0 | 4 | 1 | 0 | 2 | 0 | 0 | 0 | 0 | 0 |
| MICROSCOPE | 0 | 0 | 0 | 0 | 1 | 0 | 0 | 0 | 0 | 15 | 0 | 0 | 0 | 0 | 1 | 0 | 0 | 0 | 0 | 1 | 0 | 0 |
| Total | 104 | 200 | 241 | 79 | 1689 | 115 | 9 | 803 | 48 | 1043 | 2208 | 79 | 46 | 54 | 1259 | 0 | 541 | 61 | 122 | 26 | 4 | 270 |

**Table 7** Total number of authors for the partner-led mission publications from institutes located in the ESA Member States. Refereed publications that appeared until the end of 2021 and meet the inclusion criteria listed in Sect. 2 are included.

| Mission | AT | BE | CH | CZ | DE | DK | EE | ES | FI | FR | GB | GR | HU | IE | IT | LU | NL | NO | PL | PT | RO | SE |
|---|---|---|---|---|---|---|---|---|---|---|---|---|---|---|---|---|---|---|---|---|---|---|
| HST | 467 | 991 | 2452 | 167 | 13464 | 1500 | 28 | 6179 | 383 | 8904 | 13991 | 344 | 171 | 260 | 12223 | 0 | 4276 | 97 | 708 | 323 | 22 | 1643 |
| Cassini | 76 | 188 | 10 | 49 | 1018 | 0 | 0 | 166 | 48 | 1481 | 1959 | 111 | 50 | 4 | 589 | 1 | 72 | 25 | 10 | 18 | 1 | 319 |
| Double Star | 134 | 1 | 1 | 12 | 73 | 1 | 0 | 1 | 9 | 166 | 313 | 0 | 0 | 26 | 17 | 0 | 48 | 4 | 1 | 0 | 2 | 19 |
| Suzaku | 18 | 8 | 114 | 10 | 727 | 33 | 2 | 373 | 44 | 301 | 518 | 10 | 7 | 41 | 799 | 0 | 164 | 1 | 97 | 5 | 0 | 41 |
| AKARI | 46 | 217 | 70 | 27 | 569 | 16 | 0 | 437 | 33 | 642 | 807 | 37 | 116 | 9 | 304 | 0 | 180 | 3 | 222 | 13 | 1 | 59 |
| Hinode | 92 | 69 | 56 | 115 | 468 | 4 | 0 | 200 | 11 | 297 | 767 | 63 | 24 | 13 | 269 | 0 | 16 | 119 | 40 | 1 | 4 | 40 |
| CoRoT | 181 | 285 | 169 | 5 | 550 | 21 | 0 | 312 | 0 | 1734 | 175 | 0 | 43 | 0 | 192 | 0 | 155 | 1 | 12 | 30 | 0 | 13 |
| Chandrayaan | 0 | 1 | 45 | 0 | 33 | 0 | 0 | 3 | 23 | 30 | 131 | 0 | 0 | 1 | 5 | 0 | 16 | 1 | 1 | 0 | 0 | 90 |
| IRIS | 1 | 21 | 55 | 70 | 119 | 1 | 0 | 37 | 0 | 84 | 219 | 11 | 10 | 3 | 73 | 0 | 4 | 276 | 17 | 0 | 3 | 47 |
| Hitomi | 0 | 0 | 49 | 13 | 21 | 0 | 0 | 30 | 0 | 78 | 73 | 0 | 16 | 28 | 15 | 0 | 186 | 0 | 14 | 0 | 0 | 13 |
| MICROSCOPE | 0 | 0 | 0 | 0 | 18 | 0 | 0 | 1 | 0 | 120 | 3 | 0 | 0 | 0 | 2 | 0 | 2 | 0 | 0 | 1 | 0 | 0 |
| Total | 1015 | 1781 | 3021 | 468 | 17060 | 1576 | 30 | 7739 | 551 | 13837 | 18956 | 576 | 437 | 385 | 14488 | 1 | 5119 | 527 | 1122 | 391 | 33 | 2284 |

**Table 8** Number of authors for the partner-led mission publications from institutes located in the ESA Member States normalised so that each publication contributes 1.0 authors. Refereed publications that appeared until the end of 2021 and meet the inclusion criteria listed in Sect. 2 are included.

| Mission | AT | BE | CH | CZ | DE | DK | EE | ES | FI | FR | GB |
|---|---|---|---|---|---|---|---|---|---|---|---|
| HST | 46.7 | 111.6 | 186.7 | 14.5 | 1276.1 | 100.5 | 6.4 | 620.3 | 39.1 | 747.7 | 1523.4 |
| Cassini | 9.9 | 25.3 | 2.0 | 6.6 | 128.5 | 0.0 | 0.0 | 25.0 | 5.4 | 177.2 | 257.7 |
| Double Star | 10.2 | 0.1 | 0.5 | 1.4 | 6.4 | 0.1 | 0.0 | 0.3 | 1.0 | 16.9 | 29.7 |
| Suzaku | 0.2 | 2.2 | 7.8 | 1.0 | 47.2 | 1.8 | 0.7 | 16.2 | 3.4 | 17.9 | 63.3 |
| AKARI | 3.5 | 20.0 | 3.6 | 3.8 | 52.9 | 1.6 | 0.0 | 35.0 | 2.7 | 44.0 | 60.2 |
| Hinode | 15.9 | 10.1 | 10.7 | 22.2 | 97.6 | 0.8 | 0.0 | 48.4 | 1.2 | 50.3 | 158.3 |
| CoRoT | 10.5 | 15.9 | 4.7 | 1.0 | 24.2 | 1.3 | 0.0 | 19.0 | 0.0 | 67.3 | 9.1 |
| Chanrayaan | 0.0 | 0.0 | 4.5 | 0.0 | 4.8 | 0.0 | 0.0 | 0.3 | 1.7 | 3.4 | 13.2 |
| IRIS | 0.1 | 2.8 | 9.9 | 13.5 | 17.4 | 0.2 | 0.0 | 6.7 | 0.0 | 14.0 | 43.3 |
| Hitomi | 0.0 | 0.0 | 0.3 | 0.0 | 0.1 | 0.0 | 0.0 | 0.1 | 0.0 | 0.8 | 1.8 |
| MICROSCOPE | 0.0 | 0.0 | 0.0 | 0.0 | 1.4 | 0.0 | 0.0 | 0.5 | 0.0 | 14.0 | 0.2 |
| Total | 96.9 | 188.1 | 230.8 | 64.0 | 1656.7 | 106.2 | 7.1 | 771.9 | 54.6 | 1153.5 | 2160.2 |

| Mission | GR | HU | IE | IT | LU | NL | NO | PL | PT | RO | SE |
|---|---|---|---|---|---|---|---|---|---|---|---|
| HST | 33.7 | 13.8 | 31.7 | 1033.6 | 0.0 | 435.4 | 7.7 | 73.0 | 21.5 | 1.6 | 180.3 |
| Cassini | 11.2 | 6.7 | 0.6 | 56.5 | 0.1 | 7.9 | 3.0 | 2.8 | 1.4 | 0.1 | 35.2 |
| Double Star | 0.0 | 0.0 | 2.1 | 2.2 | 0.0 | 3.7 | 0.3 | 0.1 | 0.0 | 0.1 | 1.5 |
| Suzaku | 1.9 | 0.4 | 2.9 | 56.1 | 0.0 | 18.1 | 0.0 | 5.8 | 0.7 | 0.0 | 1.9 |
| AKARI | 4.1 | 10.5 | 0.8 | 23.5 | 0.0 | 15.1 | 0.2 | 21.4 | 0.9 | 0.1 | 4.3 |
| Hinode | 15.3 | 1.7 | 3.1 | 40.2 | 0.0 | 2.8 | 19.2 | 5.3 | 0.1 | 0.4 | 8.6 |
| CoRoT | 0.0 | 4.3 | 0.0 | 15.4 | 0.0 | 7.0 | 0.0 | 0.6 | 1.9 | 0.0 | 0.5 |
| Chanrayaan | 0.0 | 0.0 | 0.1 | 1.2 | 0.0 | 2.6 | 0.1 | 0.1 | 0.0 | 0.0 | 11.3 |
| IRIS | 2.6 | 0.8 | 0.5 | 10.3 | 0.0 | 1.0 | 40.0 | 2.8 | 0.0 | 0.2 | 9.7 |
| Hitomi | 0.0 | 0.1 | 0.1 | 0.1 | 0.0 | 1.8 | 0.0 | 0.1 | 0.0 | 0.0 | 0.1 |
| MICROSCOPE | 0.0 | 0.0 | 0.0 | 1.0 | 0.0 | 0.1 | 0.0 | 0.0 | 0.5 | 0.0 | 0.0 |
| Total | 68.7 | 38.3 | 41.9 | 1239.9 | 0.1 | 495.3 | 70.7 | 112.0 | 27.0 | 2.5 | 253.3 |

**Table 9** Number of first authors for publications using the inclusion criteria given in Sect. 2 from scientists located in (non-ESA Member State) major space powers. The "Other" column is the number of authors who are not located in an ESA Member State nor one of the countries listed in this table.

| Mission | AR | AU | BR | CA | CH | CN | IL | IN | JA | KR | MX | RU | US | Other |
|---|---|---|---|---|---|---|---|---|---|---|---|---|---|---|
| COS-B | 0 | 0 | 0 | 0 | 0 | 13 | 0 | 0 | 0 | 0 | 0 | 3 | 14 | 5 |
| IUE | 47 | 34 | 39 | 98 | 24 | 32 | 19 | 51 | 40 | 37 | 59 | 31 | 1953 | 139 |
| Exosat | 1 | 3 | 0 | 4 | 1 | 2 | 0 | 36 | 6 | 1 | 0 | 3 | 101 | 7 |
| Giotto | 0 | 0 | 0 | 0 | 0 | 0 | 9 | 0 | 0 | 0 | 0 | 4 | 58 | 1 |
| Hipparcos | 31 | 48 | 41 | 59 | 51 | 71 | 7 | 18 | 28 | 16 | 19 | 162 | 530 | 250 |
| Ulysses | 2 | 9 | 6 | 8 | 0 | 43 | 0 | 7 | 28 | 3 | 6 | 55 | 876 | 61 |
| ISO | 4 | 11 | 13 | 35 | 15 | 39 | 7 | 19 | 67 | 18 | 16 | 4 | 411 | 43 |
| SOHO | 45 | 35 | 44 | 13 | 2 | 498 | 2 | 275 | 127 | 101 | 26 | 235 | 2175 | 194 |
| Huygens | 0 | 1 | 1 | 1 | 2 | 0 | 3 | 0 | 2 | 0 | 2 | 2 | 79 | 1 |
| XMM-Newton | 38 | 61 | 31 | 121 | 48 | 229 | 30 | 145 | 280 | 28 | 36 | 83 | 1621 | 226 |
| Cluster | 1 | 12 | 10 | 27 | 1 | 377 | 3 | 43 | 79 | 14 | 11 | 174 | 663 | 54 |
| INTEGRAL | 6 | 15 | 8 | 17 | 6 | 85 | 3 | 33 | 54 | 4 | 6 | 161 | 260 | 55 |
| SMART-1 | 0 | 0 | 0 | 1 | 0 | 6 | 0 | 6 | 2 | 0 | 0 | 2 | 1 | 2 |
| MEX | 2 | 6 | 2 | 21 | 2 | 37 | 0 | 29 | 13 | 1 | 5 | 64 | 410 | 17 |
| Rosetta | 0 | 0 | 3 | 2 | 2 | 10 | 7 | 5 | 13 | 3 | 3 | 32 | 205 | 34 |
| VEX | 2 | 7 | 1 | 4 | 0 | 48 | 3 | 4 | 44 | 0 | 4 | 41 | 145 | 6 |
| Herschel | 22 | 45 | 13 | 96 | 54 | 114 | 7 | 60 | 74 | 23 | 25 | 21 | 729 | 65 |
| Planck | 4 | 32 | 49 | 77 | 30 | 215 | 13 | 114 | 83 | 43 | 10 | 34 | 532 | 139 |
| PROBA-2 | 2 | 0 | 1 | 0 | 0 | 4 | 0 | 5 | 4 | 1 | 0 | 12 | 20 | 4 |
| Gaia | 38 | 168 | 125 | 163 | 192 | 413 | 30 | 112 | 116 | 66 | 49 | 176 | 1683 | 322 |
| LISA PF | 0 | 0 | 0 | 0 | 0 | 0 | 0 | 0 | 0 | 0 | 0 | 0 | 5 | 0 |
| ExoMars 2016 | 0 | 1 | 0 | 5 | 0 | 0 | 0 | 1 | 1 | 0 | 0 | 19 | 12 | 4 |
| BepiColombo | 0 | 0 | 0 | 0 | 0 | 0 | 0 | 0 | 8 | 0 | 0 | 2 | 1 | 2 |
| CHEOPS | 0 | 0 | 0 | 0 | 0 | 0 | 0 | 0 | 0 | 0 | 0 | 0 | 1 | 1 |
| Solar Orbiter | 0 | 0 | 0 | 0 | 0 | 5 | 0 | 0 | 0 | 1 | 0 | 1 | 18 | 2 |
| HST | 78 | 406 | 191 | 418 | 264 | 363 | 93 | 146 | 351 | 138 | 162 | 212 | 8554 | 386 |
| Cassini | 7 | 7 | 6 | 9 | 0 | 19 | 4 | 9 | 32 | 9 | 6 | 11 | 1148 | 17 |
| Double Star | 0 | 1 | 0 | 0 | 0 | 73 | 0 | 0 | 4 | 0 | 0 | 3 | 16 | 0 |
| Suzaku | 2 | 2 | 1 | 13 | 0 | 31 | 0 | 39 | 491 | 2 | 2 | 9 | 271 | 33 |
| AKARI | 5 | 18 | 13 | 12 | 16 | 36 | 0 | 20 | 228 | 67 | 7 | 14 | 177 | 64 |
| Hinode | 6 | 6 | 3 | 0 | 0 | 119 | 1 | 67 | 217 | 32 | 0 | 13 | 402 | 27 |
| CoRoT | 0 | 0 | 18 | 2 | 3 | 1 | 6 | 1 | 0 | 0 | 0 | 3 | 8 | 8 |
| Chandrayaan | 0 | 0 | 0 | 2 | 0 | 10 | 0 | 56 | 3 | 0 | 0 | 1 | 39 | 8 |
| IRIS | 0 | 1 | 1 | 0 | 0 | 100 | 0 | 24 | 23 | 6 | 0 | 2 | 117 | 4 |
| Hitomi | 0 | 1 | 0 | 1 | 0 | 0 | 0 | 0 | 34 | 0 | 0 | 1 | 19 | 1 |
| MICROSCOPE | 0 | 0 | 0 | 0 | 0 | 0 | 0 | 0 | 0 | 0 | 0 | 0 | 0 | 0 |

**Table 10** Total number of authors for publications using the inclusion criteria given in Sect. 2 from scientists located in (non-ESA Member State) major space powers. The "Other" column is the number of authors who are not located in an ESA Member State nor one of the countries listed in this table.

| Mission | AR | AU | BR | CA | CH | CN | IL | IN | JA | KR | MX | RU | US | Other |
|---|---|---|---|---|---|---|---|---|---|---|---|---|---|---|
| COS-B | 0 | 7 | 0 | 3 | 0 | 41 | 0 | 0 | 6 | 0 | 0 | 20 | 118 | 21 |
| IUE | 164 | 208 | 169 | 489 | 257 | 169 | 81 | 153 | 222 | 118 | 212 | 156 | 8529 | 602 |
| Exosat | 1 | 24 | 5 | 10 | 19 | 7 | 0 | 97 | 36 | 2 | 0 | 9 | 461 | 50 |
| Giotto | 0 | 0 | 0 | 2 | 0 | 0 | 21 | 2 | 3 | 0 | 0 | 25 | 333 | 4 |
| Hipparcos | 92 | 358 | 221 | 305 | 310 | 317 | 25 | 89 | 277 | 105 | 86 | 477 | 3211 | 1061 |
| Ulysses | 13 | 61 | 23 | 50 | 6 | 167 | 6 | 40 | 183 | 17 | 18 | 279 | 4555 | 248 |
| ISO | 34 | 99 | 65 | 256 | 128 | 136 | 58 | 83 | 460 | 79 | 59 | 29 | 3277 | 235 |
| SOHO | 233 | 140 | 193 | 61 | 7 | 2518 | 9 | 890 | 742 | 561 | 131 | 1071 | 10460 | 904 |
| Huygens | 2 | 7 | 3 | 6 | 8 | 0 | 13 | 1 | 15 | 0 | 6 | 7 | 404 | 19 |
| XMM-Newton | 225 | 1242 | 377 | 920 | 869 | 1415 | 215 | 724 | 2877 | 195 | 448 | 768 | 17932 | 2177 |
| Cluster | 23 | 53 | 47 | 134 | 7 | 2283 | 15 | 140 | 501 | 72 | 36 | 827 | 4596 | 244 |
| INTEGRAL | 156 | 730 | 156 | 173 | 184 | 858 | 60 | 484 | 1143 | 175 | 178 | 1443 | 8067 | 991 |
| SMART-1 | 0 | 1 | 0 | 1 | 0 | 15 | 0 | 40 | 20 | 0 | 0 | 6 | 26 | 9 |
| MEX | 2 | 30 | 8 | 101 | 7 | 184 | 0 | 95 | 105 | 8 | 14 | 378 | 3098 | 109 |
| Rosetta | 3 | 6 | 16 | 10 | 21 | 66 | 23 | 38 | 111 | 11 | 10 | 109 | 2270 | 420 |
| VEX | 11 | 31 | 6 | 22 | 1 | 276 | 8 | 20 | 335 | 0 | 15 | 286 | 975 | 48 |
| Herschel | 92 | 882 | 120 | 2043 | 1142 | 1145 | 78 | 360 | 1435 | 420 | 356 | 170 | 12764 | 1195 |
| Planck | 14 | 598 | 425 | 1762 | 487 | 1071 | 49 | 566 | 903 | 424 | 109 | 520 | 13665 | 1409 |
| PROBA-2 | 11 | 6 | 3 | 2 | 0 | 21 | 0 | 22 | 28 | 7 | 1 | 49 | 163 | 22 |
| Gaia | 283 | 2936 | 1404 | 1571 | 3094 | 3244 | 333 | 711 | 2108 | 710 | 480 | 920 | 25023 | 3050 |
| LISA PF | 0 | 2 | 0 | 0 | 0 | 0 | 0 | 1 | 0 | 0 | 0 | 0 | 152 | 0 |
| ExoMars | 0 | 5 | 0 | 31 | 0 | 0 | 0 | 8 | 5 | 0 | 0 | 320 | 170 | 46 |
| BepiColombo | 2 | 1 | 0 | 0 | 0 | 3 | 0 | 2 | 164 | 0 | 0 | 33 | 108 | 10 |
| CHEOPS | 0 | 6 | 0 | 2 | 9 | 0 | 1 | 0 | 0 | 0 | 0 | 1 | 81 | 4 |
| Solar Orbiter | 2 | 1 | 0 | 3 | 0 | 29 | 0 | 0 | 5 | 12 | 0 | 12 | 606 | 12 |
| HST | 488 | 3849 | 1243 | 3485 | 3893 | 2693 | 998 | 883 | 4659 | 998 | 1149 | 1113 | 76189 | 3679 |
| Cassini | 28 | 40 | 30 | 44 | 3 | 84 | 20 | 26 | 157 | 32 | 19 | 71 | 8746 | 99 |
| Double Star | 0 | 1 | 1 | 8 | 0 | 482 | 0 | 0 | 27 | 0 | 0 | 20 | 182 | 22 |
| Suzaku | 9 | 36 | 3 | 75 | 40 | 169 | 6 | 151 | 4343 | 17 | 28 | 66 | 2558 | 269 |
| AKARI | 19 | 169 | 62 | 187 | 255 | 196 | 3 | 90 | 2384 | 518 | 74 | 80 | 2112 | 536 |
| Hinode | 46 | 24 | 20 | 0 | 1 | 568 | 5 | 222 | 1299 | 199 | 1 | 70 | 2339 | 118 |
| CoRoT | 4 | 15 | 187 | 23 | 26 | 5 | 98 | 6 | 4 | 0 | 3 | 9 | 180 | 43 |
| Chandrayaan | 0 | 0 | 0 | 14 | 0 | 63 | 0 | 316 | 50 | 1 | 0 | 12 | 390 | 39 |
| IRIS | 2 | 6 | 4 | 0 | 7 | 453 | 1 | 64 | 112 | 53 | 0 | 14 | 910 | 24 |
| Hitomi | 0 | 5 | 0 | 56 | 0 | 0 | 0 | 0 | 2433 | 1 | 0 | 2 | 1097 | 5 |
| MICROSCOPE | 0 | 0 | 0 | 0 | 0 | 0 | 0 | 0 | 0 | 0 | 0 | 0 | 3 | 1 |

**Table 11** Number of authors for publications using the inclusion criteria given in Sect. 2 from scientists located in (non-ESA Member State) major space powers, normalised so that each publication contributes 1.0 authors. The "Other" column is the number of authors who are not located in an ESA Member State nor one of the countries listed in this table.

| Mission | AR | AU | BR | CA | CH | CN | IL | IN | JA | KR | MX | RU | US | Other |
|---|---|---|---|---|---|---|---|---|---|---|---|---|---|---|
| COS-B | 0.0 | 0.8 | 0.0 | 0.6 | 0.0 | 11.6 | 0.0 | 0.0 | 0.0 | 0.0 | 0.0 | 3.3 | 15.6 | 5.2 |
| IUE | 43.9 | 32.3 | 37.3 | 104.5 | 31.6 | 30.3 | 16.5 | 41.6 | 43.2 | 32.0 | 50.0 | 32.0 | 2013 | 139 |
| Exosat | 0.3 | 4.8 | 1.0 | 2.9 | 3.3 | 1.6 | 0.0 | 35.4 | 7.3 | 1.0 | 0.0 | 3.8 | 109 | 8.8 |
| Giotto | 0.0 | 0.0 | 0.0 | 0.5 | 0.0 | 0.0 | 6.4 | 0.2 | 0.7 | 0.0 | 0.0 | 3.8 | 59.0 | 0.9 |
| Hipparcos | 33.7 | 46.0 | 36.0 | 60.5 | 54.5 | 66.8 | 6.4 | 19.3 | 31.8 | 15.8 | 19.0 | 157 | 545 | 240 |
| Ulysses | 1.9 | 7.7 | 4.9 | 10.7 | 0.2 | 39.1 | 0.2 | 9.4 | 29.4 | 3.0 | 4.9 | 52.3 | 868 | 58.6 |
| ISO | 4.8 | 12.1 | 11.4 | 31.9 | 13.9 | 33.4 | 9.2 | 15.4 | 68.0 | 16.7 | 13.9 | 5.2 | 450 | 36.2 |
| SOHO | 39.9 | 32.4 | 42.4 | 15.1 | 2.1 | 460 | 2.0 | 240 | 143 | 99.0 | 23.2 | 231 | 2270 | 176 |
| Huygens | 0.1 | 0.8 | 1.1 | 0.9 | 1.3 | 0.0 | 3.2 | 0.3 | 1.8 | 0.0 | 0.9 | 2.7 | 76.9 | 0.6 |
| XMM-Newton | 33.6 | 82.2 | 33.2 | 107 | 52.7 | 208 | 26.7 | 136 | 282 | 21.3 | 44.5 | 94.0 | 1759 | 211 |
| Cluster | 3.8 | 10.5 | 8.7 | 25.1 | 1.4 | 298 | 3.7 | 38.2 | 82.1 | 11.7 | 6.1 | 147 | 696 | 39.8 |
| INTEGRAL | 7.5 | 16.7 | 8.0 | 16.3 | 9.3 | 80.6 | 5.6 | 30.3 | 52.6 | 3.2 | 6.2 | 157.4 | 289 | 59.1 |
| SMART-1 | 0.0 | 0.1 | 0.0 | 1.0 | 0.0 | 6.2 | 0.0 | 5.3 | 1.9 | 0.0 | 0.0 | 0.9 | 2.7 | 1.1 |
| MEX | 0.4 | 5.4 | 2.0 | 17.6 | 1.0 | 32.7 | 0.0 | 26.0 | 12.7 | 1.3 | 3.4 | 54.7 | 425 | 18.0 |
| Rosetta | 0.5 | 0.8 | 2.6 | 1.2 | 1.7 | 10.0 | 6.6 | 5.6 | 15.1 | 2.7 | 1.4 | 27.7 | 210 | 32.3 |
| VEX | 1.9 | 5.0 | 1.0 | 3.3 | 0.2 | 35.7 | 2.3 | 4.2 | 39.1 | 0.0 | 2.8 | 40.1 | 155 | 5.2 |
| Herschel | 17.1 | 53.5 | 9.4 | 94.0 | 68.1 | 105.2 | 5.9 | 47.5 | 83.8 | 24.5 | 27.5 | 22.2 | 761 | 61.3 |
| Planck | 3.7 | 30.9 | 52.6 | 79.6 | 29.2 | 200.2 | 7.8 | 114 | 85.8 | 43.3 | 14.5 | 42.9 | 599 | 137 |
| PROBA-2 | 1.2 | 0.2 | 0.4 | 0.6 | 0.0 | 2.5 | 0.0 | 3.6 | 3.8 | 1.3 | 0.0 | 8.5 | 22.6 | 3.9 |
| Gaia | 40.0 | 185 | 111 | 161 | 213 | 379 | 28.7 | 102 | 129 | 58.7 | 53.7 | 177 | 1747 | 305 |
| LPF | 0.0 | 0.5 | 0.0 | 0.0 | 0.0 | 0.0 | 0.0 | 0.1 | 0.0 | 0.0 | 0.0 | 0.0 | 3.7 | 0.0 |
| ExoMars | 0.0 | 0.3 | 0.0 | 3.0 | 0.0 | 0.0 | 0.0 | 1.5 | 0.2 | 0.0 | 0.0 | 19.3 | 12.2 | 4.0 |
| BepiColombo | 0.0 | 0.0 | 0.0 | 0.0 | 0.0 | 0.0 | 0.0 | 0.1 | 6.6 | 0.0 | 0.0 | 2.1 | 2.6 | 1.0 |
| CHEOPS | 0.0 | 0.0 | 0.0 | 0.0 | 0.1 | 0.0 | 0.0 | 0.0 | 0.0 | 0.0 | 0.0 | 0.2 | 0.9 | 0.1 |
| Solar Orbiter | 0.2 | 0.1 | 0.0 | 0.0 | 0.0 | 3.2 | 0.0 | 0.0 | 0.4 | 0.2 | 0.0 | 0.4 | 18.8 | 0.2 |
| HST | 64.5 | 414 | 168 | 423 | 335 | 317 | 94.2 | 129 | 361 | 123 | 176 | 202 | 8798 | 338 |
| Cassini | 5.4 | 6.3 | 6.2 | 6.0 | 0.6 | 12.5 | 4.2 | 7.9 | 24.6 | 6.4 | 6.4 | 13.9 | 1182 | 16.4 |
| DoubleStar | 0.0 | 0.0 | 0.1 | 0.6 | 0.0 | 55.5 | 0.0 | 0.0 | 2.5 | 0.0 | 0.0 | 1.4 | 19.4 | 1.9 |
| Suzaku | 0.7 | 1.9 | 0.3 | 11.6 | 2.3 | 27.3 | 0.5 | 39.5 | 470 | 2.6 | 2.8 | 9.4 | 271 | 30.5 |
| AKARI | 3.3 | 18.3 | 11.6 | 11.9 | 20.4 | 35.8 | 0.2 | 18.5 | 245 | 62.9 | 7.0 | 15.2 | 191 | 60.9 |
| Hinode | 7.0 | 4.1 | 2.4 | 0.0 | 0.3 | 110.8 | 0.7 | 58.0 | 218 | 30.3 | 0.3 | 14.6 | 438 | 23.1 |
| CoRoT | 0.2 | 0.7 | 14.3 | 2.5 | 2.4 | 1.0 | 5.0 | 0.3 | 0.5 | 0.0 | 0.1 | 2.5 | 9.0 | 7.8 |
| Chanrayaan | 0.0 | 0.0 | 0.0 | 2.4 | 0.0 | 7.8 | 0.0 | 52.6 | 5.1 | 0.1 | 0.0 | 1.8 | 43.8 | 5.8 |
| IRIS | 0.2 | 1.3 | 0.7 | 0.0 | 0.2 | 87.9 | 0.2 | 17.3 | 18.0 | 7.5 | 0.0 | 2.2 | 125 | 5.4 |
| Hitomi | 0.0 | 1.0 | 0.0 | 0.8 | 0.0 | 0.0 | 0.0 | 0.0 | 41.9 | 0.1 | 0.0 | 0.0 | 22.5 | 0.5 |
| MICROSCOPE | 0.0 | 0.0 | 0.0 | 0.0 | 0.0 | 0.0 | 0.0 | 0.0 | 0.0 | 0.0 | 0.0 | 0.0 | 0.2 | 0.1 |

The author's locations were analysed in the three ways given above. The names of the ESA Member States together with selected major space powers are listed using their ISO 3166 country codes along the top of the tables, while the number of publications are given separately for each mission and country. The results for the first authors of the publications from the ESA-led missions (Table 3) are shown graphically in Figs. 3, 4, and 5.

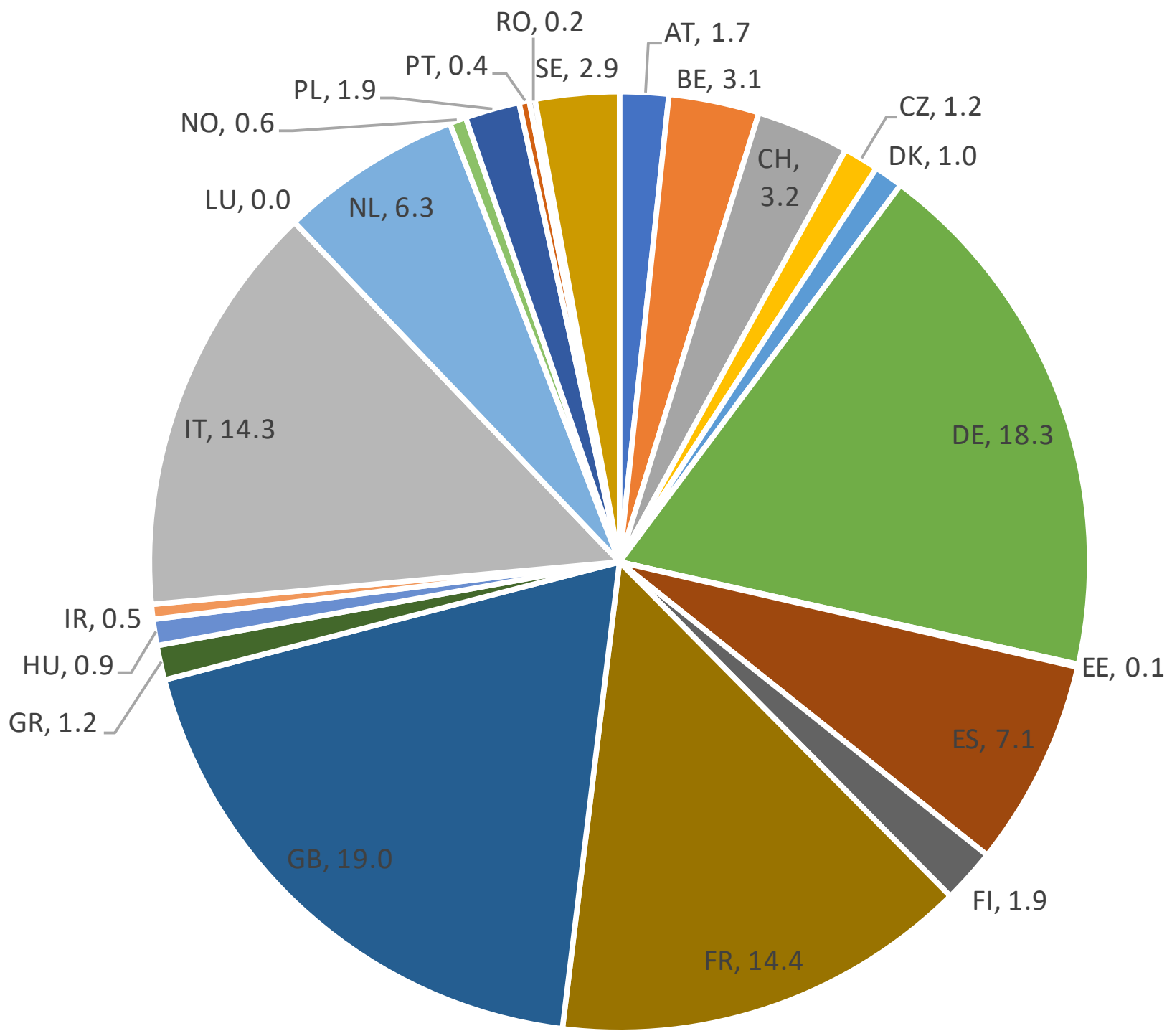

**Fig. 3** Percentages of first author publications for the ESA-led missions for each ESA Member State.

Figure 3 displays the percentage of publications from each ESA Member State with respect to the total from all Member States and it reveals that the bulk of the publications are associated with first authors located in Germany, France, United Kingdom, and Italy. To better understand any trends or correlations, we also calculated the average financial contributions to the ESA Science Programme by each Member State between 2000 and 2021. For Member States that joined ESA after 2000, the average values from the year of joining to 2021 were used. These are indicated as "% Contrib" in Table 3. Since the contributions are proportional to the Gross Domestic Product (GDP) of each Member State, they give an indication of the relative economic weights of the Member States. Dividing the relative publication numbers by the financial contributions gives a measure of how a Member State's scientists were able to exploit and lead (as first authors) the opportunities afforded by the Science Programme compared to the financial contribution that their country makes to the Programme.

Figure 4 shows, for each Member State, the relative fraction of first-author publications as a function of the fraction of their contribution to the ESA Science Programme. The distri-bution appears to follow a linear correlation in agreement with the straight line, which shows equal contributions to publications and finances, although some scatter is clearly present. This is easier to see in Fig. 5, where we take the ratio of the two quantities shown in Fig. 4 to better highlight the scatter as well as the uncertainties (which, as already mentioned, are assumed to follow a Poisson distribution on the number of publications).

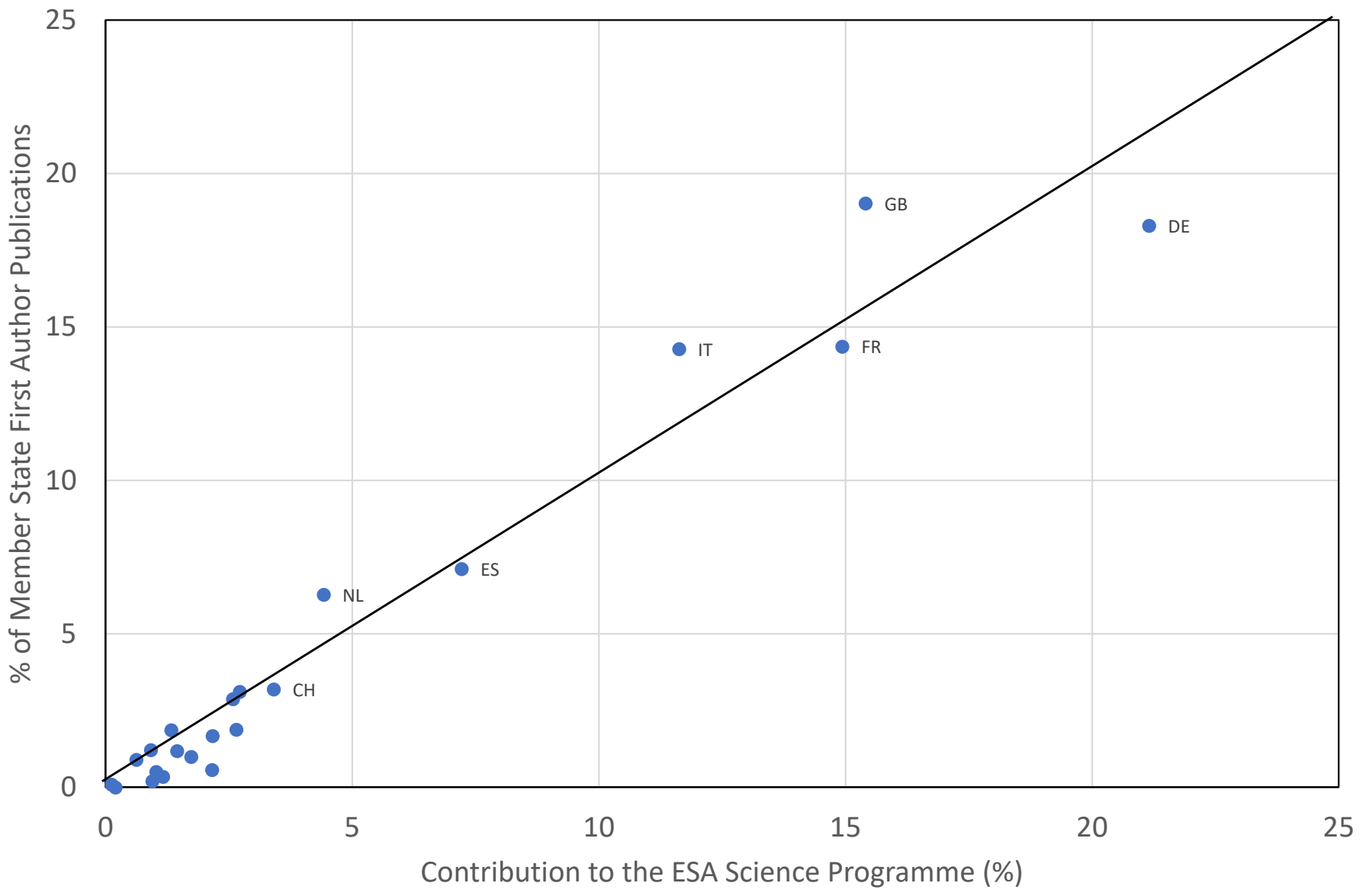

**Fig. 4** Percentages of first author publications for each ESA Member State as a function of the 2000–2021 financial contributions of those countries to the Science Programme. The seven countries making the largest financial contributions to the ESA Science Programme are labelled . Error bars are too small to be visible.

Examining the four countries that each contributed more than 10 % of the Science Programme's finances reveals that scientists located in the United Kingdom, Germany, France, and Italy had 4362, 4195, 3293 and 3275 first author publications, respectively (see Table 3). Comparing these numbers with the financial contributions to the Science Programme (as we do in Figs. 4 and 5) shows that scientists located in the United Kingdom and Italy both produced 23% more first author publications than predicted by these countries contributions. Similarly, scientists located in France and Germany produced 4% fewer and 14% fewer first author papers than predicted by their contributions, respectively.

Of the other ESA Member States, three stand out as having exceptionally high publication-to-contribution ratios:

1. The Netherlands has 1439 first author publications which is 42% more than predicted.
2. Finland has 428 publications which is 40% more than predicted.
3. Hungary has 206 publications which is 45% more than predicted.

We note that of the 206 first author publications from Hungary a high fraction (56 publications) use results from Gaia and 31 use Herschel results. It is noticeable that some of the smaller Member States have low first author publication-to-contribution ratios (see Fig. 5):

1. Denmark with 229 first author publications which is 43 % less than predicted.

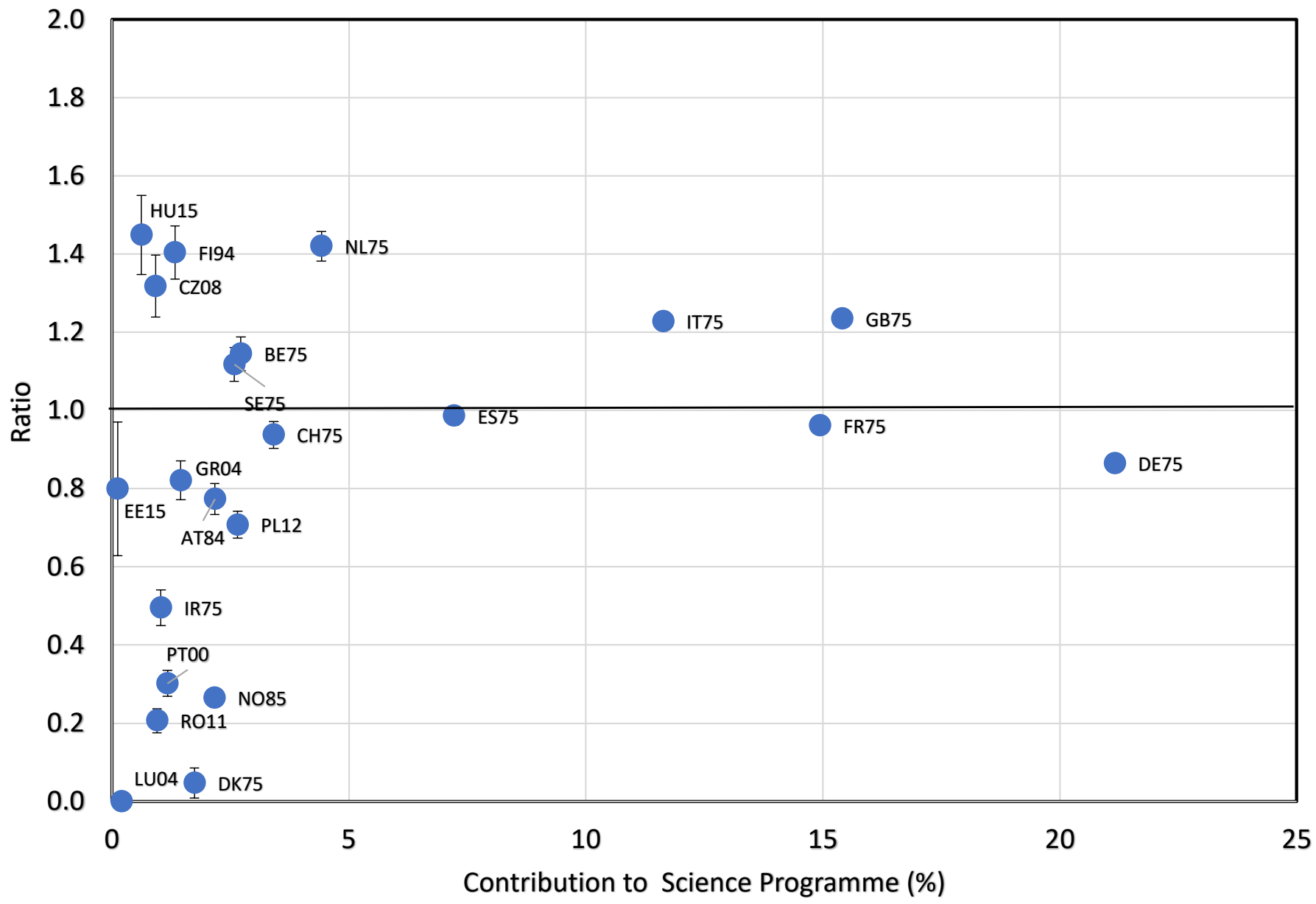

**Fig. 5** Ratio between the number of first authors located in each ESA Member State and the average 2000–2021 financial contributions of those countries to the ESA Science Programme. Error bars are only to "guide the eye" as they assume that the publication numbers follow a Poisson distribution. Next to the ISO 3166 country codes of each Member State, we also indicate the last two digits of the jear in which they joined ESA.

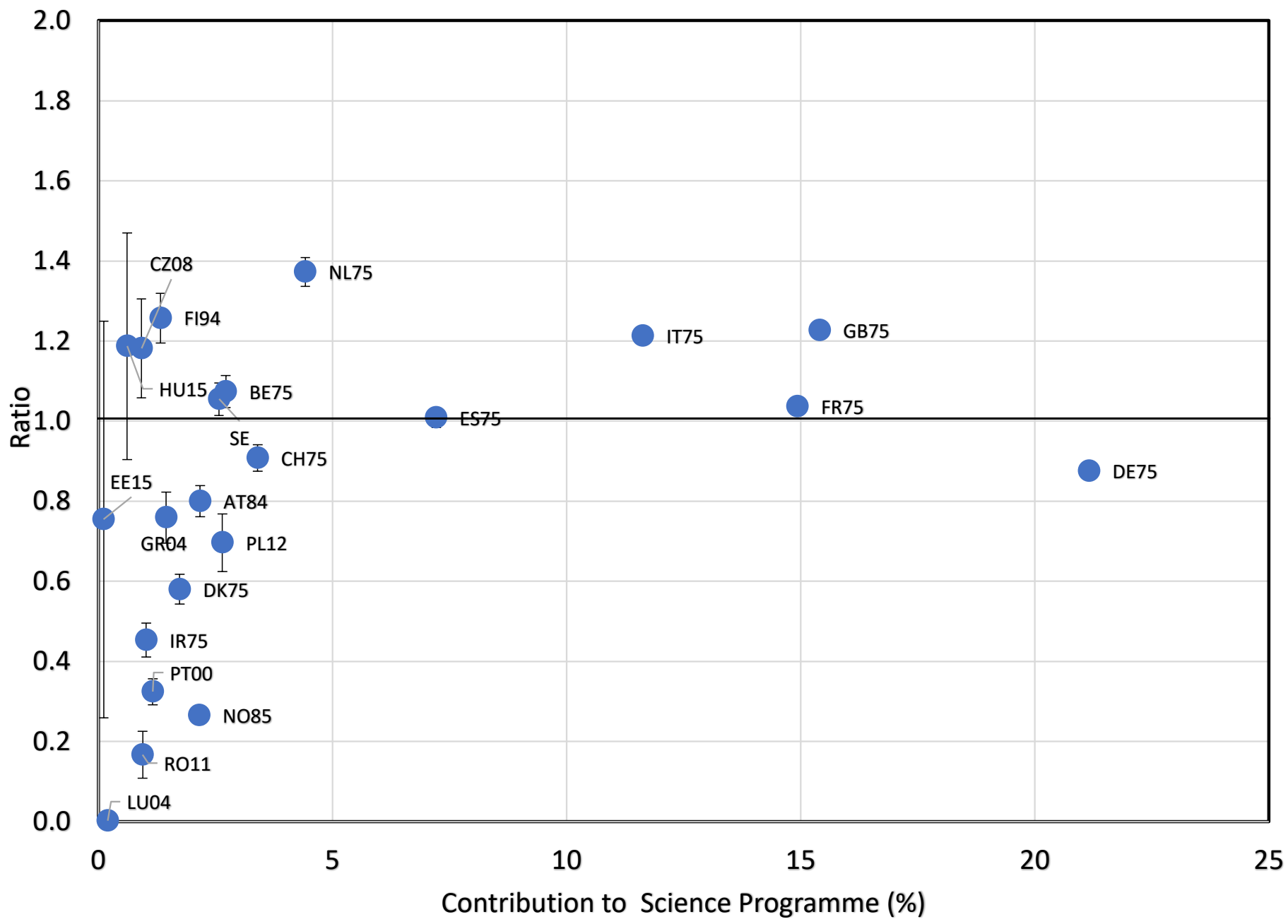

**Fig. 6** Ratio between the number of all authors located in each ESA Member State and the average 2000–2021 financial contributions of those countries to the ESA Science Programme. Error bars are only to "guide the eye" as they assume that the publication numbers follow a Poisson distribution. Next to the ISO 3166 country codes of each Member State, we also indicate the last two digits of the jear in which they joined ESA.

2. Norway with 131 first author publications which is 74 % less than predicted.
3. Ireland with 117 first author publications which is 50 % less than predicted.
4. Portugal with 81 first authors which is 70 % less than expected.

We have also examined the publication statistics using all authors normalised so that each publication contributes 1.0 authors (Tables 4 and 5). This is our preferred method of measuring the *overall* contribution of a country whereas the number of first authors provides a better measure of "leadership" in data exploitation. The above conclusions are not changed significantly for the four largest Member States if this metric is used.

Indeed, the normalised fraction of papers per Member State divided by their contribution to ESA shown in Fig. 6 reveals that the large scatter remains, particularly for smaller Member States. The largest ratio changes are seen from Hungary where the ratio decreases from 1.45 for first author publications to 1.19 when all authors are considered and for the Czech Republic where this ratio decreases from 1.31 to 1.18 (see also Tables 3 and 5).

Considering first-author papers, most of the founding Member States (all of which joined ESA when it was created in 1975) have ratios within ~ 20% of one. Larger variations are seen for two of the founding Member States, namely Denmark and Ireland, with ratios of ~ 0.05 and ~ 0.5, respectively. This might appear unusual if one were to assume that, after almost 40 years, all founding Member States should have reached a similar level of engagement by their communities in space science research and technology. However, it is also clear that for smaller Member States there is a very large scatter and there does not appear to be a direct correlation between the year in which a Member State joined and the ratio of first- or all-authors in publications. To clarify this, we have added to the country codes in Figs. 5 and 6 the last two digits of the year in which they joined ESA. The lack of a correlation between ratio of first-authors and "space maturity" can be seen for instance by comparing Hungary, the Czech Republic, and Finland, which joined, respectively, in 2015, 2008, and 1994: these three countries have some of the highest rates, comparable to that of the Neherlands, which as a founding member joined ESA in 1975.

Therefore, the evaluations of the number of expected papers in the analysis above, particularly for smaller Member States, are at best an indication, because they are normalised by the GDP of each country. To be sure, a number of additional factors besides the GDP can affect the productivity of a scientific community in a particular country and field of research. These could include for instance the scientific and technical priorities of the country, the size and specialisation of its scientific communities, as well as the historical heritage. Therefore, the statistics provided here are not meant to be used to assess or judge the scientific performance of the space science communities in the ESA Member States. Instead, they might serve as a long term tool to help identify potential untapped areas of space research and technology development. Additionally they can also serve as a reference for future studies.

## 3 Publication Metrics

One of the advantages of using ADS is that it provides a number of powerful metrics that can be easily used to provide insights into the publications in a library [3]. For each of the ESA- and partner-led missions, the following were extracted from the ADS:

1. The numbers of refereed publications with >100 citations (i100). This is often used as a measure of the number of highly influential publications.
2. The number of citations (from refereed and non-refereed) publications to the publications in the publication libraries

3. The Hirsch [11] or h-index, which is the position in a citation ranked list where the rank equals the number of citations. For example a facility or individual with an h-index of 100 would have 100 publications each with 100 or more citations. This index gives an estimate of the importance and broad impact of a mission's cumulative results.
4. The m-index. Since the h-index can only increase with time, the m-index, defined as the h-index divided by the time between the first and last publications of a mission is also used to provide an estimate of the productivity of a mission.
5. The 2-year impact factor for each mission with sufficient publications in 2019 and 2020. The impact factor used is the number of citations in 2021 to the mission's refereed publications in 2019 and 2020, divided by the number of such publications in ADS. It may be regarded as a measure of the importance of the scientific output of a mission in the years immediately following publication.

In addition, a dedicated software tool was used to calculate:

1. The percentage of publications whose first authors are located in institutes in the ESA Member States, compared to the total.
2. The number of unique author names from the publications for each mission. This is not the same as the number of first authors as e.g., A.N Other and A. Other will be counted as two authors, while multiple authors who have the same name will be under-counted.

Table 12 shows the metrics for the ESA-led missions. It can be seen that the largest numbers of highly influential publications (i100) are from XMM-Newton [5] and SOHO [6] with 500 and 478, respectively. Of the older missions, IUE [12] has 399 such publications and Hipparcos [9] 264. Publications from the more recently launched Herschel [13] and Planck [14] missions (both launched in 2009) and Gaia [7] (launched in 2013), have had less time to accumulate citations, so their values of i100 of 235, 278 and 178, respectively are also very impressive.

There are large variations in the percentage of ESA Member-State first authors compared to the total first authors for the different missions with the average Member state/total first authors of 59 % for the ESA-led missions. This illustrates the global nature of the exploitation of the results from ESA's Science Programme missions. As expected, missions that are major collaborations with other agencies, such as IUE [12], SOHO [6], and Ulysses [15] have lower ESA Member State first author percentages of 39 %, 35 % and 38 %, respectively. In terms of absolute numbers of publications, from the ESA-led missions that are not in collaboration, then Gaia [7] and XMM-Newton [5] have provided 3778 and 3133 non-ESA Member State first author publications through to 2021.

The number of unique author names is a measure of the size of the community exploiting data from a mission. The numbers given in Table 12 need to be interpreted with caution. An author with publications under A. Other and A.N. Other will appear twice, while authors with the same name will be only counted once. To obtain an estimate of how big this effect is, the 1120 unique names for Exosat [16] were found to correspond to 929 authors, which is 21 % higher than the number of unique names. Thus, this metric is best used for comparison between missions, rather than an absolute measure of community size. Indeed, missions such as INTEGRAL [17] have benefited from contributing to several multi-messenger publications with $\sim$ 1000 authors each, and it could be argued that this does not reliably reflect the users of the mission itself. Unsurprisingly, Gaia [7] and XMM-Newton [5] have the largest numbers of unique author names (25,107 and 18,648), respectively.

The long-lived XMM-Newton [5] and SOHO [6] missions dominate the citation statistics with 261,004 and 229,472, citations respectively. There are also strong showings from IUE [12], Hipparcos [9], Herschel [13], Planck [14] and Gaia [7].

**Table 12** Publication metrics for the ESA-led missions arranged in order of launch date. For missions still in operation, only the launch date is given under "Operations". The total number of refereed publications is given since launch. The number of citations is the number of citations from both refereed and non-refereed publications to the mission's publications. MS/Total is the number of first authors from institutes located in the ESA Member States divided by the number of first authors from all countries. The 2-year impact factor is the number of citations in 2021 to the publications from a mission in 2019 and 2020 divided by the number of such publications.

| ESA-led | Operations | Refereed Publications | | | Authors | Citations | Indicators | | |
|---|---|---|---|---|---|---|---|---|---|
| Mission | | Total No. | No. with >100 citations | MS/Total First Authors | No. Unique Names | | h-index | m-index | 2-year Impact Factor |
| COS-B | 1975-1982 | 146 | 12 | 81% | 380 | 5872 | 12 | 0.8 | - |
| IUE | 1978-1996 | 5105 | 399 | 39% | 8355 | 204250 | 169 | 3.9 | 4.2 |
| Exosat | 1983-1986 | 743 | 52 | 76% | 1122 | 26770 | 79 | 2.2 | 2.0 |
| Giotto | 1985-1992 | 266 | 40 | 71% | 513 | 14988 | 62 | 2.1 | - |
| Hipparcos | 1989-1993 | 2692 | 264 | 46% | 6724 | 121977 | 153 | 4.8 | 6.6 |
| Ulysses | 1990-2005 | 1950 | 118 | 38% | 3302 | 61066 | 107 | 3.7 | 6.8 |
| ISO | 1995-1998 | 2062 | 218 | 60% | 5516 | 106466 | 136 | 5.4 | 3.9 |
| SOHO | 1995- | 6252 | 473 | 35% | 8905 | 229472 | 172 | 6.6 | 4.0 |
| Huygens | 1997-2005 | 222 | 12 | 52% | 802 | 7772 | 46 | 2.2 | 1.0 |
| XMM-Newton | 1999- | 6963 | 500 | 55% | 18648 | 261004 | 178 | 8.5 | 6.3 |
| Cluster | 2000- | 2962 | 132 | 40% | 4736 | 76682 | 109 | 4.4 | 4.1 |
| INTEGRAL | 2002- | 1904 | 145 | 61% | 11213 | 78067 | 115 | 6.1 | 6.9 |
| SMART-1 | 2003-2006 | 70 | 0 | 70% | 352 | 795 | 17 | 1.0 | 0.3 |
| Mars Express | 2003- | 1464 | 70 | 54% | 3793 | 44729 | 88 | 4.9 | 3.6 |
| Rosetta | 2004-2016 | 1314 | 48 | 67% | 3831 | 29049 | 77 | 4.5 | 3.5 |
| Venus Express | 2005-2014 | 737 | 15 | 53% | 2135 | 15797 | 55 | 3.7 | 3.2 |
| Herschel | 2009-2013 | 3284 | 235 | 58% | 11801 | 125619 | 137 | 12.5 | 5.1 |
| Planck | 2009-2013 | 2742 | 278 | 43% | 9424 | 158333 | 155 | 15.5 | 17.4 |
| PROBA-2 | 2009- | 137 | 2 | 59% | 747 | 2644 | 28 | 2.8 | 3.2 |
| Gaia | 2013- | 6403 | 178 | 41% | 25107 | 144049 | 123 | 17.6 | 8.1 |
| LISA Pathfinder | 2015-2017 | 33 | 2 | 81% | 275 | 753 | 9 | 1.8 | 1.8 |
| ExoMars 2016 | 2016- | 138 | 0 | 56% | 1032 | 1432 | 20 | 2.9 | 7.8 |
| BepiColombo | 2018- | 71 | 0 | 76% | 947 | 419 | 10 | 3.3 | 3.4 |
| CHEOPS | 2019- | 38 | 0 | 95% | 448 | 266 | 9 | 9.0 | - |
| Solar Orbiter | 2020- | 114 | 1 | 72% | 1684 | 1206 | 17 | 17.0 | - |

For the three indicators, the h-indices are dominated by the long-lived missions XMM-Newton [5], SOHO [6], and IUE [12], but the more recent Herschel [13], Planck [14], and Gaia [7] missions do well explaining their high m values. The mission with the highest 2-year impact factor is Planck [14] (17.4) followed by Gaia [7] and ExoMars 2016 [18], with impact factors of 8.1 and 7.8, respectively. The impact factors of Hipparcos [9] (1989–1993) and Ulysses [15] (1990–2005) remain an impressive 6.6 and 6.8, respectively. The INTEGRAL [17] impact factor of 6.9 is in part due to a number of multi-messenger publications to which the mission contributed.

Table 13 shows the same metrics as above for the partner-led missions. HST dominates the statistics with 2474 publications with more than 100 citations (i100) followed by Cassini [8] with 133. If the missions that are led by an agency from an ESA Member State are ignored (CoRoT [19] and MICROSCOPE [20]) then 29 % of the first authors from the remaining missions are located in the ESA Member States. In terms of absolute numbers, the 34 % of HST first authors located in the ESA Member States means a total of 6573 pub-lications led by authors from ESA Member State institutes through to 2021. Cassini [8] has provided 757 such publications and Hinode [21] 519, also illustrating their importance to ESA Member States scientists. HST also dominates the citation numbers, with more than a million citations to the publications from the mission and it has the highest h-index of any of the missions (324).

**Table 13** Publication metrics for the partner-led missions arranged in order of launch date. For missions still being operated, only the launch date is given under "Operations". The total number of refereed publications is given since launch. The number of citations is the number of citations from both refereed and non-refereed publications to the mission's publications. MS/Total is the number of first authors from institutes located in the ESA Member States divided by the number of first authors from all countries. The 2-year impact factor is the number of citations in 2021 to the publications from a mission in 2019 and 2020 divided by the number of such publications.

| Partner-Led | Operations | Refereed Publications | | | Authors | Citations | Indicators | | |
|---|---|---|---|---|---|---|---|---|---|
| Mission | | Total No. | No. with >100 citations | MS/Total First Authors | No. Unique Names | | h-index | m-index | 2-year Impact Factor |
| HST | 1990- | 19333 | 2474 | 34% | 38800 | 1057706 | 324 | 10.8 | 9.6 |
| Cassini | 1997-2017 | 2227 | 133 | 34% | 4027 | 79268 | 111 | 4.6 | 2.9 |
| Double Star | 2003-2007 | 161 | 3 | 39% | 628 | 2835 | 30 | 1.8 | 4.7 |
| Suzaku | 2005-2015 | 1139 | 54 | 19% | 4053 | 35458 | 81 | 5.4 | 4.0 |
| AKARI | 2006-2011 | 1086 | 32 | 30% | 5320 | 25888 | 70 | 4.7 | 4.3 |
| Hinode | 2006- | 1442 | 75 | 36% | 2441 | 51382 | 92 | 6.6 | 3.6 |
| CoRoT | 2006-2013 | 238 | 20 | 78% | 995 | 10477 | 55 | 3.9 | 3.7 |
| Chandrayaan-1 | 2008-2009 | 164 | 1 | 24% | 723 | 3186 | 30 | 2.3 | 2.4 |
| IRIS | 2013- | 449 | 8 | 35% | 1007 | 10719 | 49 | 6.1 | 5.4 |
| Hitomi | 2016-2016 | 74 | 2 | 13% | 821 | 1109 | 14 | 2.8 | 2.4 |
| MICROSCOPE | 2016-2018 | 17 | 1 | 94% | 71 | 491 | 9 | 1.8 | 7.0 |

Figures 7 to 26 show the evolution of selected metrics against year after launch for ESA-led missions[1]. The time when each mission was operational is also shown in the upper panels. For each mission, as well as the annual number of refereed publications, the number of citations per year, the annual number of i100 publications and the 2-year impact factor (the number of citations in year $n$ to the publications in years $n-1$ and $n-2$ divided by the number of such publications) are shown when there are sufficient statistics to evaluate these parameters.

These figures reveal a number of interesting insights:

1. For many missions the number of publications per year increases rapidly after the first ~2–3 years and remains high during operations before decreasing slowly after the end of operations. This "long tail" of publications can last for many years, e.g., 25 years after the end of operations there are still ~25 refereed publications per year from IUE [12], some 20% of the peak publication rate (Fig. 8). This emphasises the importance of long-term scientific archives that provide continued access to relevant data.
2. In contrast, the publication rates from Hipparcos [9], Planck [14] and Gaia [7] took longer to increase towards their peaks. This is likely due to it being necessary to accumulate and analyse all-sky data sets before detailed scientific analysis could be performed (see Figs. 11, 25 and 26).

[1] The figures are arranged in launch chronological order, except for Proba-2, which is shown on the same page as SMART-1

3. For the first 11 years of the Rosetta [22], [23] mission the publication rate remained relatively low (Fig. 22) during and after the encounters with the Steins and Lutetia asteroids. This changed once the mission encountered its main target comet 67P/Churyumov-Gerasimenko which provided for a marked increase in the publication rate.
4. The scientific importance of a number of missions can be seen in the very high number of citations many years after operations started. In particular, the citation rates for four missions still in operations have grown to impressive numbers: XMM-Newton [5] has around 25,000 citations per year after 22 years of operations, SOHO [6] has around 14,000 citations per year after 26 years of operations, Cluster [10] has around 8000 citation per year after 22 years of operations as does INTEGRAL [17] after 19 years of operations.
5. The longevity of the scientific relevance in IUE [12], Hipparcos [9] and Infrared Space Observatory (ISO) [24] results is remarkable; IUE [12] and Hipparcos [9] continue to have around 5000 citations per year some 25 years after operations ended. Similarly, there are around 4000 citations per year to the ISO [24] publications some 26 years after operations ended. Both Planck [14] and Herschel [13] have impressive citation statistics having some 30,000 and 15,000 citations per year some nine years after the ends of operations.

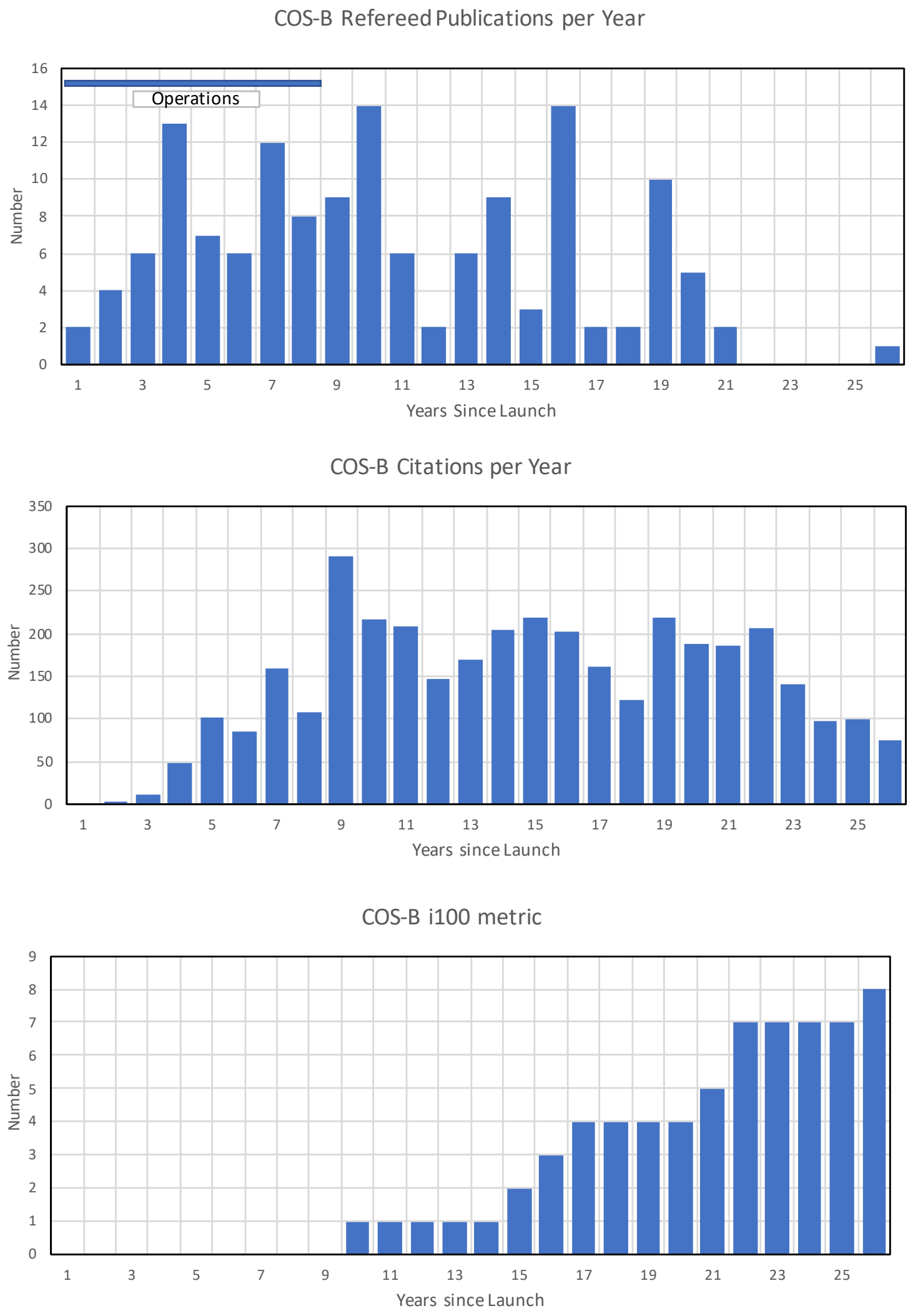

**Fig. 7** Cos-B Publication Metrics

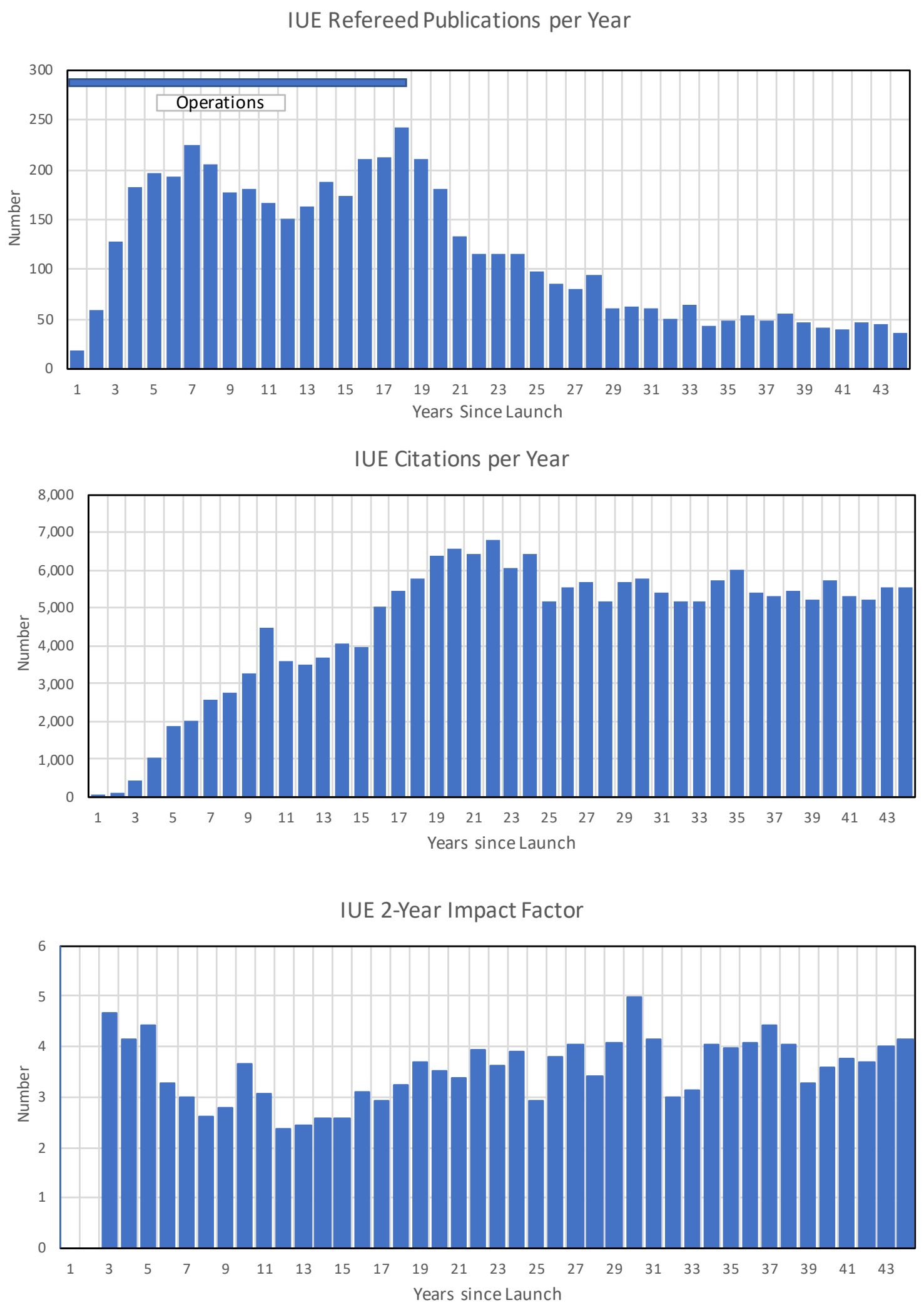

**Fig. 8** IUE Publication Metrics

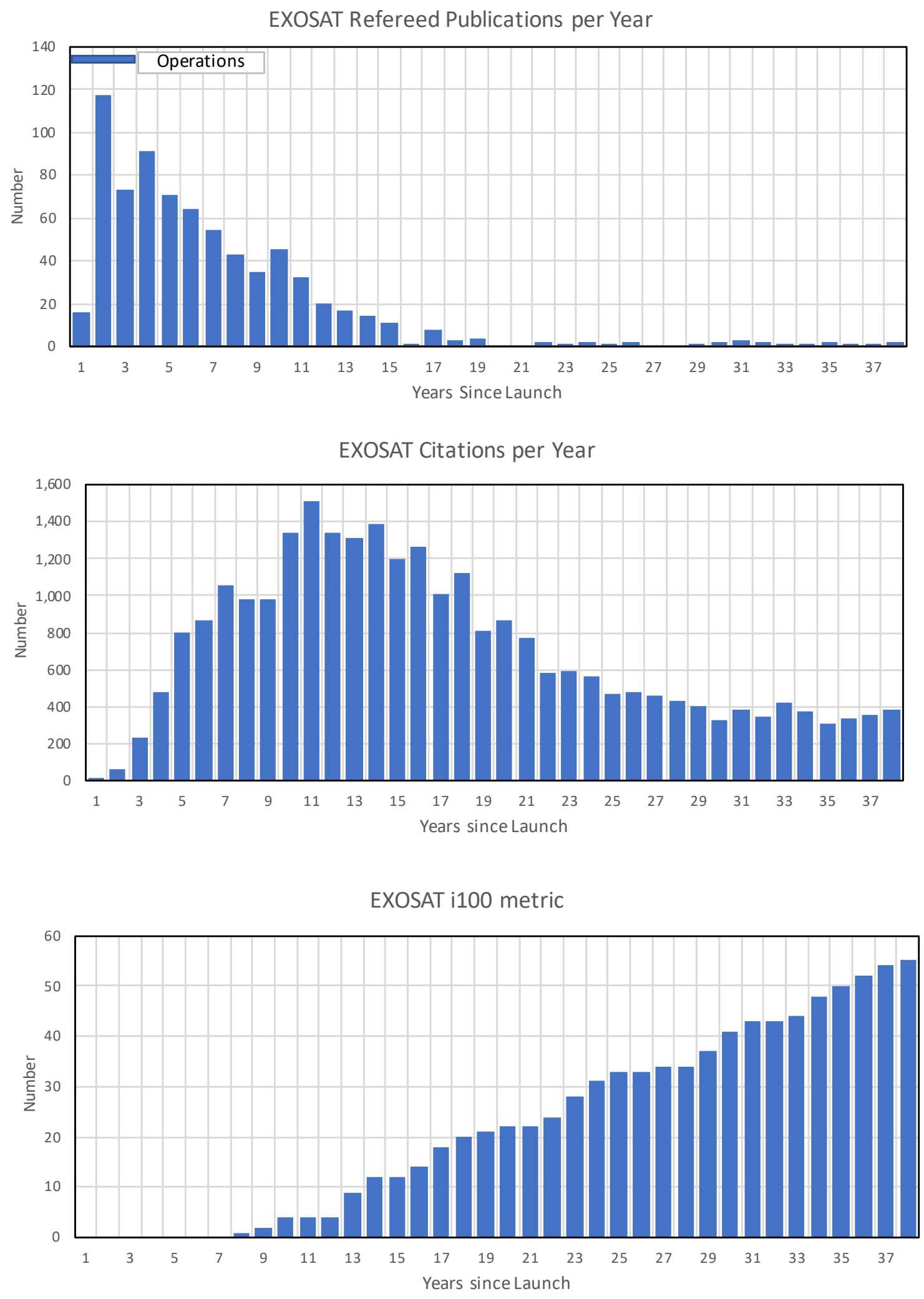

**Fig. 9** Exosat Publication Metrics

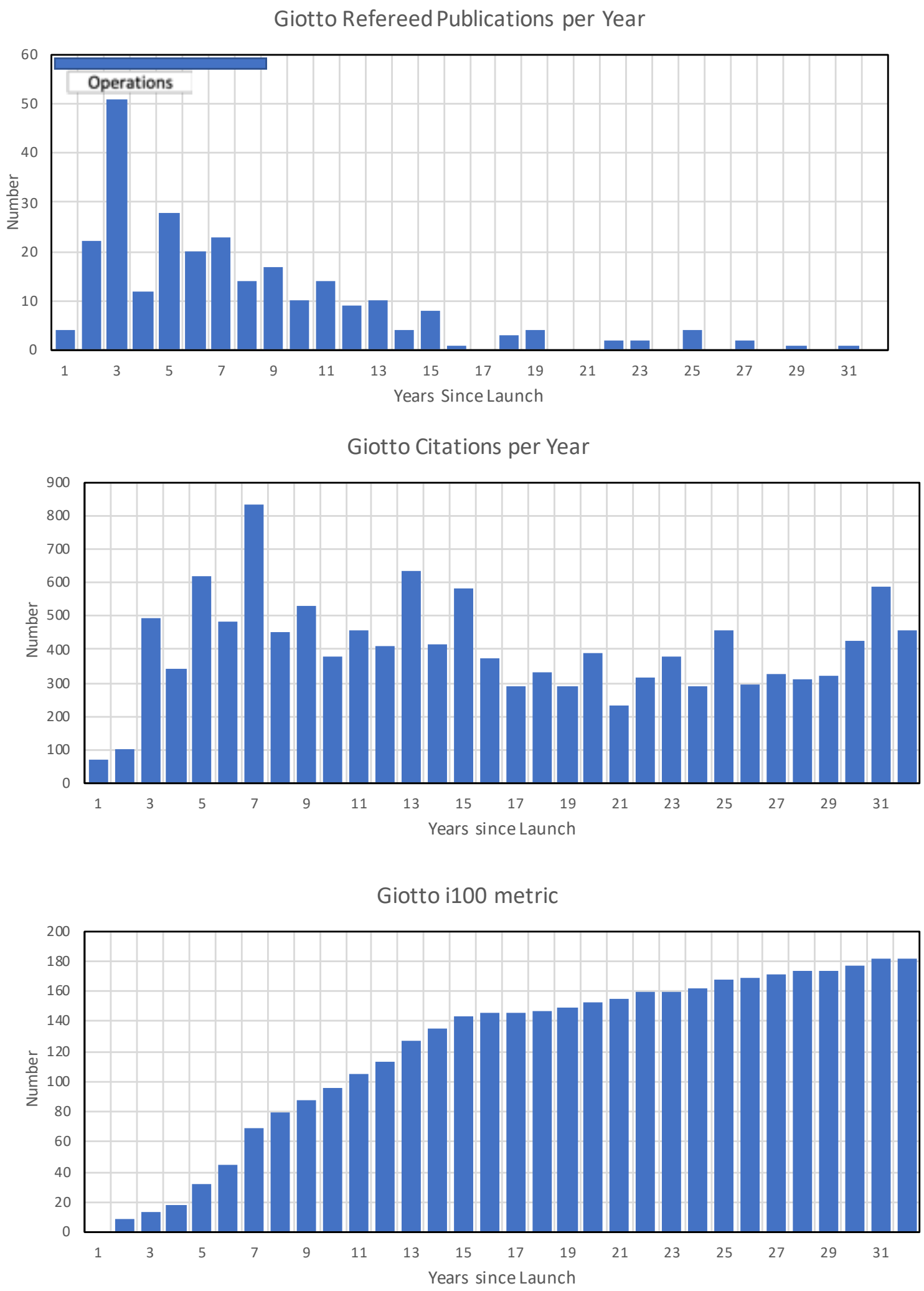

**Fig. 10** Giotto Publication Metrics

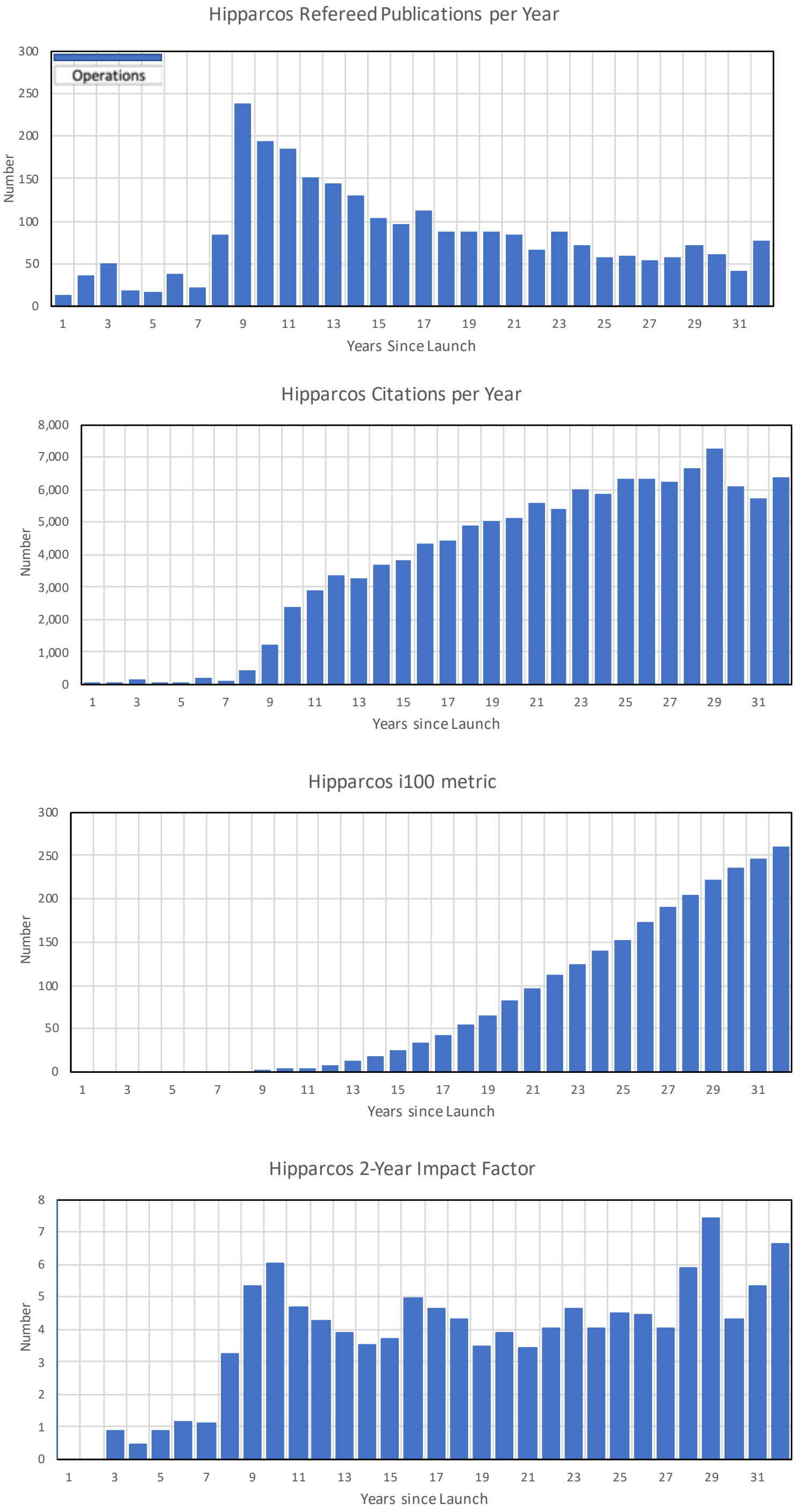

**Fig. 11** Hipparcos Publication Metrics

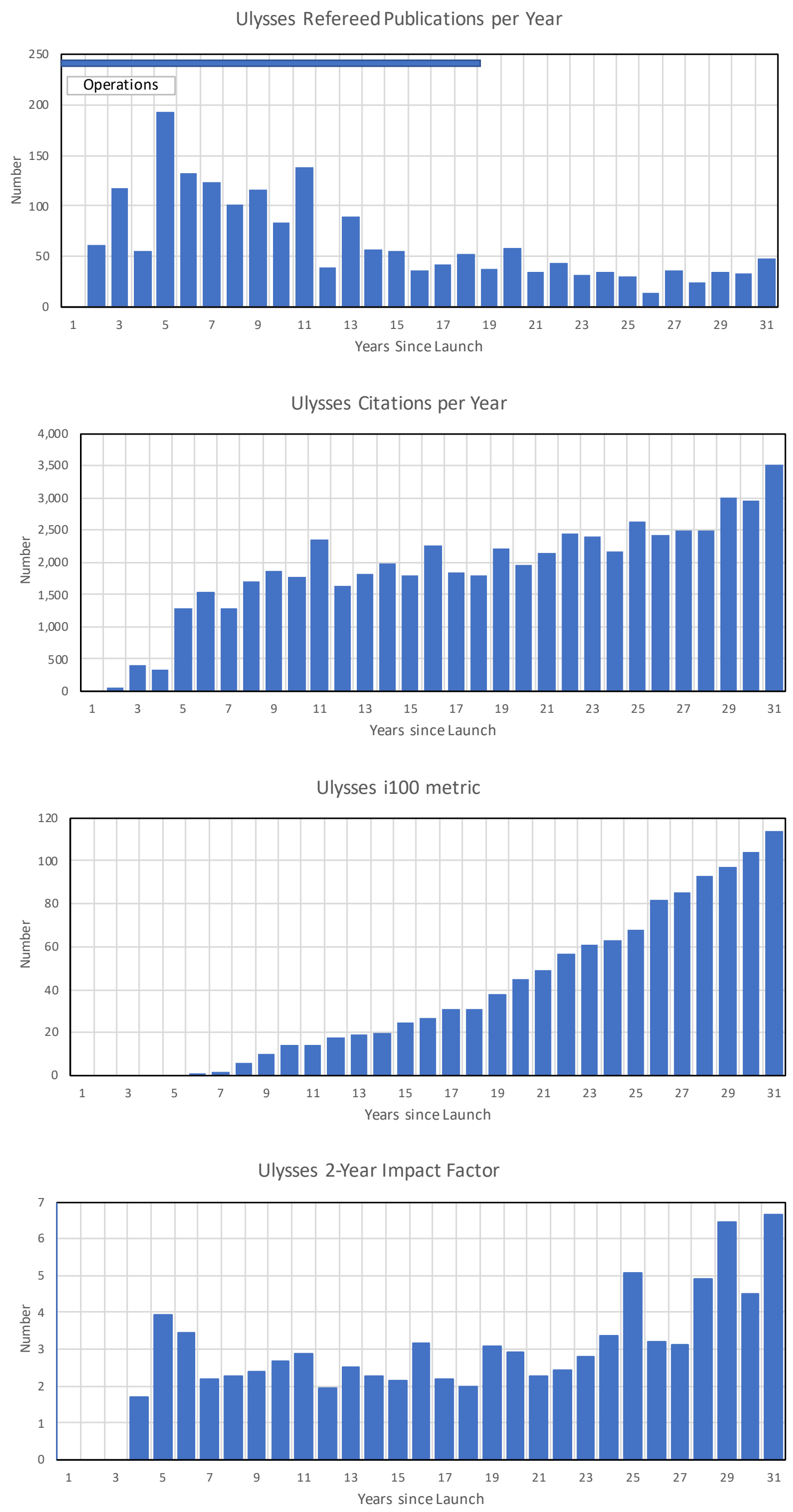

**Fig. 12** Ulysses Publication Metrics

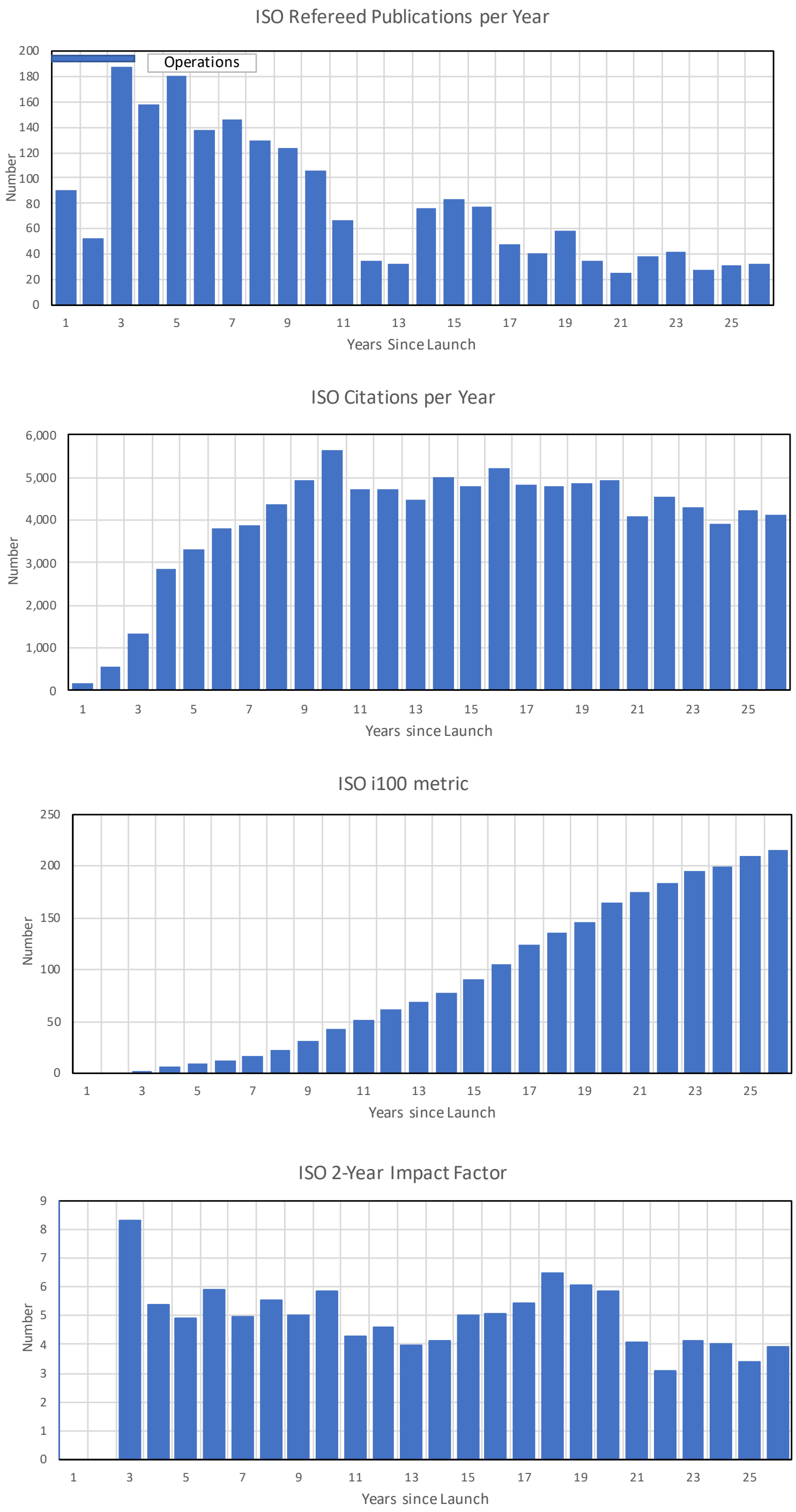

**Fig. 13** ISO Publication Metrics

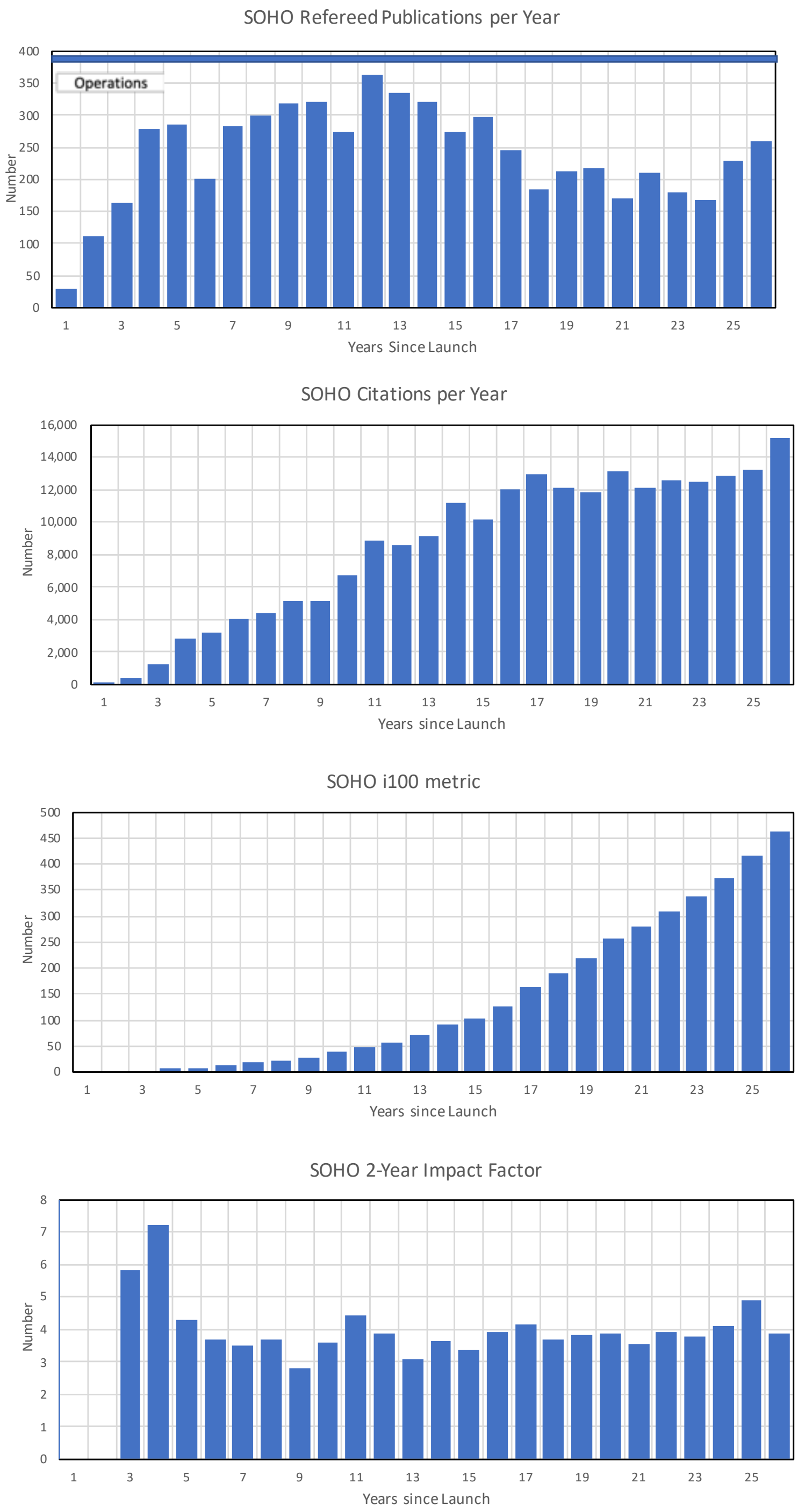

**Fig. 14** SOHO Publication Metrics

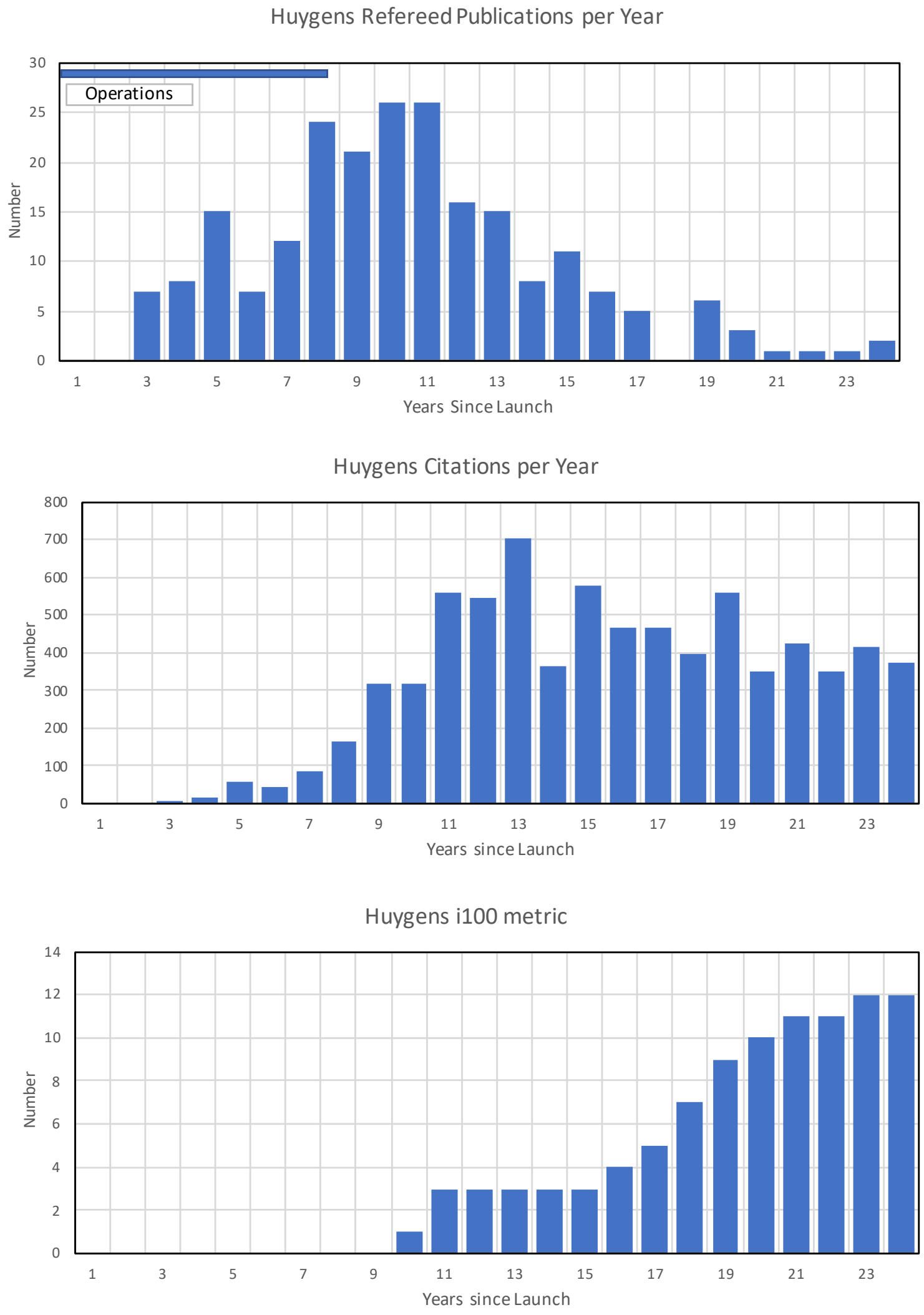

**Fig. 15** Huygens Publication Metrics

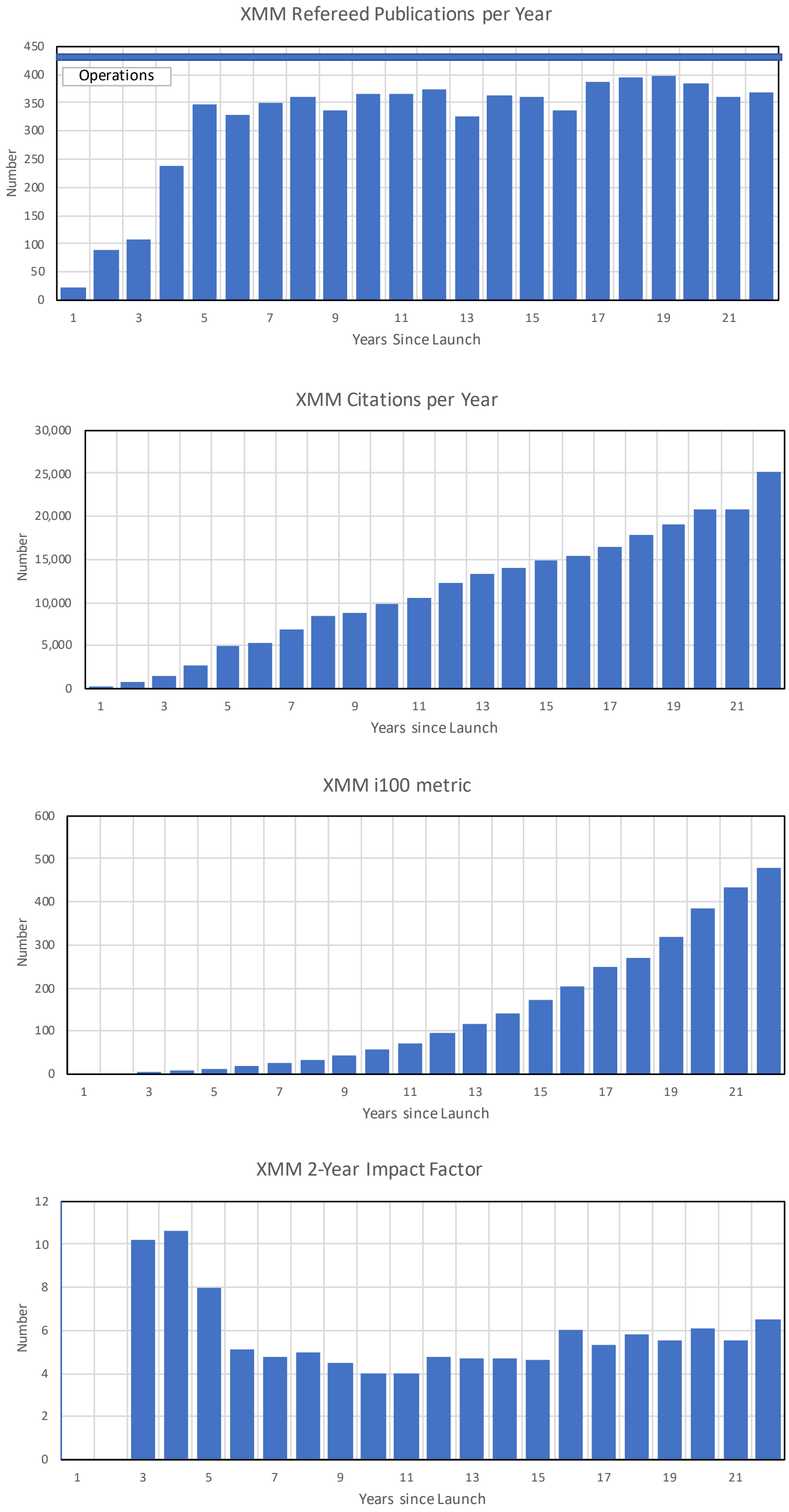

**Fig. 16** XMM-Newton Publication Metrics

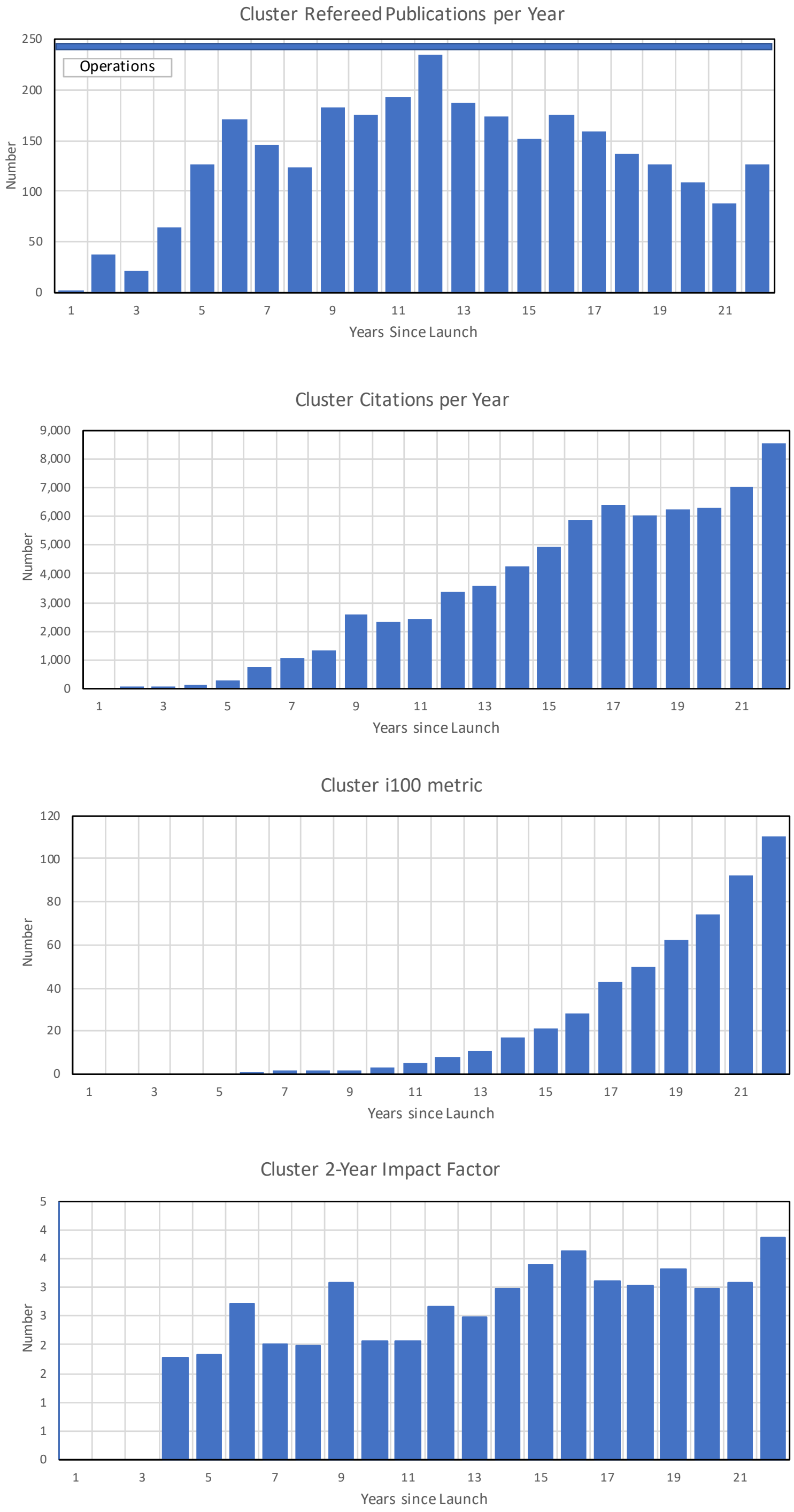

**Fig. 17** Cluster Publication Metrics

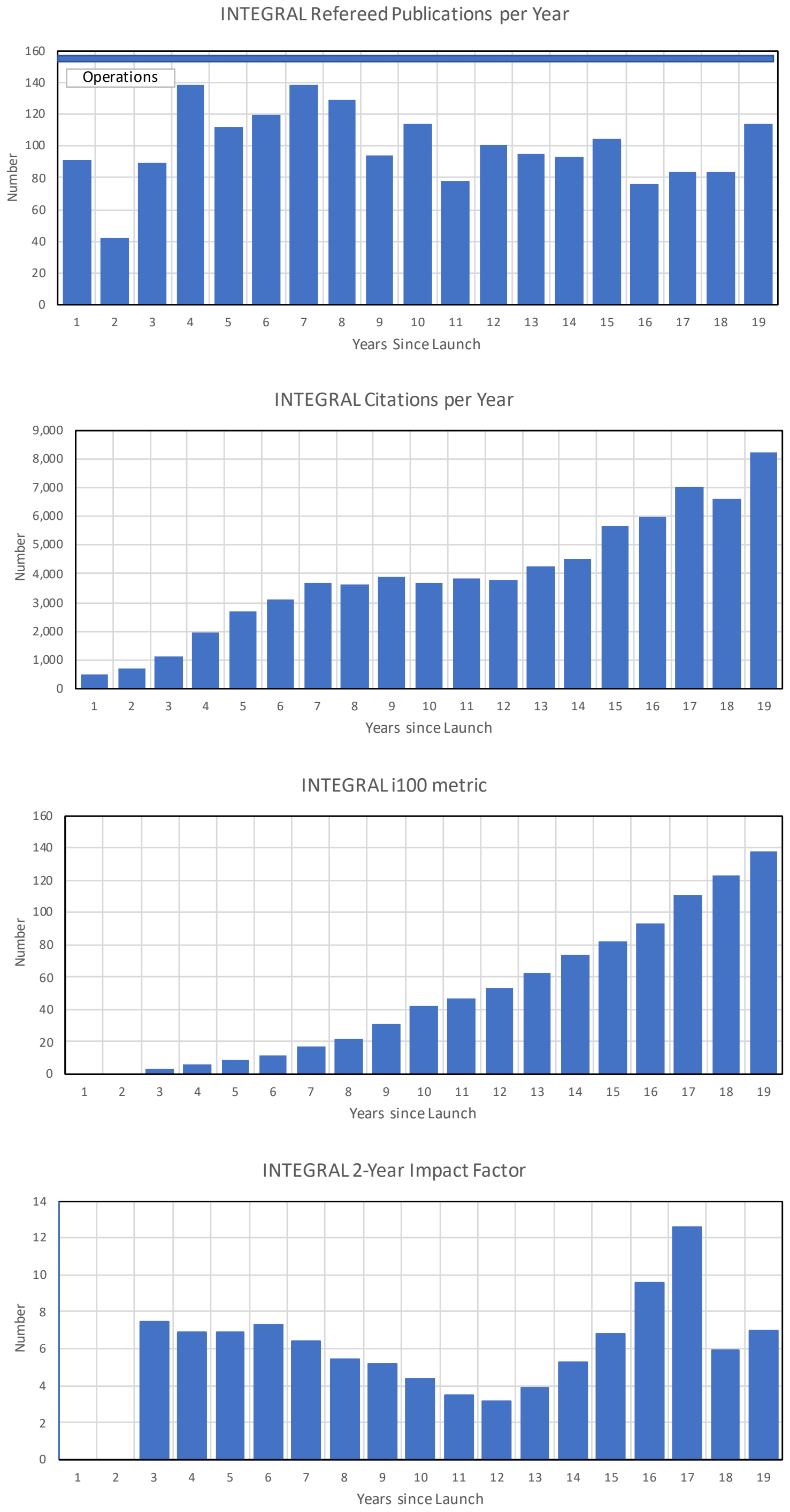

**Fig. 18** INTEGRAL Publication Metrics

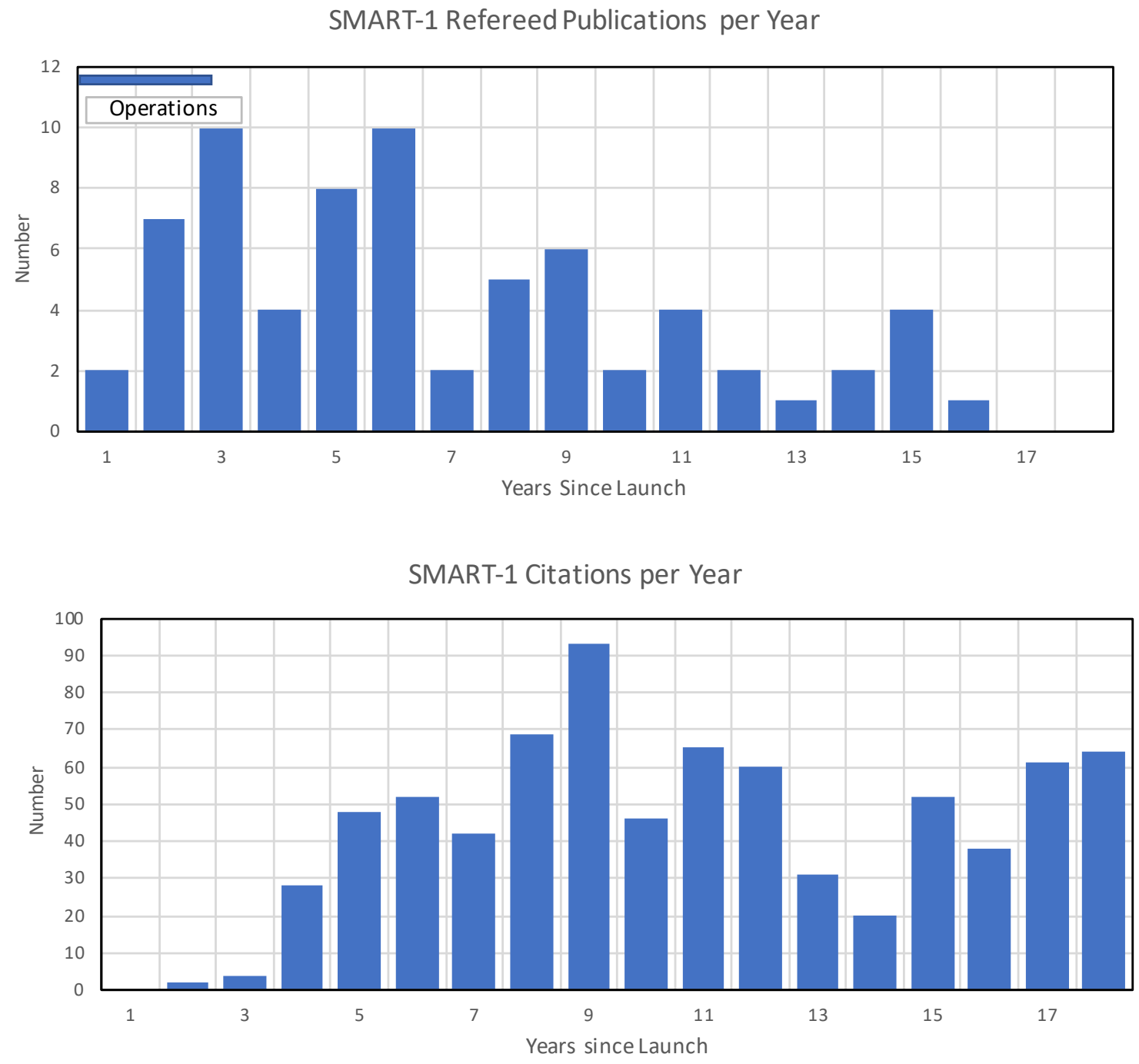

**Fig. 19** SMART-1 Publication Metrics

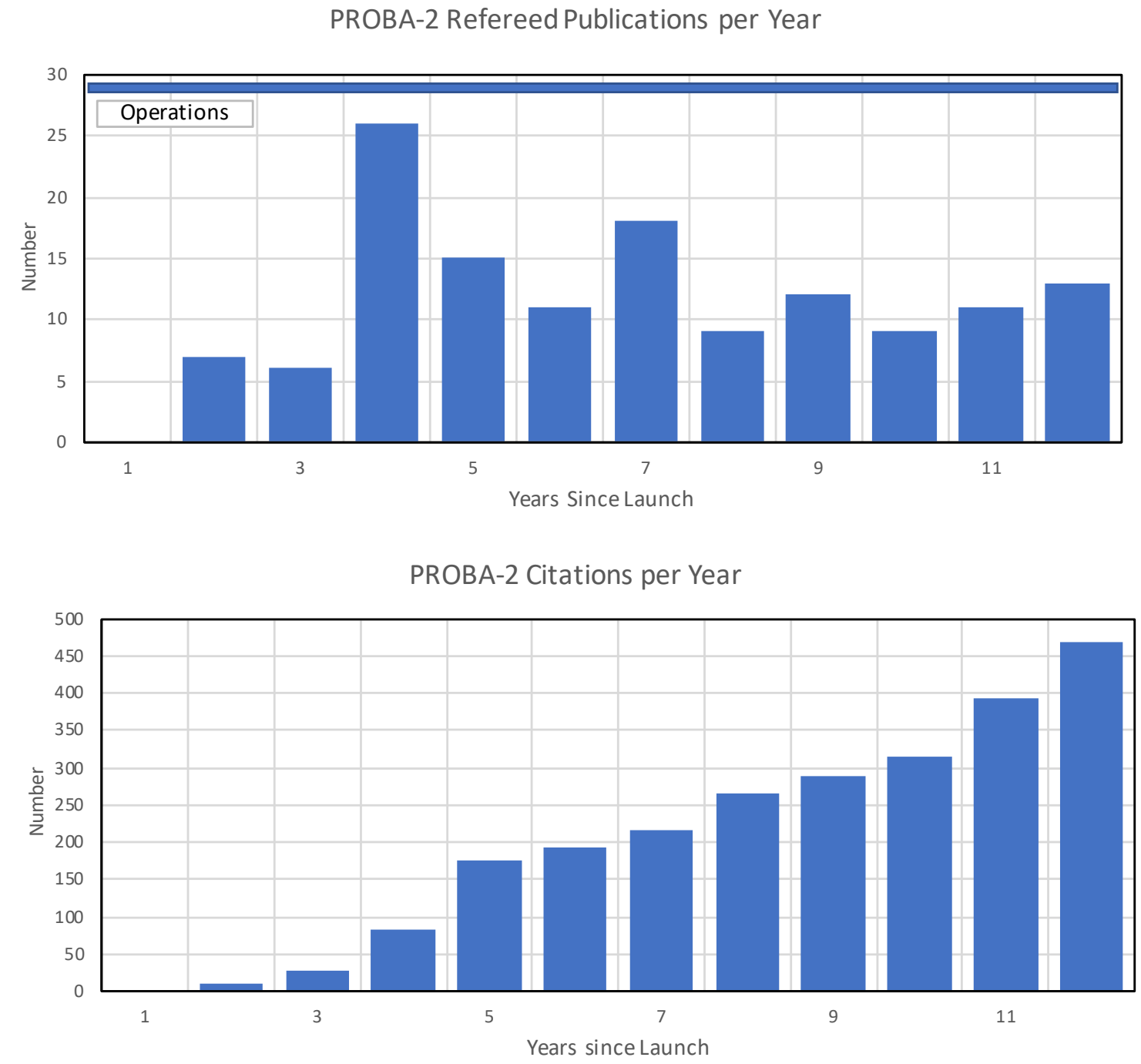

**Fig. 20** Proba-2 Publication Metrics

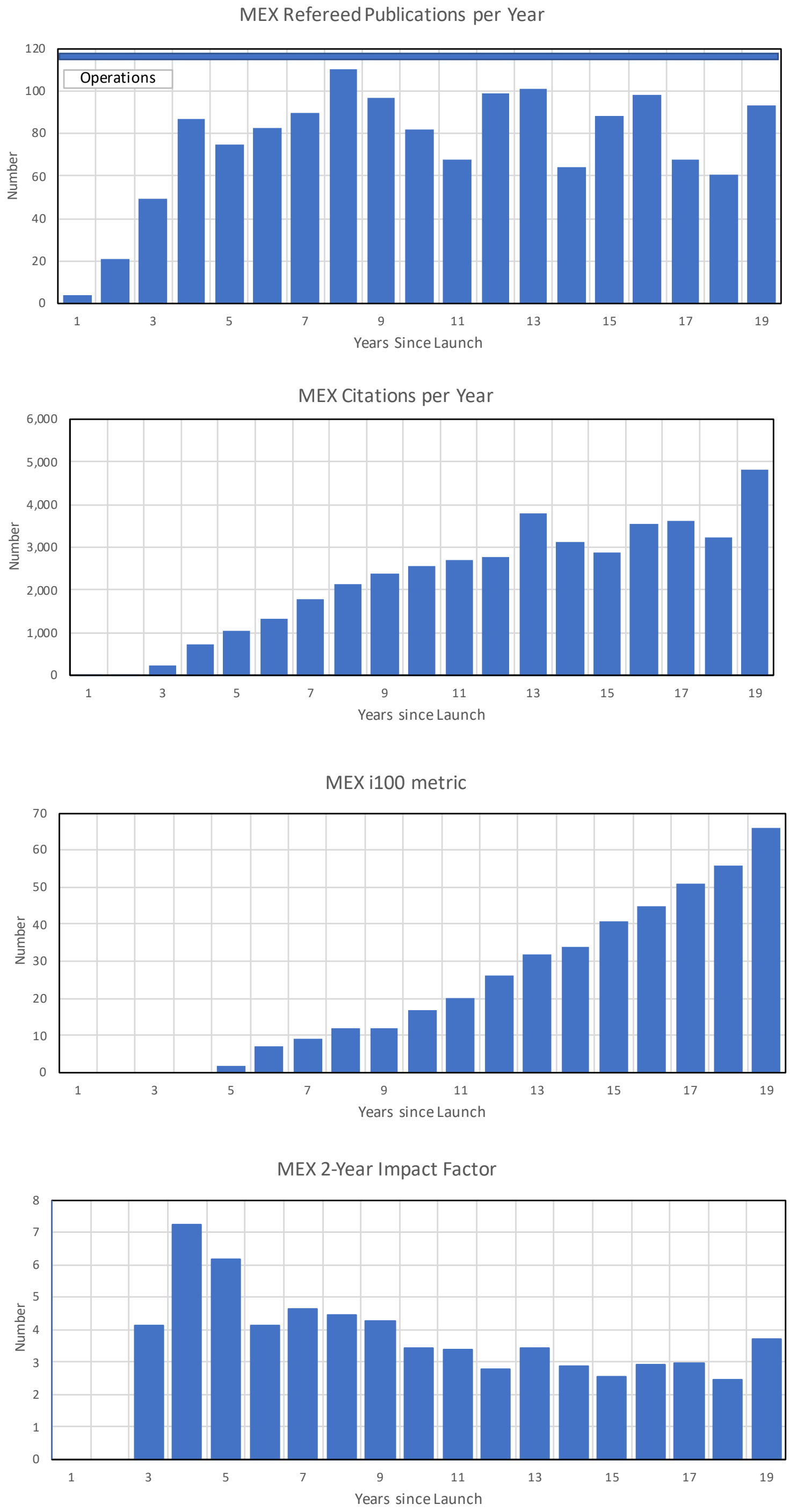

**Fig. 21** Mars Express Mission (MEX) Publication Metrics

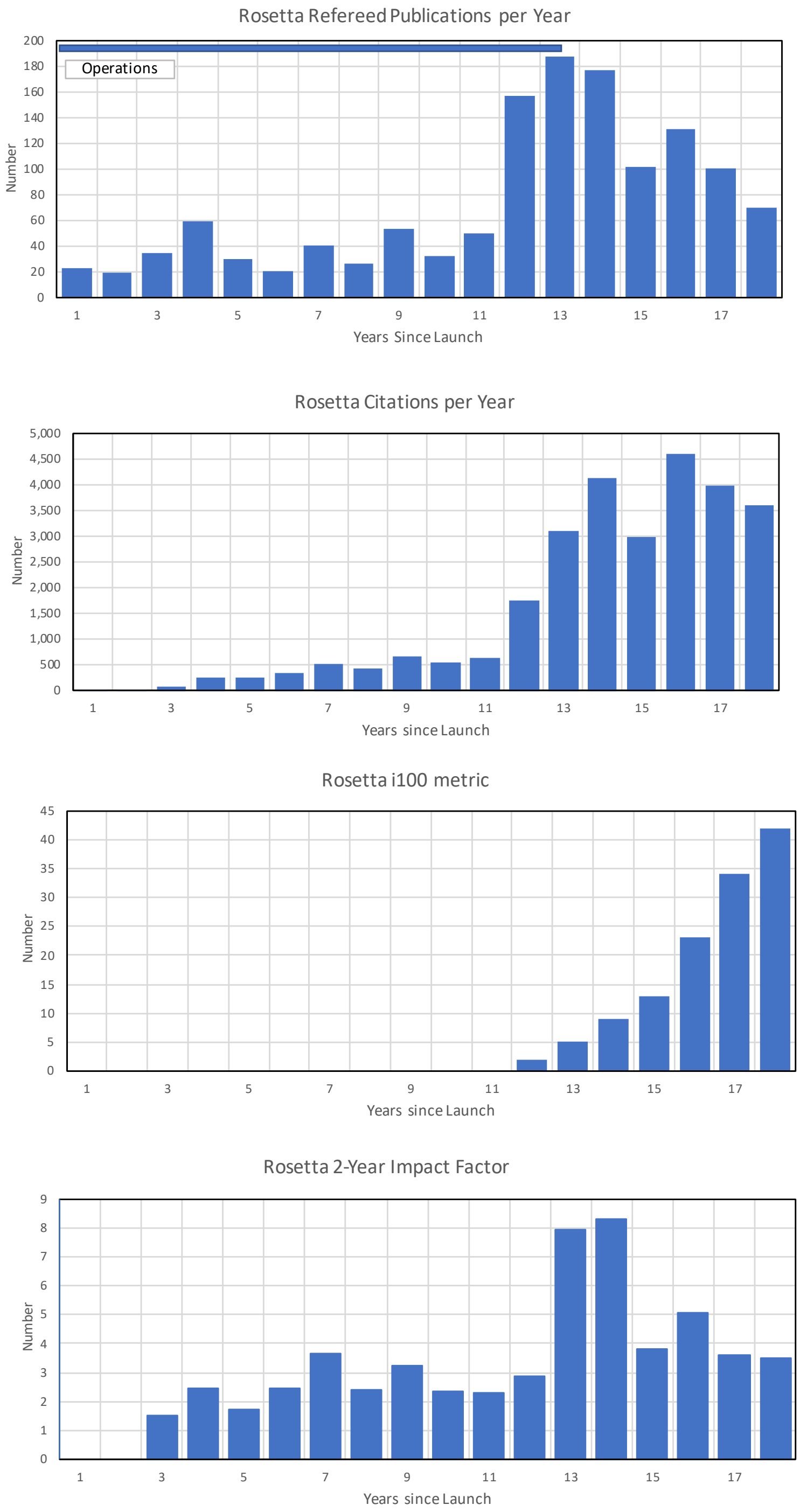

**Fig. 22** Rosetta Publication Metrics

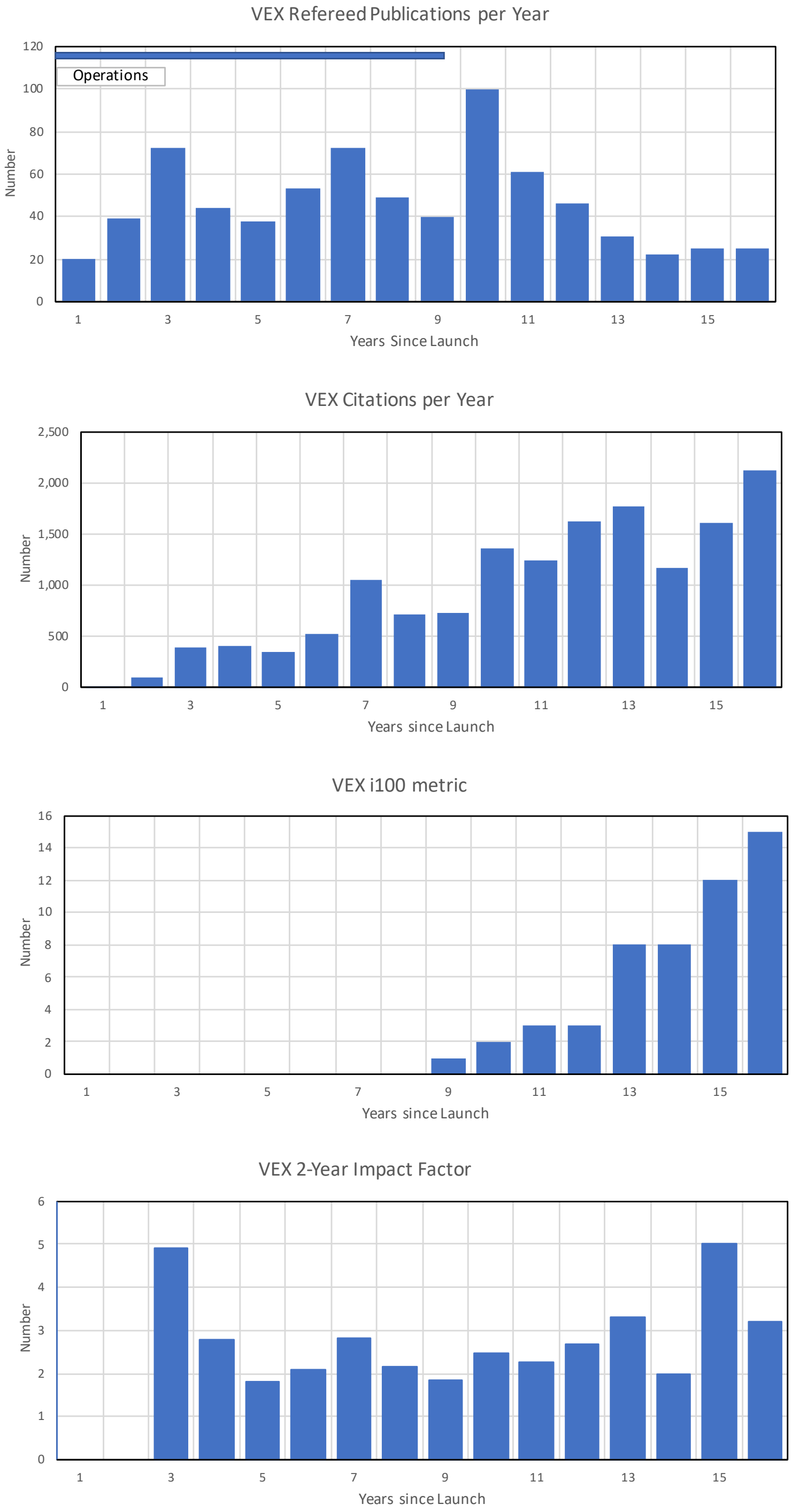

**Fig. 23** Venus Express Mission (VEX) Publication Metrics

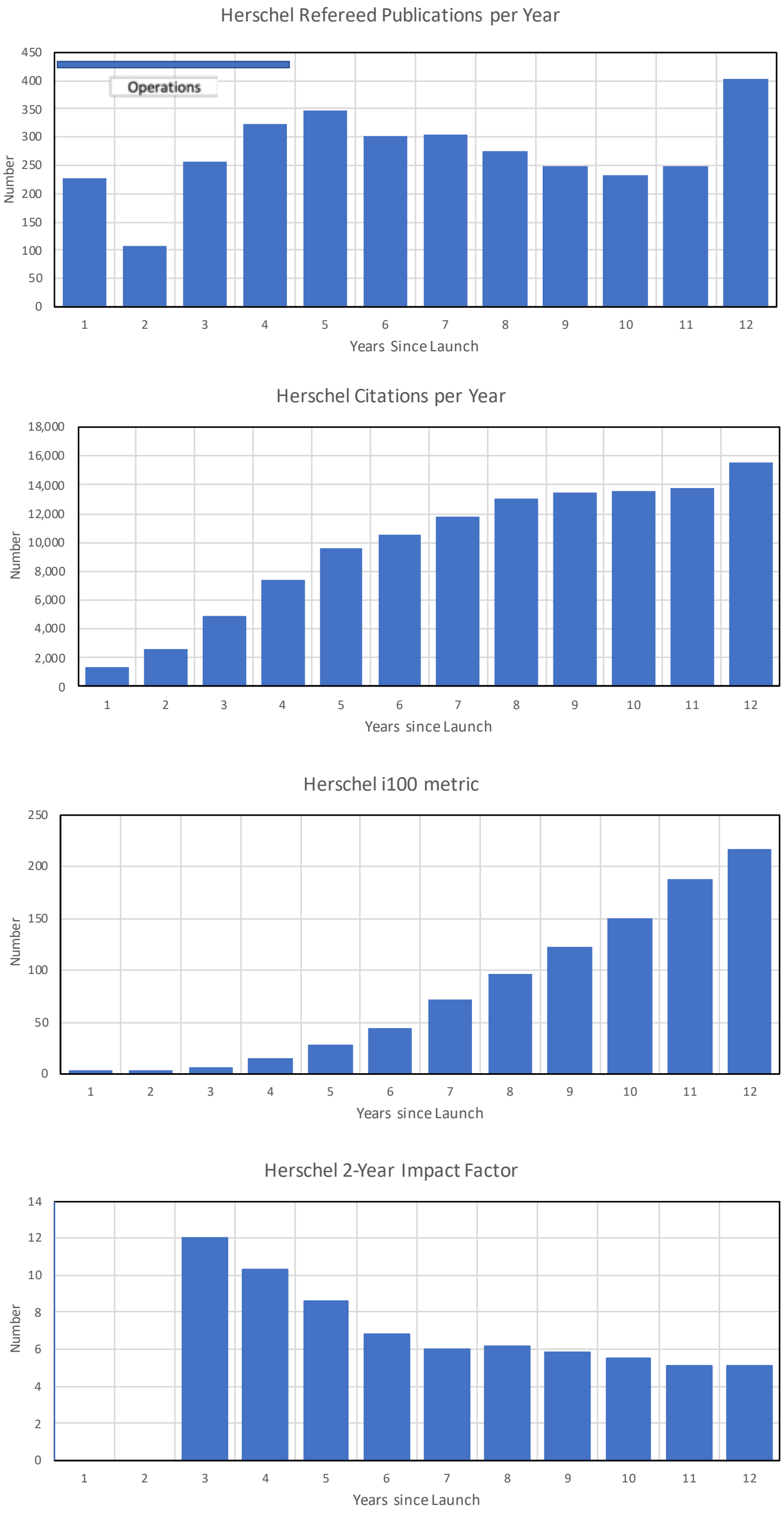

**Fig. 24** Herschel Publication Metrics

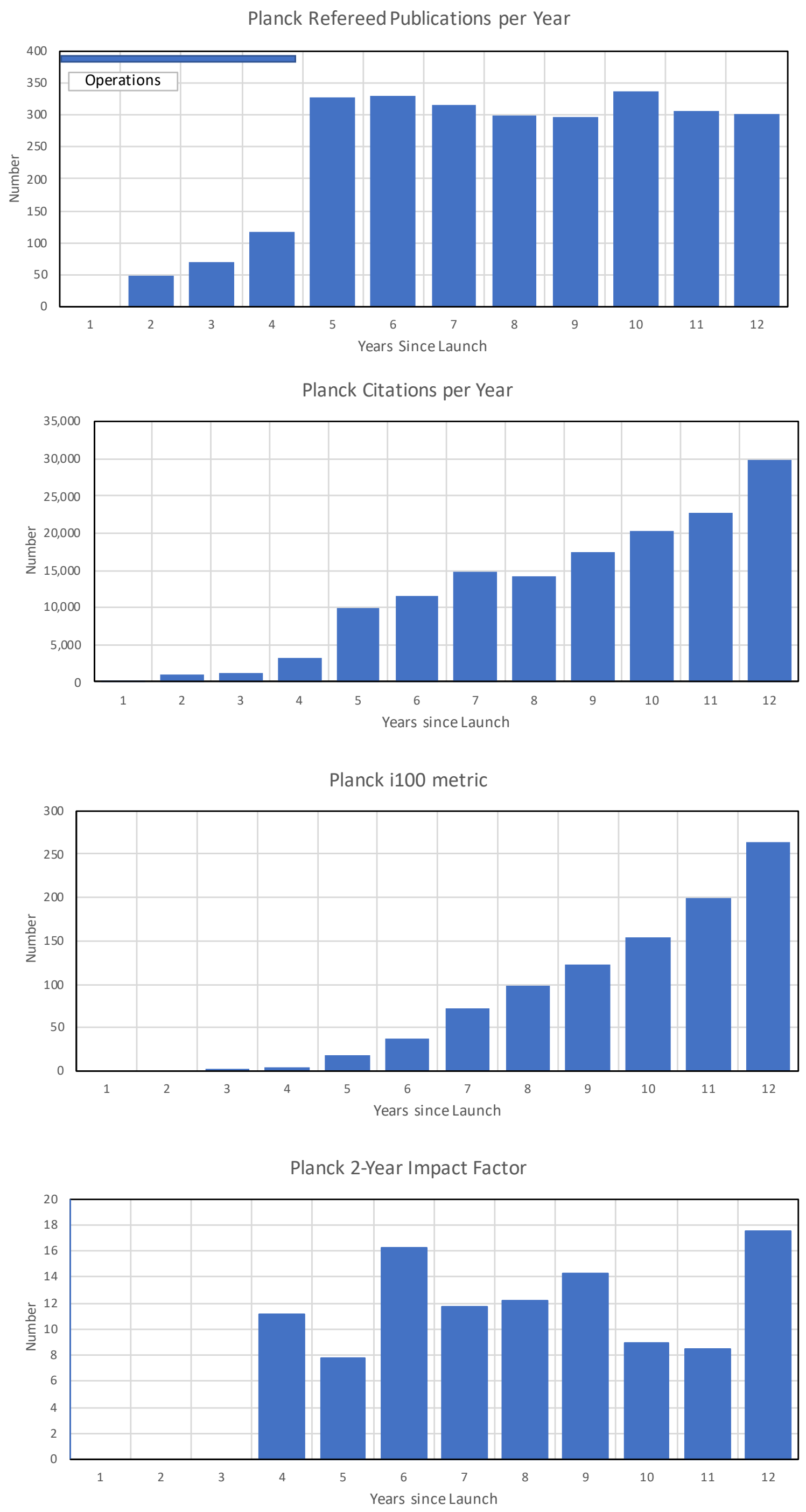

**Fig. 25** Planck Publication Metrics

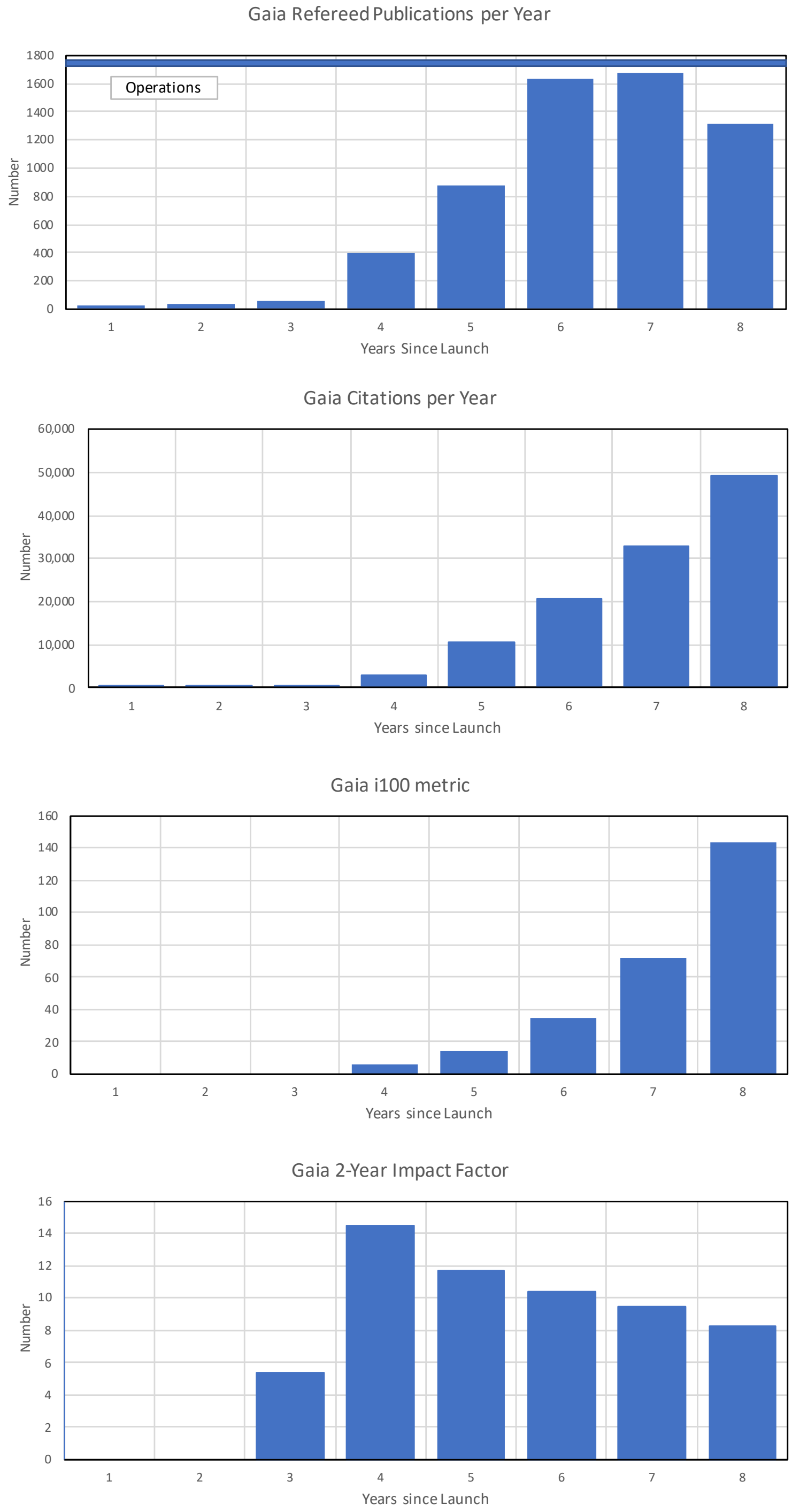

**Fig. 26** Gaia Publication Metrics

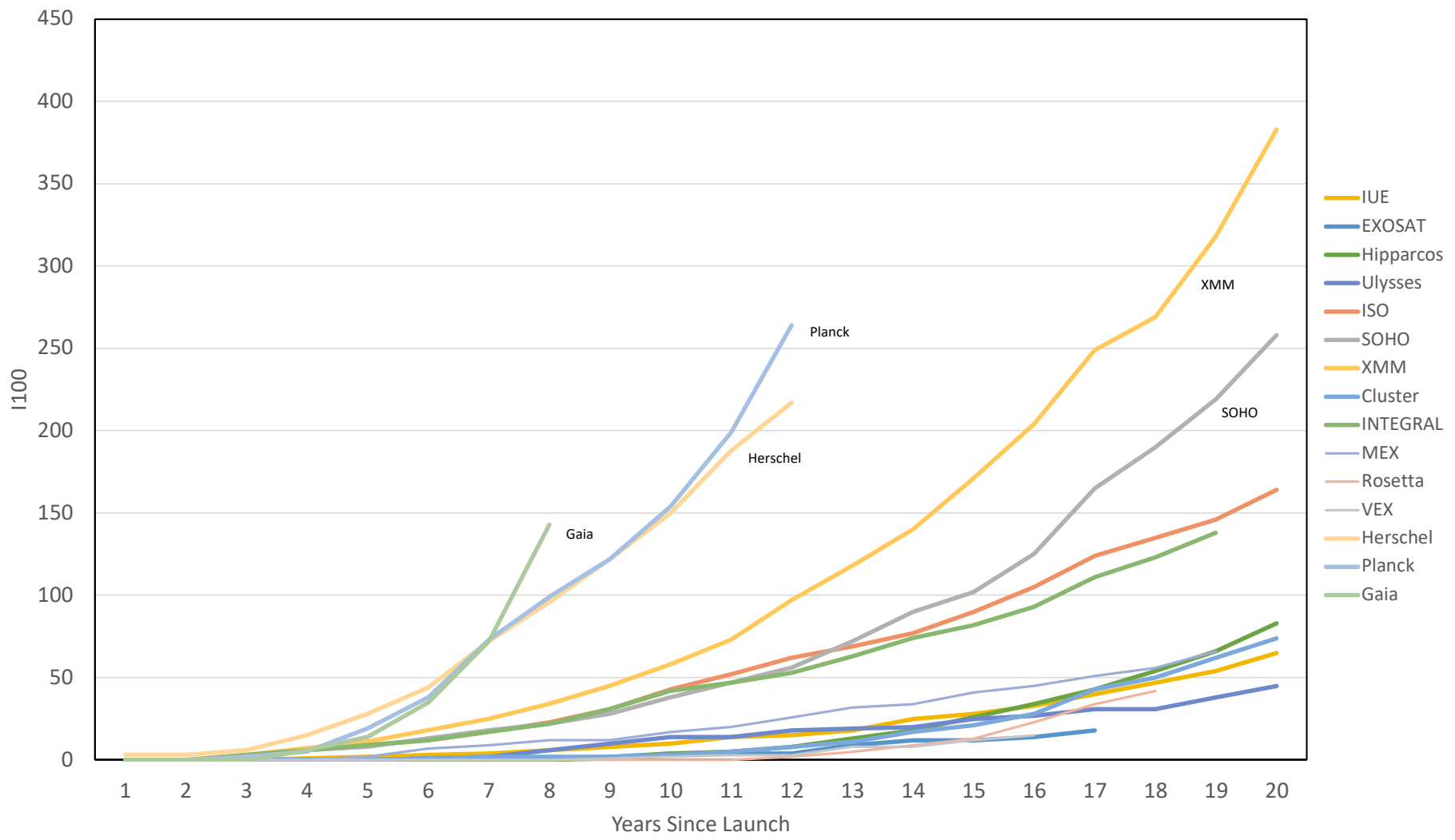

**Fig. 27** Evolution of the i100 metric with years after launch for selected ESA-led missions. Different missions are represented by different coloured lines. Selected missions are individually labelled.

It is interesting to compare how i100, the number of citations, the h-indices and the 2-year impact factor evolve with time for the ESA-led missions. These were extracted from ADS and are shown plotted in Figs. 27 to 30. In order to compare missions, the values are shown against years since launch up to a maximum of 20 years. Solar Orbiter is not included as it was launched too recently to have meaningful data.

Figure 27 shows how the number of publications with >100 citations (i100) evolves with time after launch for selected ESA-led missions. It can be seen that this metric increased most rapidly for Gaia [7], Planck [14] and Herschel [13] followed by XMM-Newton [5] and SOHO [6] with XMM-Newton having the highest absolute value. The evolution of this metric illustrates the high interest in the results from these missions as well as illustrating that it takes typically three to four years from launch before significant numbers of such publications start to appear.

Figure 28 shows how the number of citations for selected ESA-led missions evolved with time. The rapid increase in the number of citations from Gaia [7] is evident as are the contributions of Planck [14], Herschel [13], XMM-Newton [5] and SOHO [6]. Unsurprisingly these same missions had also the highest numbers of i100 publications (see Fig. 27).

Figure 29 shows the evolution of the h-indices with time for selected ESA-led missions. The rapid increase in h-indices for Gaia [7], Planck [14] and Herschel [13] is evident, as are the high absolute values for XMM-Newton [5] and SOHO [6].

Figure 30 shows how the 2-year impact factors evolve with time for a selection of ESA-led missions. The impact factors typically rise quickly to maxima between 3 to 5 years after launch and then slowly decay. The effect on the impact factors of the cometary encounter for Rosetta [22], [23] and the advent of multi-messenger astronomy for INTEGRAL [17] are clearly evident in this figure.

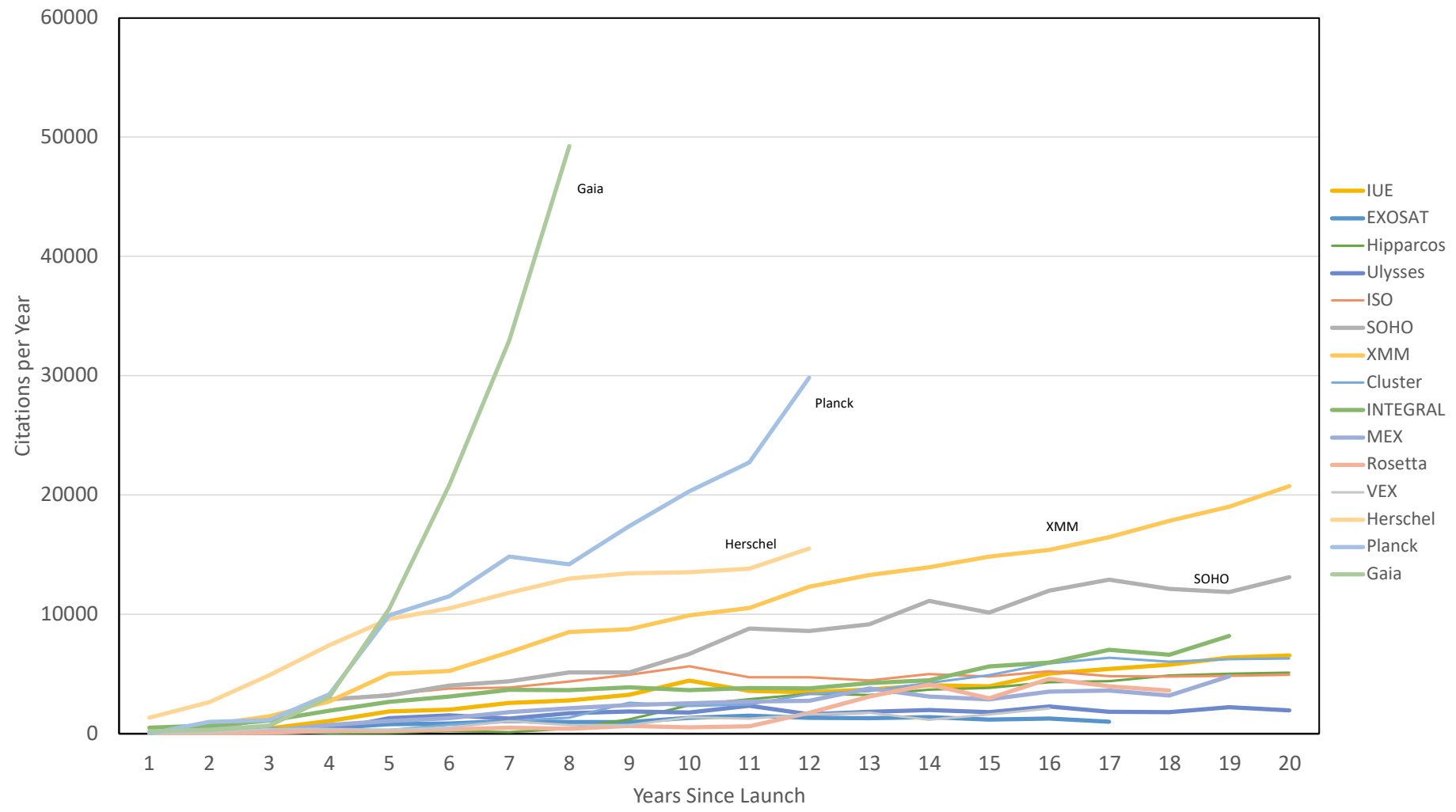

**Fig. 28** Evolution of the number of citations with years after launch for selected ESA-led missions. Different missions are represented by different coloured lines. Selected missions are individually labelled.

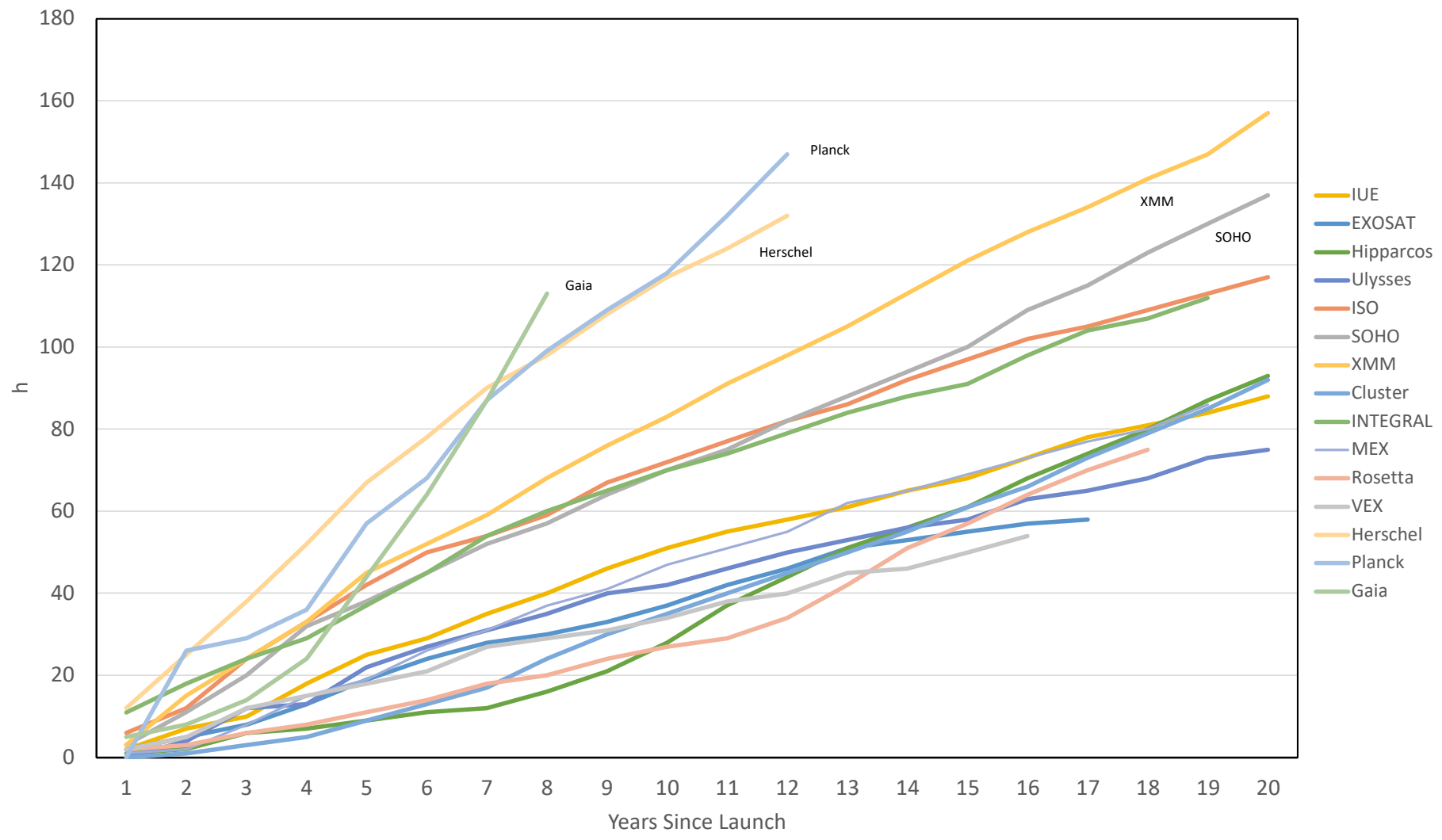

**Fig. 29** Evolution of the h-index metric with years after launch for selected ESA-led missions. Different missions are represented by different coloured lines. Selected missions are individually labelled.

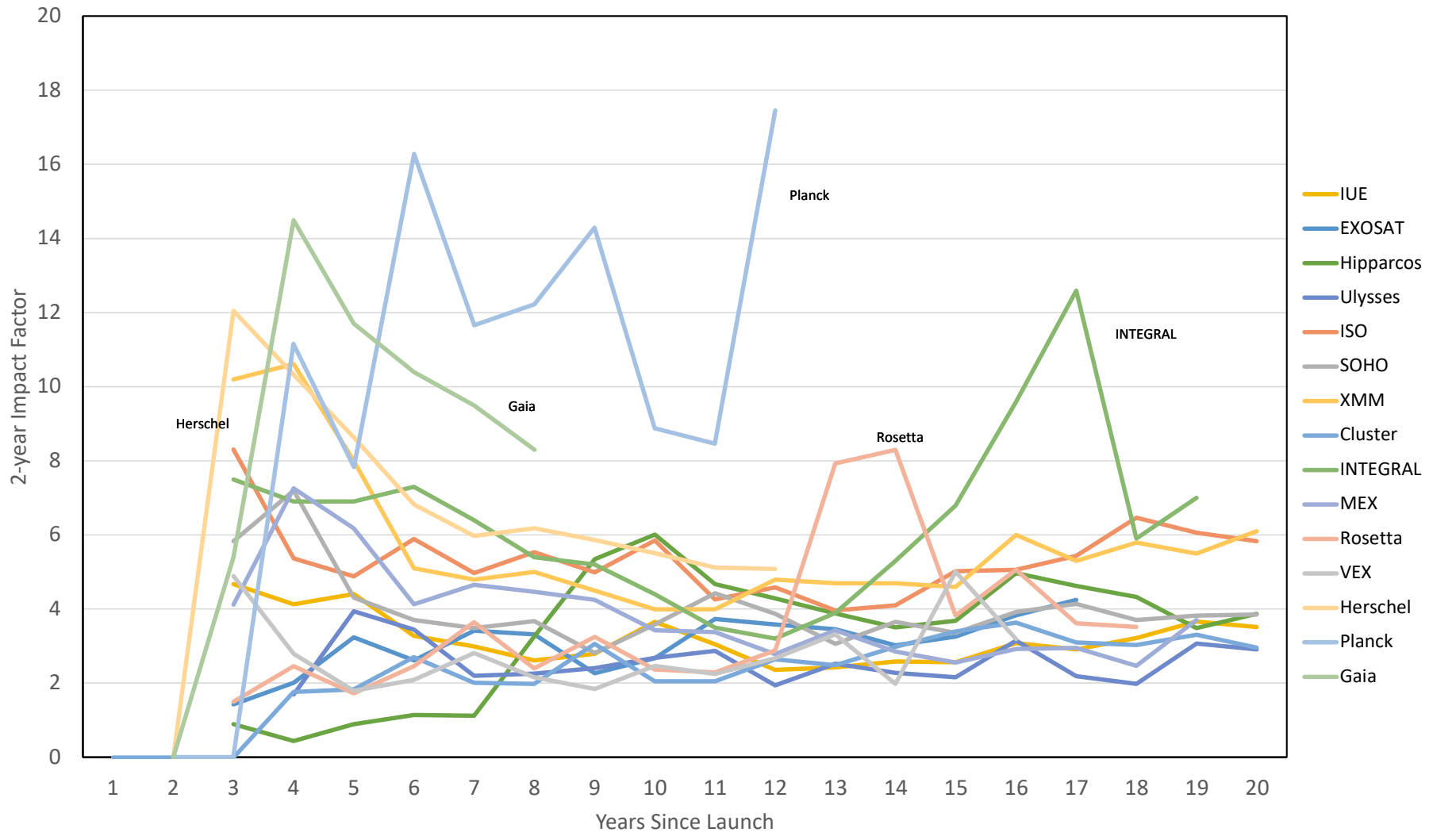

**Fig. 30** Evolution of the 2-year impact factor with years after launch for selected ESA-led missions. Different missions are represented by different coloured lines. Selected missions are individually labelled.

## 4 Cross Mission Publications

The publications libraries are build per mission, using uniform search criteria (see Sect. 2), so publications making use of data from more than one mission can be identified by simply cross-matching the individual libraries. In Fig. 31 we show the results of this analysis in graphical form for a sample of astronomy and solar system missions, both ESA-led and partner-led. In each panel, a specific colour is assigned to each mission. The total number of papers for each mission, in a five year period between 2016 and 2021, is indicated at the top of the corresponding mission's column. The fraction of papers in common with other missions is indicated by the height of the other colour bars in each mission column. For instance, of the 4620 Gaia papers in that period, about 7.6% also used data from HST, 2.6% from XMM-Newton, 1.5% from Herschel, 1.0% from Planck, and 0.3% from ISO. For the astronomy missions in Fig. 31, the average fraction of cross-mission papers, weighted by the number of papers, is about 18%, but can be considerably larger for legacy missions, namely 28 % for Herschel and over 50% for ISO. One possible interpretation is that one should expect more overlap between missions that cover similar wavelengths ranges (e.g., ISO and Herschel), but it is also possible that the overlap with legacy missions is larger because thanks to the ESA Space Science Archives their data remain available and are used as a reference or comparison stone when data from new missions become available in a similar or complementary wavelength range. The fraction of papers in common for Solar System missions is somewhat lower, on average 9% for heliophysics and 5% for planetary science missions, but again it can be very large for missions covering the same or similar fields (e.g. ExoMars and MEX, or Hinode and SOHO). In general, this underlines the importance of preserving data from legacy missions, which can enrich the scientific interpretation of more recently acquired data.

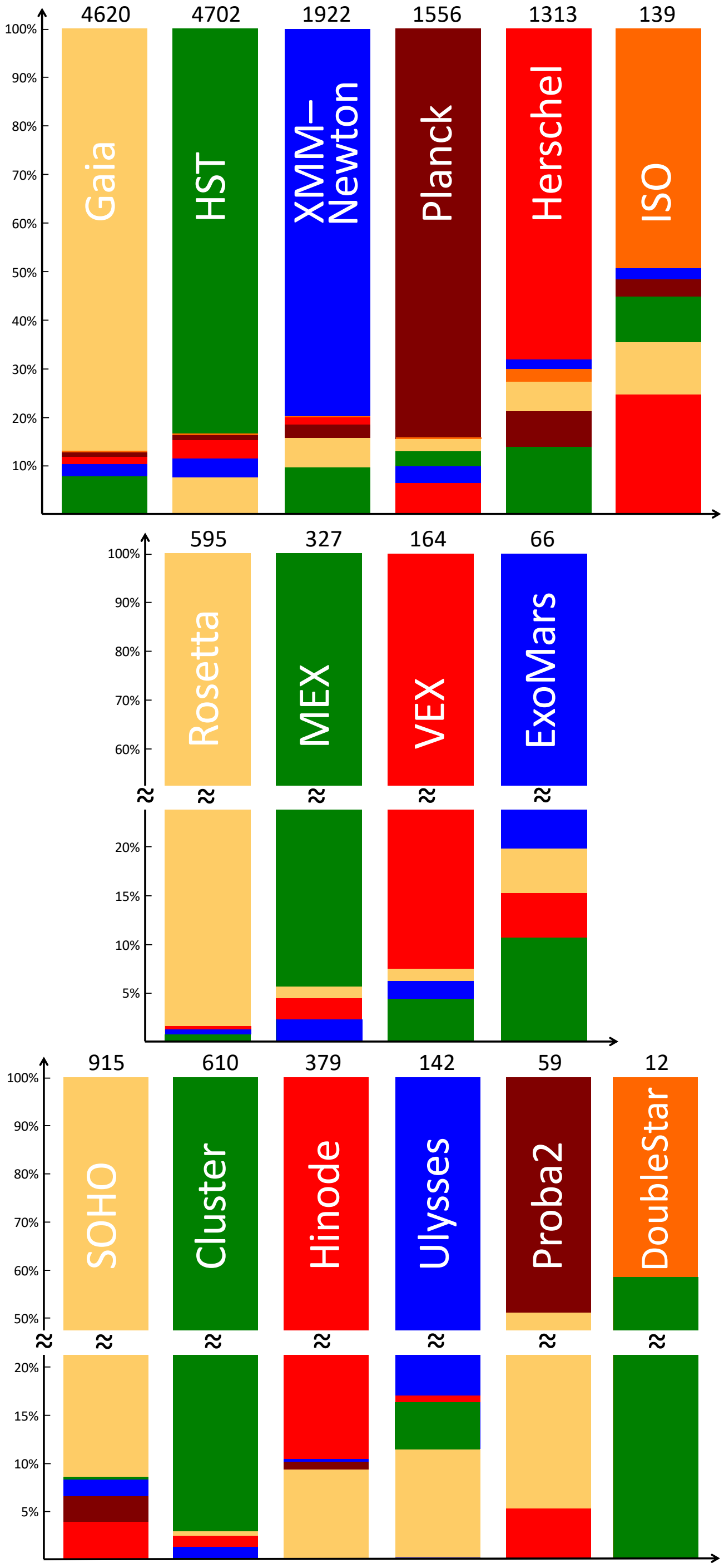

**Fig. 31** Fraction of cross-mission papers for a number of representative missions over the period between 2016 and 2021. Astronomy missions are in the top panel, Planetary missions in the middle, and Heliophysics at the bottom. In each panel, a different colour is assigned to each mission and the total number of papers in that time frame is indicated at the top of each mission column. The typical fraction of cross-mission papers, weighted by the number of papers, is about 18% for astronomy, 5% for planetary science, and 9% for heliophysics.

## 5 Archival Publications

The ESA Space Science Archives serve as the single access point for all scientists interested in observations and measurements collected by current and legacy ESA Science Programme missions. When scientists who are not involved in the original research projects or in the provision of on-board instruments retrieve those data and use them for new investigations, they help increase the value of the missions. They contribute to the advancement of science through independent analysis of existing data to answer a range of scientific questions. In this way, investigations based on archival data effectively augment the scientific return of the ESA missions. We present in this section the statistics concerning such "archival publications" and their impact.

It is useful to point out that the phrases "archival publication", "archival paper", and "archival research" might be perceived differently in different communities. In the astronomy and heliophysics areas, science archives have been for decades the main tool to access scientific data, both for recent data, still covered by time-limited restricted access, and for public data, no longer subject to such restrictions. On the other hand, at ESA archives for Planetary Science were introduced about 20 years ago, and until recently data would still be delivered first directly to the teams of the Principal Investigator (PI)s and only later would they reach the ESA Planetary Science Archive. Here, the adjective "archival" will refer not so much to the place where the data are located (even though most of them are now physically in the ESA Space Science Archives) but rather to indicate data that have already been used to address the original scientific questions posed by the researchers who first obtained or requested them.

In this light, in order to allow for meaningful comparisons across areas of research, we adopted a working definition of archival publications that is as uniform as possible for different missions. The adopted classification of a publication is based on the relationship between its authors and the scientists who lead the research project generating the data used in that paper. The inspiring principle is that, in general, when a publication is written by scientists who built the instruments that produced those data or who were awarded observing time to collect those data, the publication is non-archival. If no such relationship exists, the publication is archival.

We searched for a definition applicable to all missions served by the ESA Space Science Archives, which can be roughly sorted in two main groups:

1. Observatory-type missions, where time-limited restricted access to the data of the on-board instruments is guaranteed to scientists who successfully qualify following an Announcement of Opportunity (AO).
2. Survey-type missions, where selected scientists provide on-board instruments to which they receive privileged, but time-limited, access during the mission.

Astronomy missions can fall into both groups, while Solar System missions typically belong to the second group. For some missions both possibilities exist. In all cases, however, after a period of restricted access, all data become available to all scientists in the community through the ESA Space Science Archives.

The definition of archival publication is as follows: a publication based on ESA mission data is considered fully archival when the list of authors does not include the PI of the proposal, instrument, or experiment producing those data. This definition draws on the classification so far adopted by ESA for XMM-Newton and Herschel publications, as well as by NASA for Chandra publications. The definition has been further adapted to cover all astronomy, heliophysics, and planetary science missions. Similarly, non-archival publication would be those that contain in the list of authors the name of the data or instrument PI. A third possibility is for partly-archival publications, which include data from more than one proposal, instrument or experiment without including all the corresponding PIs in the list of authors.

We point out that these definitions might slightly overestimate the number of archival publications, because they would not detect publications authored by some Co-Investigator (Co-I)s without including the PI. It would appear unlikely that the PI be specifically excluded, yet one does see this happening, particularly after several years of operations, when collaborations between team members become somewhat looser. Another, unrelated reason why the number of archival publications determined in this way might be slightly overestimated, particularly in the early years of operations, is the presence of publications not directly making use of the data (see Sect. 2). Examples are articles addressing technical aspects of the mission or of its instruments, scientific expectations, and simulations. Some of these publications do not include the PI as an author. These publications are conceptually easy to identify, but this cannot be done in an automated way and it requires the work of a scientifically trained librarian, which is not always possible for all ESA missions. On the other hand, and perhaps more importantly, as long as the same definitions are applied consistently across all missions and over time, they will allow for meaningful comparisons and for the identification and monitoring of trends. We consider this an important element of this preliminary investigation.

For refereed publications based on XMM-Newton and Herschel data, the classification (archival versus non-archival) is simplified by the work done by members of the missions teams, who identified the specific data used in each paper and the corresponding observing proposal number. It is then simple to look for a match (or lack thereof) between the paper's authors and the proposal PI.

The relevant statistics concerning these two missions are shown in Tables 14 and 15 and in graphical form in Figs. 32 and 33. The fraction of fully archival publications is different between the two missions, being higher for XMM-Newton most probably because of its longer history and, most importantly, its wider user community. In both cases, however, the fully archival publications show a steady growth over time, reaching $\sim$ 35 % for Herschel 10 years after launch and $\sim$ 60 % for XMM-Newton after 20 years of operations. For both missions, archival (fully and partly) publications are the majority: in recent years they exceed 60 % of the total for Herschel and 80 % for XMM-Newton.

The situation of publications based on HST data is also relevant in this context. In that case, however, the definition adopted by NASA also takes into account the role of Co-Is, so the statistics are not directly comparable with those of Herschel or XMM-Newton. Nevertheless, in the past ten years archival publications (fully and partly) have consistently represented over 60 % of the total HST publications, showing an increase similar to that of the large ESA observatories.

---

[1] An alternative approach is the use of natural language processing and machine learning algorithms to identify and classify the publications automatically. This work is currently planned, as we briefly discuss in Sect. 6

[2] For the first few years of operations of XMM-Newton, a sizeable fraction of the papers cannot be classified with certainty because we lack specific knowledge of the members of the instrument teams. Furthermore, Guaranteed Time Observations and Target of Opportunity Observations were coarsely grouped in a small number of observing proposals, all under the name of the Project Scientist, with no information about the names of the scientists who were actually entrusted those data. Unfortunately, no digital records exist. Fortunately, these problems appear to be limited to the first few years of operations, and after 2006 the fraction of unclassified papers drops to close to zero.

**Table 14** Temporal evolution of non-archival, partly-archival, and fully-archival publications based on XMM-Newton data. Values are given both as actual number of papers and in percentage.

| Year | Total | Non-archival | | Partly-archival | | Fully-archival | | Unclassified | |
|---|---|---|---|---|---|---|---|---|---|
| 2001 | 82 | 35 | 43% | 5 | 6% | 0 | 0% | 42 | 51% |
| 2002 | 120 | 75 | 62% | 7 | 6% | 6 | 5% | 32 | 27% |
| 2003 | 253 | 143 | 57% | 50 | 20% | 35 | 14% | 25 | 10% |
| 2004 | 351 | 147 | 42% | 85 | 24% | 93 | 26% | 26 | 7% |
| 2005 | 331 | 104 | 31% | 86 | 26% | 124 | 37% | 17 | 5% |
| 2006 | 369 | 93 | 25% | 83 | 22% | 186 | 50% | 7 | 2% |
| 2007 | 371 | 108 | 29% | 78 | 21% | 185 | 50% | 0 | 0% |
| 2008 | 339 | 85 | 25% | 49 | 14% | 205 | 60% | 0 | 0% |
| 2009 | 390 | 113 | 29% | 62 | 16% | 215 | 55% | 0 | 0% |
| 2010 | 375 | 85 | 23% | 62 | 17% | 228 | 61% | 0 | 0% |
| 2011 | 389 | 99 | 25% | 83 | 21% | 207 | 53% | 0 | 0% |
| 2012 | 339 | 82 | 24% | 80 | 24% | 177 | 52% | 0 | 0% |
| 2013 | 363 | 93 | 26% | 73 | 20% | 197 | 54% | 0 | 0% |
| 2014 | 359 | 97 | 27% | 70 | 19% | 192 | 53% | 0 | 0% |
| 2015 | 325 | 84 | 26% | 49 | 15% | 192 | 59% | 0 | 0% |
| 2016 | 367 | 94 | 26% | 91 | 25% | 182 | 50% | 0 | 0% |
| 2017 | 387 | 88 | 23% | 80 | 21% | 219 | 57% | 0 | 0% |
| 2018 | 462 | 105 | 23% | 85 | 18% | 272 | 59% | 0 | 0% |
| 2019 | 443 | 91 | 21% | 87 | 20% | 265 | 60% | 0 | 0% |
| 2020 | 387 | 75 | 19% | 77 | 20% | 235 | 61% | 0 | 0% |
| Total | 6802 | 1896 | 28% | 1342 | 20% | 3415 | 50% | 149 | 2% |

**Table 15** Temporal evolution of non-archival, partly-archival, and fully-archival publications based on Herschel data. Values are given both as actual number of publications and in percentage.

| Year | Total | Non-archival | | Partly-archival | | Fully-archival | | Unclassified | |
|---|---|---|---|---|---|---|---|---|---|
| 2010 | 228 | 4 | 2% | 200 | 88% | 20 | 9% | 4 | 2% |
| 2011 | 109 | 1 | 1% | 100 | 92% | 6 | 6% | 2 | 2% |
| 2012 | 256 | 10 | 4% | 220 | 86% | 19 | 7% | 7 | 3% |
| 2013 | 323 | 6 | 2% | 269 | 83% | 43 | 13% | 5 | 2% |
| 2014 | 347 | 8 | 2% | 278 | 80% | 54 | 16% | 7 | 2% |
| 2015 | 301 | 5 | 2% | 230 | 76% | 62 | 21% | 4 | 1% |
| 2016 | 305 | 3 | 1% | 215 | 70% | 81 | 27% | 6 | 2% |
| 2017 | 276 | 2 | 1% | 196 | 71% | 71 | 26% | 7 | 3% |
| 2018 | 251 | 12 | 5% | 150 | 60% | 84 | 33% | 5 | 2% |
| 2019 | 246 | 86 | 35% | 75 | 30% | 78 | 32% | 7 | 3% |
| Total | 2642 | 137 | 5% | 1933 | 73% | 518 | 20% | 54 | 2% |

For all other missions, the classification of publications as archival or non-archival is performed by checking the list of authors against the names of the PIs of onboard instruments and experiments. Applying the adopted definitions to the papers based on Planck data published in 2020 and earlier, results in 178 non-archival publications (i.e. those authored by the instrument PIs) out of 2652 publications in total. The evolution over time of archival and non-archival Planck publications is shown in Fig. 34. This implies a fraction of archival publications exceeding 93 %. Alternatively, one could consider as non-archival publications those authored by the Planck Collaboration, which include the phrase "Planck Collaboration" in the list of authors. These amount to 159 articles in the same period and result in a slightly higher fraction of archival publications, namely almost 94 %. Regardless of the exact metrics adopted, there is no doubt about the overwhelmingly archival nature of the literature based on Planck data.

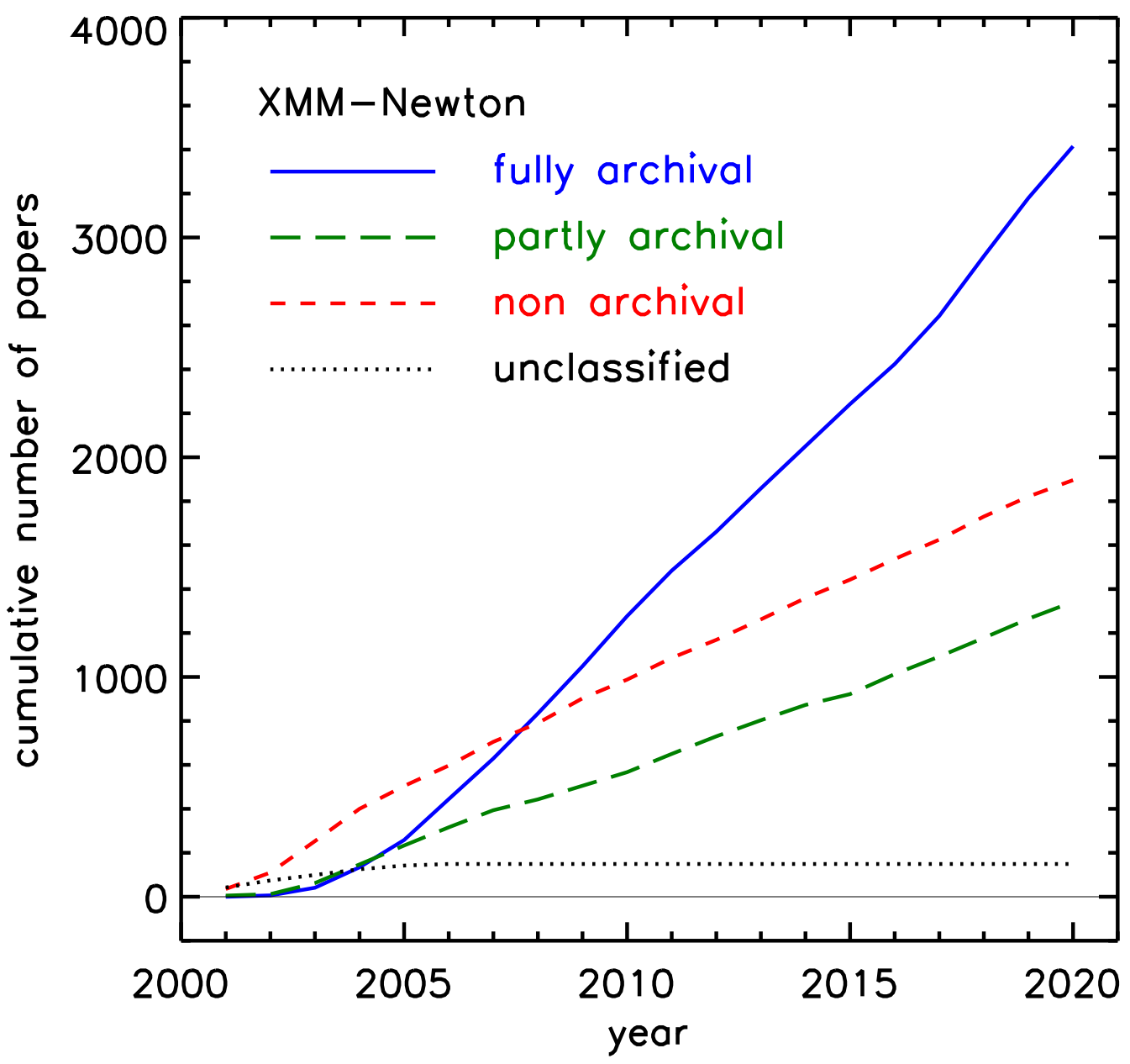

**Fig. 32** Cumulative distribution of XMM-Newton refereed papers as a function of time for different categories (see legend).

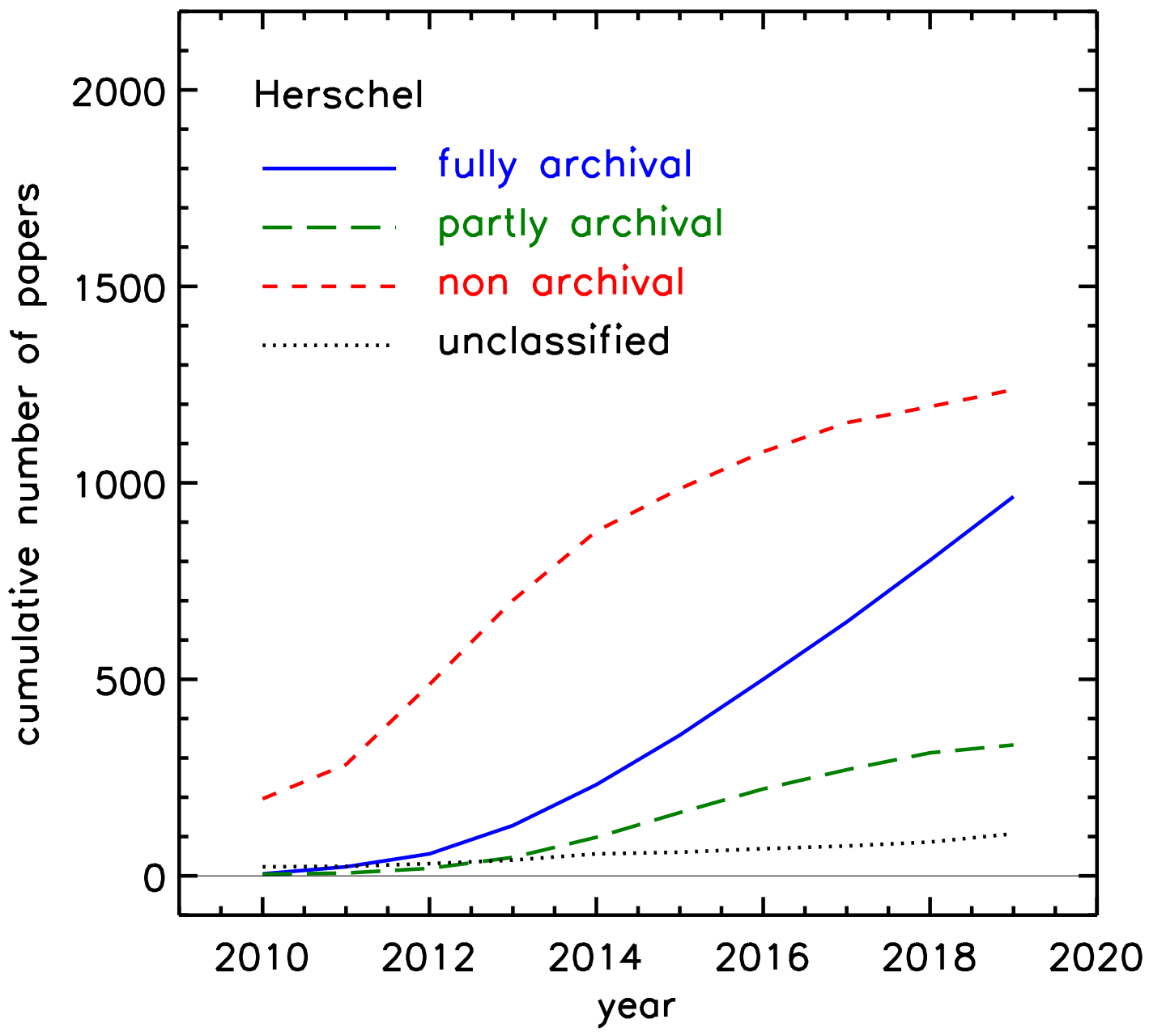

**Fig. 33** Cumulative distribution of Herschel refereed papers as a function of time for different categories (see legend).

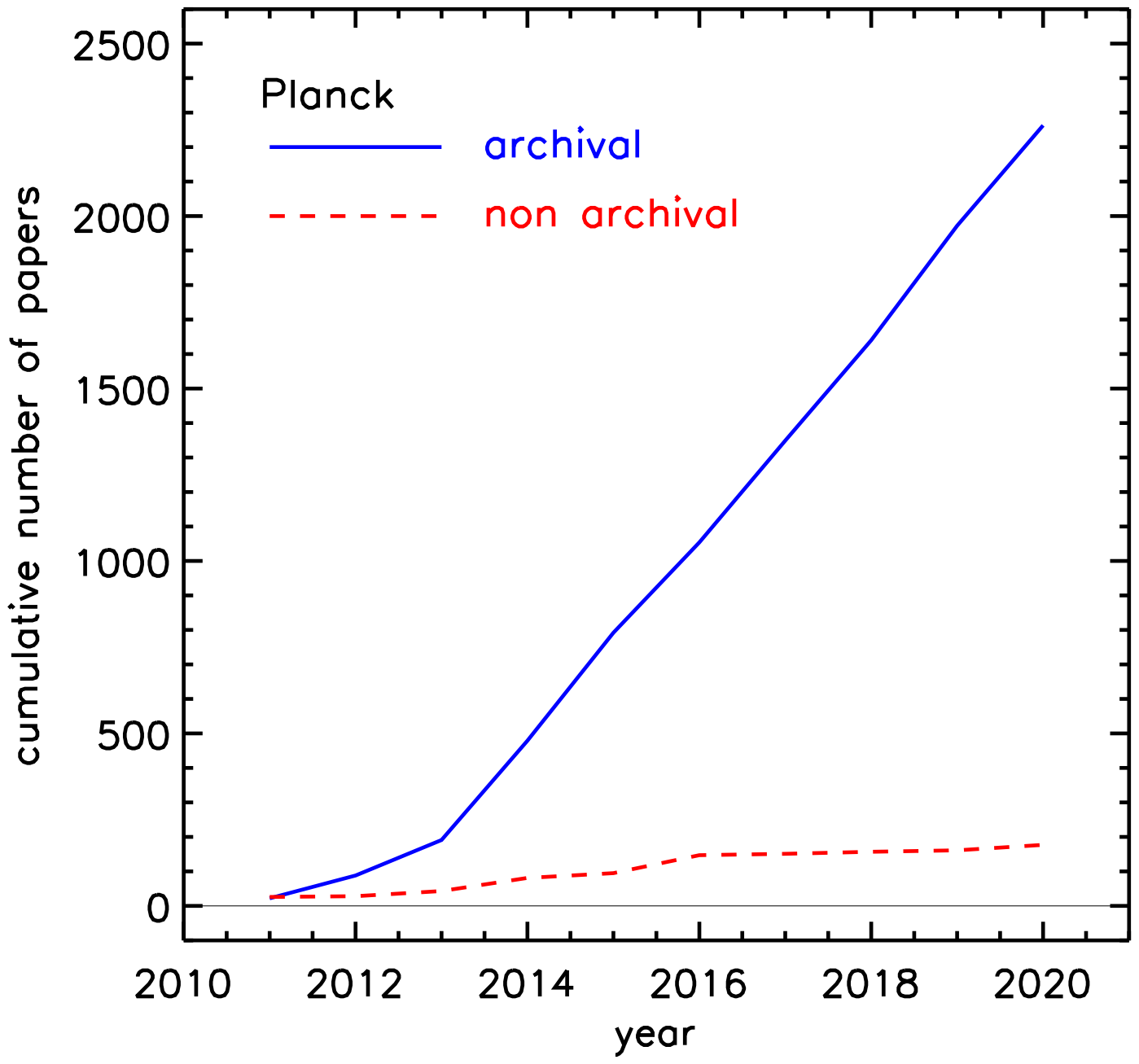

**Fig. 34** Cumulative distribution of Planck refereed publications as a function of time. Archival and non-archival publications are shown (see legend).

The situation of Gaia, one of the most prolific space science missions so far, is even more extreme: in this case there is no time-limited restricted access to the data and all Gaia Data Releases are immediately available to the entire community. Therefore, all publications based on Gaia data are by definition fully archival. The publication numbers in Table 1 above include a small amount of calibration and processing papers too, but in the case of Gaia those represent less than 2 % of the total.

Archival publications represent a sizeable fraction of the literature also for ESA Solar System missions (planetary science and heliophysics). Also here is the classification based on the existence of a match (or lack thereof) between the list of authors in a publication and the names of the PIs of the instrument or experiments onboard those missions. As representative examples, we show in Figs. 35 and 36 the cases of MEX and Rosetta, respectively. While for Rosetta the proportion of archival and non-archival publications has been relatively similar throughout the years, for MEX the number of archival publications has consistently surpassed that of non-archival publications starting in 2016. The preponderance of archival publications is even more marked for heliophysics missions, as we show in Fig. 37 for Cluster and Fig. 38 for SOHO, undoubtedly also due to the very long operational phase of the latter mission. With a total fraction of over 85 % of archival papers by 2020, SOHO is after Gaia and Planck the mission with the highest fraction of archival papers and, together with Planck, it is one of the missions with the fastest growth of archival papers over time.

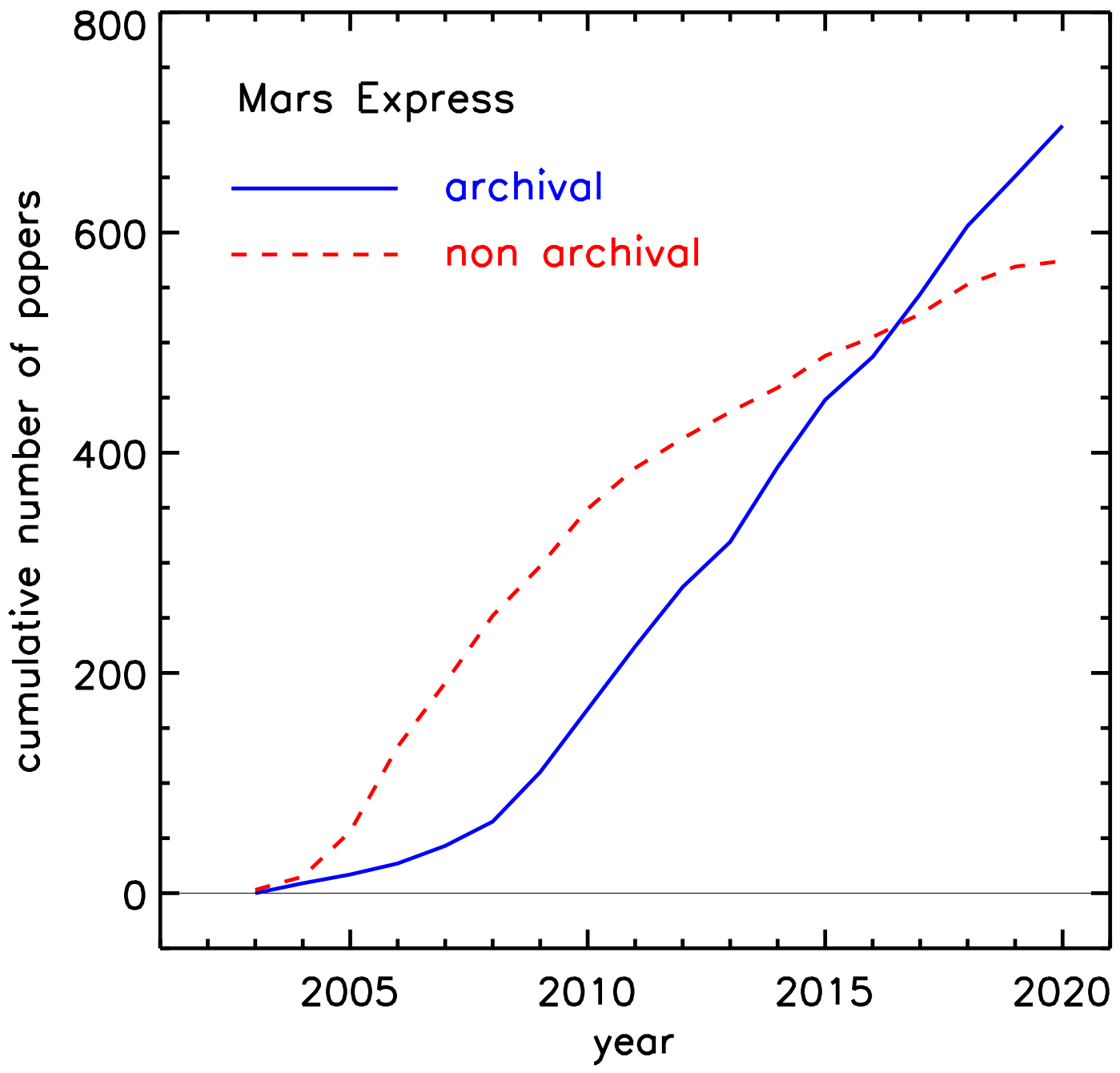

**Fig. 35** Cumulative distribution of MEX refereed publications as a function of time. Archival and non-archival publications are shown (see legend).

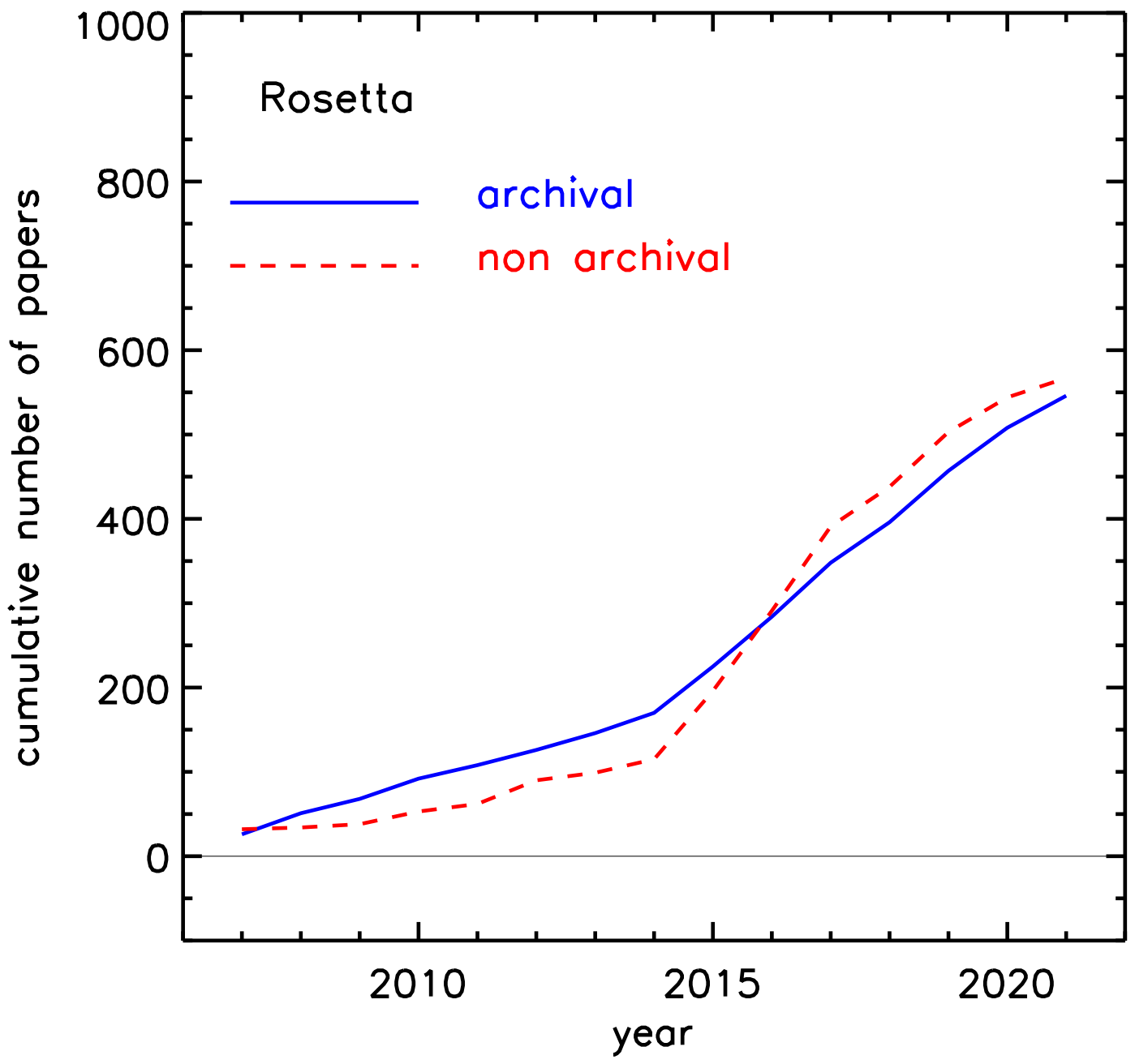

**Fig. 36** Cumulative distribution of Rosetta refereed publications as a function of time. Archival and non-archival publications are shown (see legend).

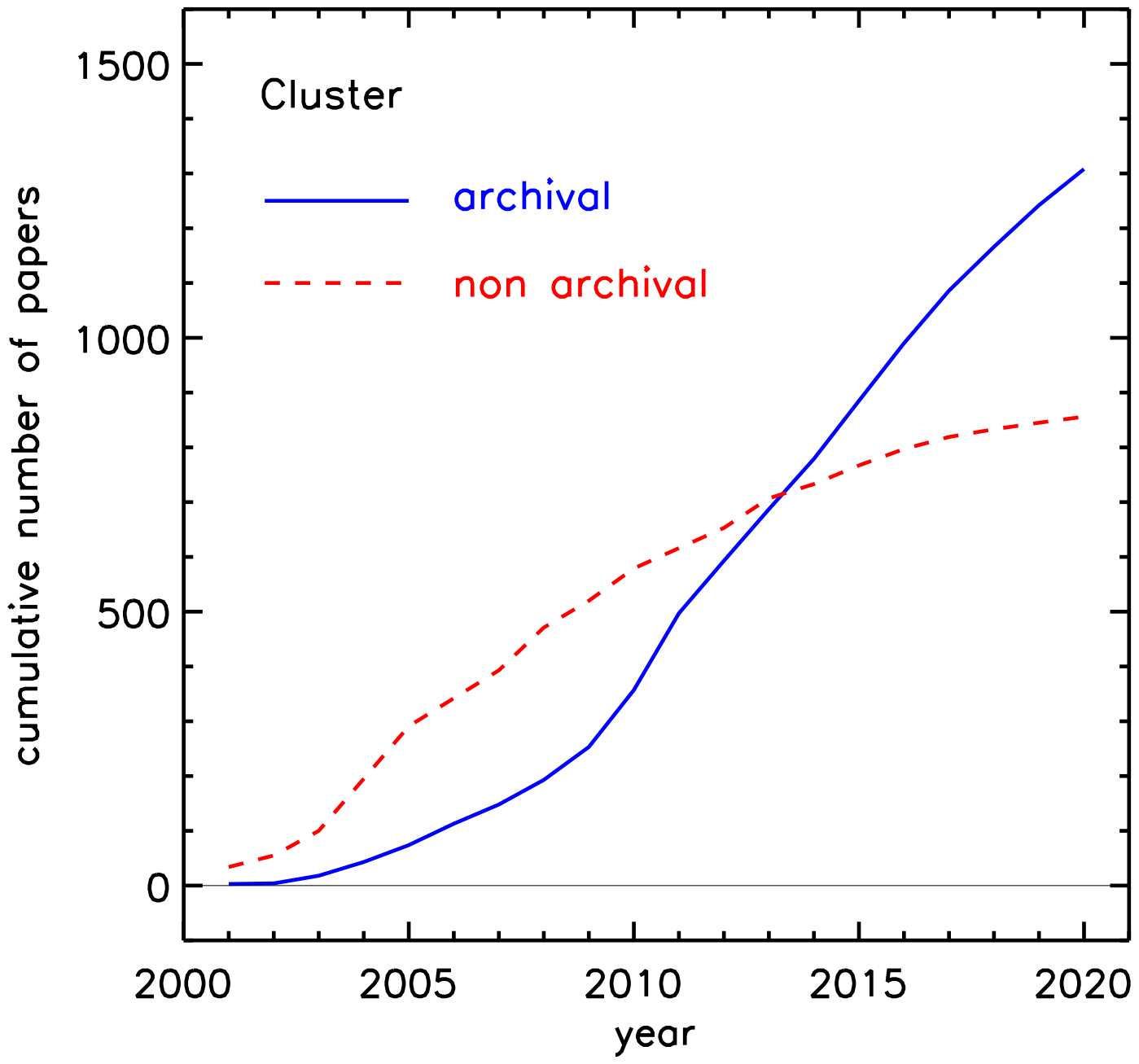

**Fig. 37** Cumulative distribution of Cluster refereed publications as a function of time. Archival and non-archival publications are shown (see legend).

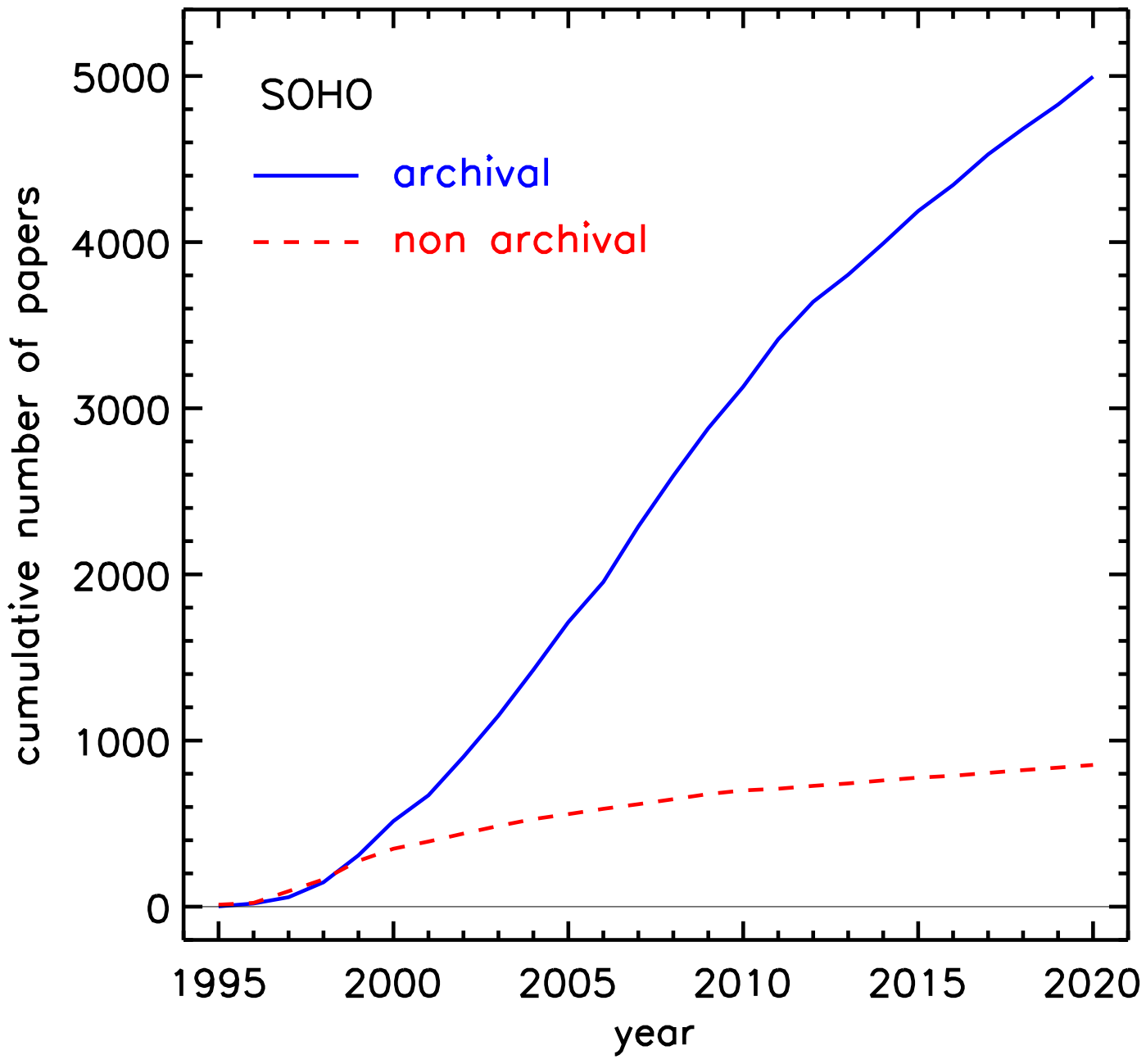

**Fig. 38** Cumulative distribution of SOHO refereed publications as a function of time. Archival and non-archival publications are shown (see legend).

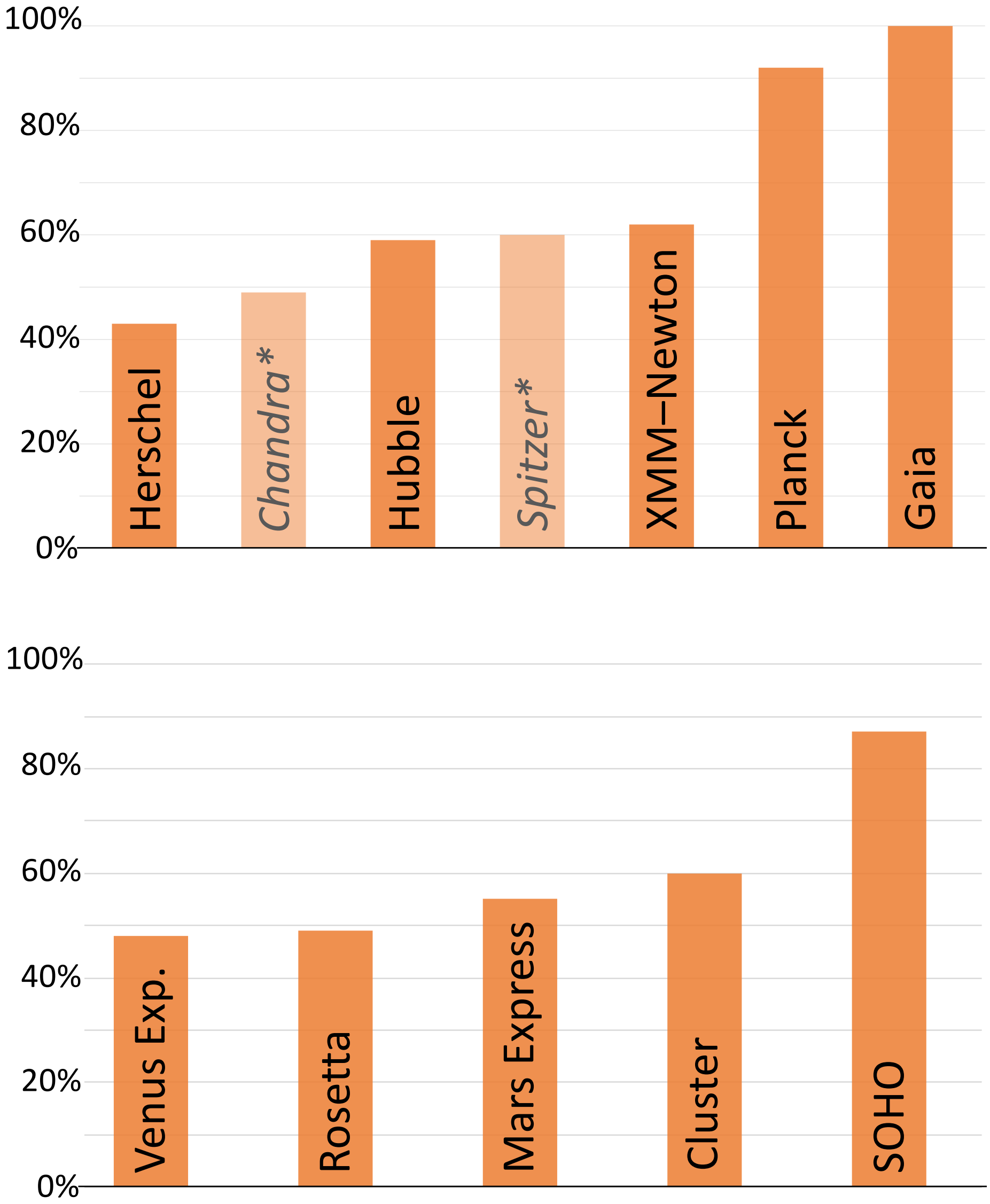

**Fig. 39** Cumulative fractions of archival papers for a sample of ESA Astronomy (top) and Solar System (bottom) missions. (*) Data about Chandra and Spitzer are from [25].

The cumulative fractions of archival papers for a sample of ESA Astronomy and Solar System missions are shown in Fig. 39. The statistics cover different time spans and the various missions have rather different lifetimes, so it is difficult to directly compare the mission to one another. However, the purpose of this figure is to show that ESA missions in general have produced a sizeable fraction of archival papers, which appears to be fully in line with that generated by NASA missions. For reference, we have included in the graph the statistics about two of NASA's Great Observatories, namely Chandra and Spitzer (from [25]).

A relevant question to ask is whether the impact of archival papers is similar to that of non-archival publications. To address this question, we explored the temporal evolution of the number of citations of both types of papers from four representative missions: an astronomical observatory (XMM-Newton), an astronomical survey mission (Planck), a planetary mission (MEX), and a heliophysics mission (Cluster). We show in the lower panels of Fig. 40 the median number of citations that papers written over the years have received until the date the calculations were made (late 2022). Dots and squares are used, respectively, for non-archival and archival papers and error bars assume Poisson statistics.

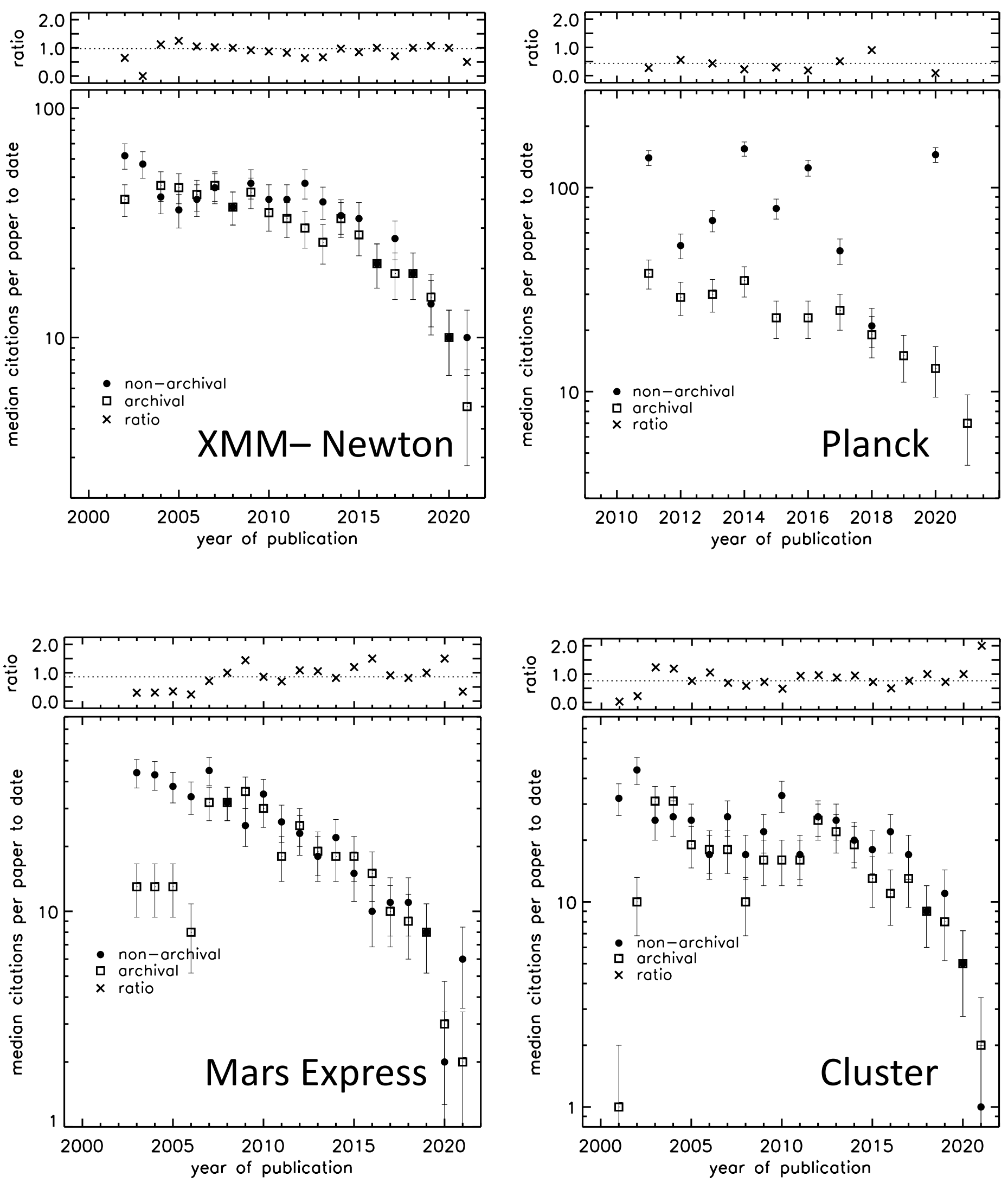

**Fig. 40** Median number of citations to date for a sample of four representative ESA Science Programme missions. Non-archival papers are indicated with dots, archival papers with squares. Error bars assume Poisson statistics. Crosses in the upper panels display the ratio of archival and non-archival papers.

Understandably, less recent papers have had more time to accumulate citations over the year, so the general decreasing trend is expected (although this is less obvious in the case of Planck's publications, as discussed below). Not surprisingly, in the first years of operations non-archival papers receive typically more citations because of the novel data that they present and because it takes some time before those data become available to other scientists not originally involved in the experiments or proposals. However, the difference becomes progressively smaller and in some instances archival papers receive on average more citations than non-archival papers. The top panels of Fig. 40 show the ratio of archival and non-archival papers as a function of time, where the dotted lines indicate the median value of the ratio, which is 0.97 for XMM-Newton, 0.43 for Planck, 0.86 for MEX, and 0.76 for Cluster. This comparison shows that, in general, archival papers have an impact only slightly smaller than that of non-archival works, and the difference decreases further after the first few years of operations.

The large difference between the Planck and the other three missions can probably be understood considering that the Planck non-archival papers are the 159 refereed articles authored by the Planck Collaboration. Many of those provide catalogues and fundamental cosmological parameters that have become primary references in the field. The fact that they are highly cited attests to the major contribution that the work done by the Planck PIs and Planck Collaboration has given to cosmology. But the impact, success, and contribution of this and other missions is also measured by the total number of citations, which in the case of Planck is clearly dominated by non-archival papers. As we show in Fig.41, about 3/4 of the citations that Planck publications receive are to archival papers, while for the other missions in the sample archival and non-archival papers receive a similar number of citations.

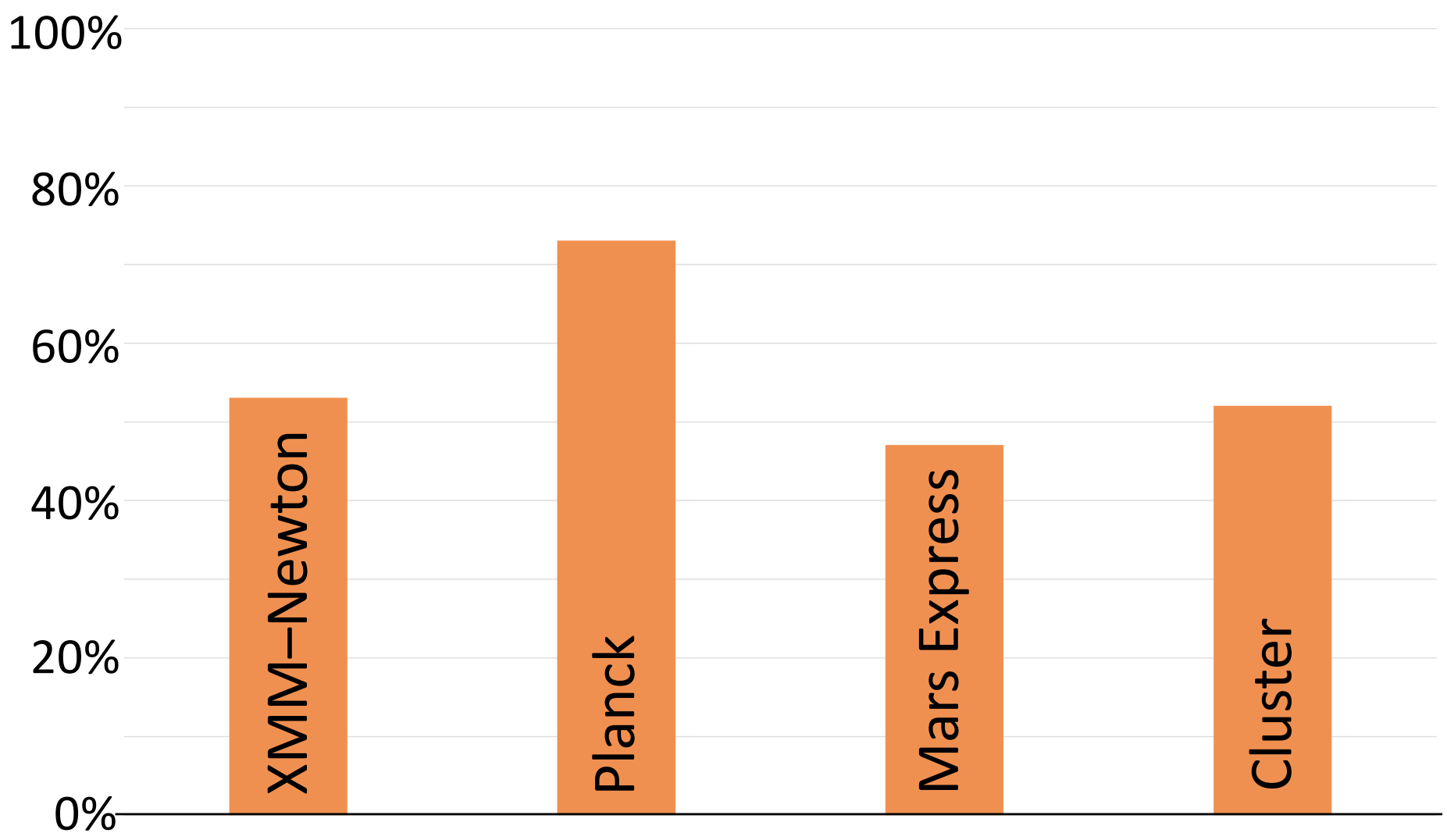

**Fig. 41** Cumulative fractions of citations to archival papers for the same missions as in Fig. 40.

In summary, this analysis has revealed that non-archival and archival papers contribute in similar ways to the visibility of the missions and, more importantly, to the advancement of science. For this to be possible, it is crucial that all the data from the missions, including the calibration and associated documentation, are shared with the community worldwide and preserved, including when the missions are no longer in operation.

## 6 Conclusions

We have examined almost half a century of scientific publications resulting from data generated by the ESA Science Programme's missions. The number of papers has increased significantly over time, matching the growth and reach of the ESA-led missions, from less than ten papers per year in the mid-to-late 1970s, when COS-B was the only mission in the programme, to over 4,000 publications per year at the end of 2021, when the ESA Science Programme counted 11 active missions and 14 more in their legacy phases, all of which are still contributing to new discoveries.

At first sight, one might attribute the increase in the number of publications from ESA's Science Programme missions to the general increase in scientific publications across all areas of space science over the past decades. Indeed, the collective number of papers in the fields of astronomy, planetary science, and heliophysics, as tracked by the ADS, has grown from $\sim 20,000$ publications per year in 2000, to $\sim 26,000$ in 2010, and $\sim 35,000$ in 2020. It is interesting that the fraction of those papers that make use of data from ESA mission has also grown with time, but at an even faster rate: In 2000, ESA mission papers accounted for 4.1% of all the astronomy, planetary science and heliophysics papers published that year; in 2014 the fraction had almost doubled, to 7.6%, and in 2021 that fraction is about three times higher, namely 11.5% for the ESA-led Science Programme missions alone, and 15.2% when the partner-led missions are also included. Thus, in a world in which the number of scientific papers keeps increasing, the fraction of papers based on ESA's Science Programme missions increases even more rapidly, witnessing an ever growing contribution by these missions to the advancement of science.

Besides the steady growth in the number of publications, our analysis shows that the scientific productivity of the missions remains high even after the end of operations. In fact, we have discovered that, in all fields of space science, archival papers represent the majority of the publications based on ESA data, and that their impact, as judged by the number of citations that they receive, is typically as large as that of non-archival papers. This highlights the importance of maintaining the missions' data available for the scientific community to explore and investigate well beyond the end of mission operations, potentially for as long as the data remain relevant. This is the role that ESA's Science Directorate has entrusted to the ESA Space Science Archives.

Our analysis shows that the geographical distribution of the publications appears generally in line with the size of the communities in the various countries. For the ESA Member States, it roughly scales with the financial contribution of each country to the ESA Science Programme, which in turn is linked to the GDP of each State. Some countries, both large and small, stand out in that their scientific communities appear to be very effective at turning mission data into scientific discoveries. Furthermore, the cases of some of the smaller countries (e.g., Hungary, Finland, Czech Republic) show that scientific research and investigations, all of which lead to scientific papers, are an effective way in which countries can become involved in space science, even before they have developed significant space technology infrastructure.

We close with some considerations about the future of this research. This investigation was possible thanks to the meticulous work done over the years by the many ESA Project and Contact Scientists who regularly survey, parse, and read the relevant scientific literature to identify papers that make use of data from the ESA Science Programme missions. No matter how refreshing and interesting it might be for scientists to learn about new discoveries and advancements in their fields, these tasks are labour intensive, particularly for very productive missions. Utilising advancements in the field of machine learning, the ESA Science Directorate is transitioning from manual identification and selection of literature papers to a semi-automated approach, whereby machine learning and natural language processing algorithms are used to sift through the literature and identify papers likely to make use of data from the ESA Science Programme missions and that fulfil the conditions indicated above. This is simplified by the availability of extensive training sets made up of the thousands of literature papers that were inspected by a scientist and eventually accepted or discarded. The papers automatically identified with these tools are then still validated by the Project and Contact Scientists. Preliminary analysis has shown that this might introduce small uncertainties in the number of papers, at the level of 10%, which are however comparable with those encountered when different scientists search and identify papers for the same mission. This holds the promise of even more detailed investigations, not just on the sheer number and impact of the papers, but also for instance on the types of data and on the methods that are used in the papers. We plan to report on the continuation of this study in a future work.

**Acknowledgements.** We thank the ESA project and contact scientists for their careful curation of the publication libraries. We also thank Edwin Henneken from the NASA Astrophysics Data System Bibliographic Services for his invaluable support.

## References

[1] A. Accomazzi, E. A. Henneken, C. Erdmann, et al. "Telescope bibliographies: an essential component of archival data management and operations". In: *Observatory Operations: Strategies, Processes, and Systems IV*. Ed. by Alison B. Peck, Robert L. Seaman, and Fernando Comeron. Vol. 8448. Soc. of Photo–Optical Instrumentation Engineers (SPIE) Conf. Ser. 2012, p. 84480.

[2] A. Accomazzi, M. J. Kurtz, E. A. Henneken, et al. "ADS: The Next Generation Search Platform". In: *Open Science at the Frontiers of Librarianship*. Ed. by A. Holl, S. Lesteven, D. Dietrich, et al. Vol. 492. Astronomical Society of the Pacific Conference Series. Apr. 2015, p. 189.

[3] E. A. Henneken, A. Accomazzi, M. J. Kurtz, et al. "Computing and Using Metrics in the ADS". In: *Open Science at the Frontiers of Librarianship*. Ed. by A. Holl, S. Lesteven, D. Dietrich, et al. Vol. 492. Astron. Soc. of the Pacific Conf. Ser. 2015, p. 80.

[4] A. H. Rots, S. L. Winkelman, and G. E. Becker. "Chandra Publication Statistics". In: *Publ. of Astr. Soc. Pacific* 124.914 (Apr. 2012), p. 391.

[5] F. Jansen, D. Lumb, B. Altieri, et al. "XMM-Newton observatory. I. The spacecraft and operations". In: *Astronomy & Astrophys.* 365 (Jan. 2001), pp. L1–L6.

[6] B. Fleck, V. Domingo, and A. I. Poland. "The SOHO mission". In: *Solar Physics* 162.1 (Dec. 1995).

[7] M. A. C. Perryman. "The GAIA mission". In: *GAIA Spectroscopy: Science and Technology*. Ed. by U. Munari. Vol. 298. Astronomical Society of the Pacific Conference Series. Jan. 2003, p. 3.

[8] T. Owen. "The Cassini mission." In: *NASA Conference Publication*. Vol. 2441. NASA Conference Publication. Jan. 1986, pp. 231–237.

[9] M. A. C. Perryman. "Hipparcos: astrometry from space". In: *Nature* 340.6229 (July 1989), pp. 111–116.

[10] C. P. Escoubet, M. Fehringer, and M. Goldstein. "Introduction: The Cluster mission". In: *Annales Geophysicae* 19 (Oct. 2001), pp. 1197–1200.

[11] J. E. Hirsch. "An index to quantify an individual's scientific research output". In: *Proceedings of the National Academy of Science* 102.46 (Nov. 2005), pp. 16569–16572.

[12] P. Benvenuti. "The IUE Mission". In: *Memoirs of the Italian Astron. Soc.* 54 (Jan. 1983), p. 359.

[13] G. L. Pilbratt, J. R. Riedinger, T. Passvogel, et al. "Herschel Space Observatory. An ESA facility for far-infrared and submillimetre astronomy". In: *Astronomy & Astrophys.* 518 (July 2010), p. L1.

[14] J. A. Tauber, N. Mandolesi, J. -L. Puget, et al. "Planck pre-launch status: The Planck mission". In: *Astronomy & Astrophys.* 520 (Sept. 2010), A1.

[15] K. P. Wenzel, R. G. Marsden, D. E. Page, et al. "The ULYSSES Mission". In: *Astronomy & Astrophys. Supplement* 92 (Jan. 1992), p. 207.

[16] B. G. Taylor, R. D. Andresen, A. Peacock, et al. "The EXOSAT Mission". In: *Space Science Reviews* 30.1-4 (Mar. 1981), pp. 479–494.

[17] E. Kuulkers, C. Ferrigno, P. Kretschmar, et al. "INTEGRAL reloaded: Spacecraft, instruments and ground system". In: *New Astronomy Reviews* 93 (2021), p. 101629.

[18] J. L. Vago, B. Gardini, P. Baglioni, et al. "Science objectives of ESA's ExoMars mission". In: *European Planetary Science Congress 2006*. Jan. 2006, p. 76.

[19] J. Schneider, M. Auvergne, A. Baglin, et al. “The COROT Mission: From Structure of Stars to Origin of Planetary Systems”. In: *Origins*. Ed. by Charles E. Woodward, J. Michael Shull, and Jr. Thronson Harley A. Vol. 148. Astronomical Society of the Pacific Conference Series. Jan. 1998, p. 298.

[20] J. Bergé, P. Touboul, M. Rodrigues, et al. “Status of MICROSCOPE, a mission to test the Equivalence Principle in space”. In: *Journal of Physics Conference Series*. Vol. 610. Journal of Physics Conference Series. May 2015, p. 012009.

[21] S. Tsuneta. “The Hinode Mission”. In: *First Results From Hinode*. Ed. by S. A. Matthews, J. M. Davis, and L. K. Harra. Vol. 397. Astronomical Society of the Pacific Conference Series. Sept. 2008, p. 3.

[22] K.-H. Glassmeier, H. Boehnhardt, D. Koschny, et al. “Rosetta – ESA’s mission to the origin of the solar system”. In: (2009).

[23] S. McKenna-Lawlor, G. Schwehm, R. Schulz, et al. “ESA’s Comet Orbiter Rosetta and Lander Philae”. In: *Outstanding Problems in Heliophysics: From Coronal Heating to the Edge of the Heliosphere*. Ed. by Q. Hu and G. P. Zank. Vol. 484. Astronomical Society of the Pacific Conference Series. May 2014, p. 149.

[24] M. F. Kessler, J. A. Steinz, M. E. Anderegg, et al. “The Infrared Space Observatory (ISO) mission.” In: *Astronomy & Astrophys.* 315.2 (Nov. 1996), p. L27.

[25] J. Peek, V. Desai, R. L. White, et al. “Robust Archives Maximize Scientific Accessibility”. In: *Bulletin of the American Astronomical Society*. Vol. 51. Sept. 2019, p. 105.

## Acronym List

| | |
|---|---|
| **ADS** | Astrophysics Data Service |
| **AO** | Announcement of Opportunity |
| **Co-I** | Co-Investigator |
| **ESA** | European Space Agency |
| **GDP** | Gross Domestic Product |
| **HEASARC** | High-Energy Astrophysics Archive Centre |
| **HST** | Hubble Space Telescope |
| **ISO** | Infrared Space Observatory |
| **IUE** | International Ultraviolet Explorer |
| **MEX** | Mars Express Mission |
| **NASA** | National Aeronautics and Space Administration |
| **PhD** | Doctor of Philosophy Degree |
| **PI** | Principal Investigator |
| **SAO** | Smithsonian Astrophysical Observatory |
| **SOC** | Science Operations Centre |
| **SOHO** | Solar and Heliospheric Observatory |
| **STScI** | Space Telescope Science Institute |
| **URL** | Uniform Resource Locator |
| **VEX** | Venus Express Mission |